\documentclass[10pt,journal,compsoc]{IEEEtran}

  \usepackage{hhline}
\usepackage{slashbox}
\usepackage[singlelinecheck=false]{caption}
\ifCLASSOPTIONcompsoc
  \usepackage[nocompress]{cite}
\else
  \usepackage{cite}
\fi
\usepackage{amsmath, amsfonts, amssymb}
\usepackage{subfigure}
\usepackage{tikz}
\usepackage{soul}
\soulregister\cite7
\soulregister\ref7
\soulregister\pageref7

\newcommand{\bb}[1]{\mathbb #1}
\newcommand{\cl}[1]{\mathcal #1}

\newcommand{\n}{\mathbf n}

\renewcommand{\tt}{\mathbf t}
\renewcommand{\n}{\mathbf n}
\renewcommand{\v}{\mathbf v}
\renewcommand{\u}{\mathbf u}

\DeclareMathOperator{\lpf}{LPF}

\graphicspath{{figures/}{plot_data/}{more/}}

\begin{document}

\title{Sparse Geometric Representation Through Local Shape Probing}

\author{Julie~Digne, 
S\'ebastien~Valette, 
and Rapha\"elle~Chaine 
\IEEEcompsocitemizethanks{\IEEEcompsocthanksitem Julie Digne and Rapha\"elle Chaine are
with Univ Lyon - LIRIS - CNRS UMR 5205 - Universit\'e Lyon 1}
\IEEEcompsocitemizethanks{\IEEEcompsocthanksitem S\'ebastien Valette is with CREATIS -
CNRS UMR 5220 - INSA de Lyon}
  \IEEEcompsocitemizethanks{\IEEEcompsocthanksitem  Corresponding author: Julie Digne (julie.digne@liris.cnrs.fr)}
  \thanks{}}

%
%

\markboth{Journal of \LaTeX\ Class Files,~Vol.xx, No.xx,2017 }%
{Shell \MakeLowercase{\textit{et al.}}: Bare Demo of IEEEtran.cls for Computer Society Journals}
%


\IEEEtitleabstractindextext{%

\begin{abstract}

We propose a new shape analysis approach based on the non-local analysis
of local shape variations. Our method relies on a novel description of shape variations, called Local
Probing Field (LPF), which describes how a local probing operator transforms a pattern onto the shape.
By carefully optimizing the position and orientation of each descriptor, we are able to
capture shape similarities and gather them into a geometrically relevant dictionary over which
the shape decomposes sparsely.
This new representation permits to handle shapes with mixed intrinsic dimensionality (e.g. shapes containing both surfaces and curves) and to encode various shape features such as boundaries.
Our shape representation has several potential applications; here we demonstrate its efficiency
for shape resampling and point set denoising for both synthetic and real data.
\end{abstract}

}

\maketitle

\IEEEdisplaynontitleabstractindextext

%
\IEEEpeerreviewmaketitle


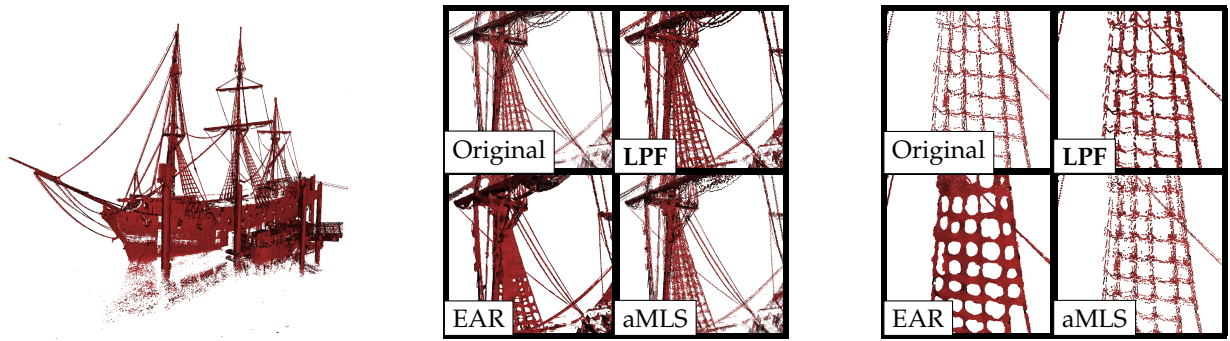
\begin{figure*}
\begin{center}
\begin{tikzpicture}
\setlength{\fboxsep}{0pt}
\setlength{\fboxrule}{2pt}
\node[anchor=south west,inner sep=0] at (0, 0) 
{\includegraphics[width=0.25\linewidth]{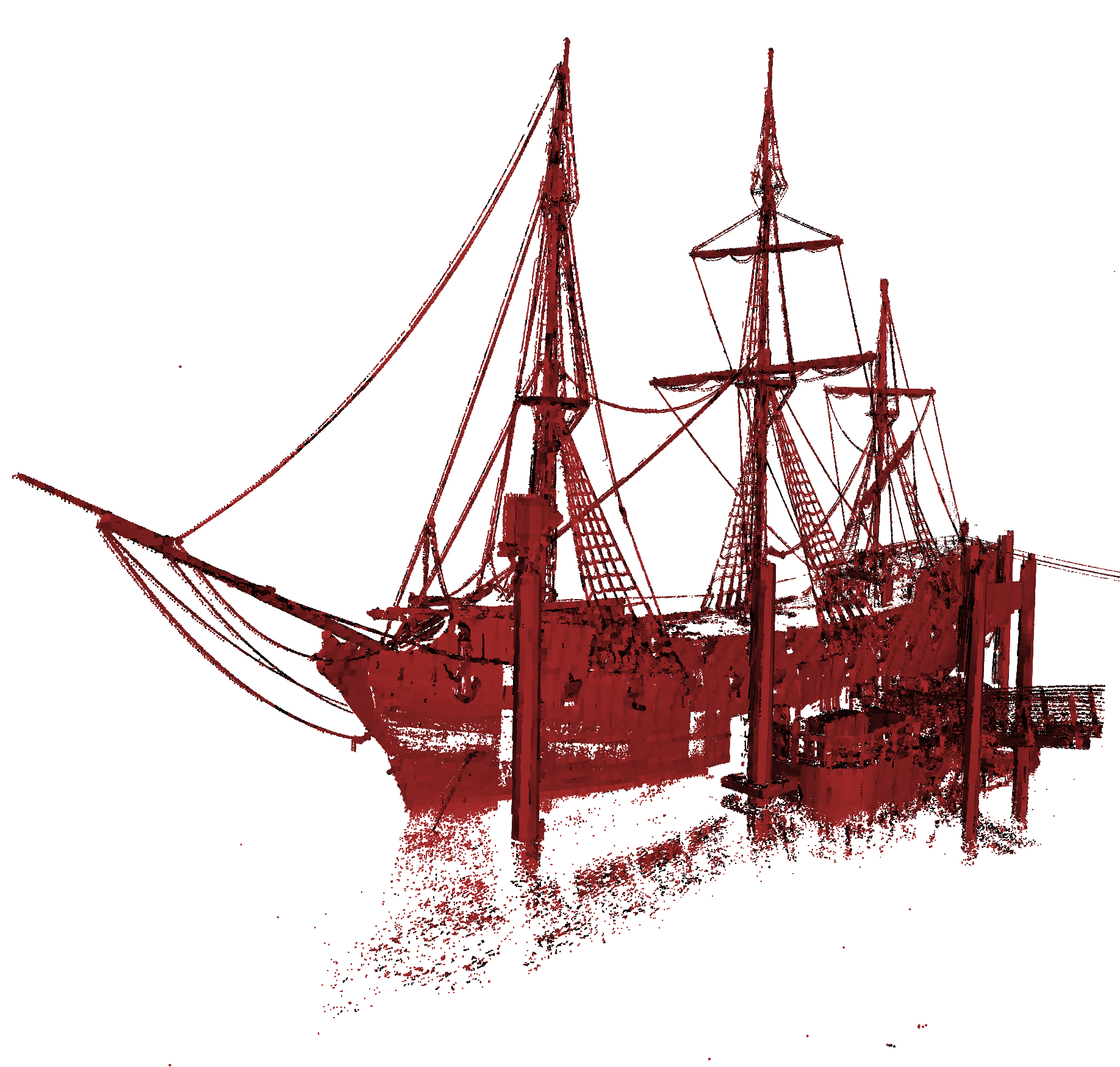}};

\node[anchor=south west,inner sep=0] at (0.32\linewidth, 0)
{\fbox{\includegraphics[width=0.12\linewidth]{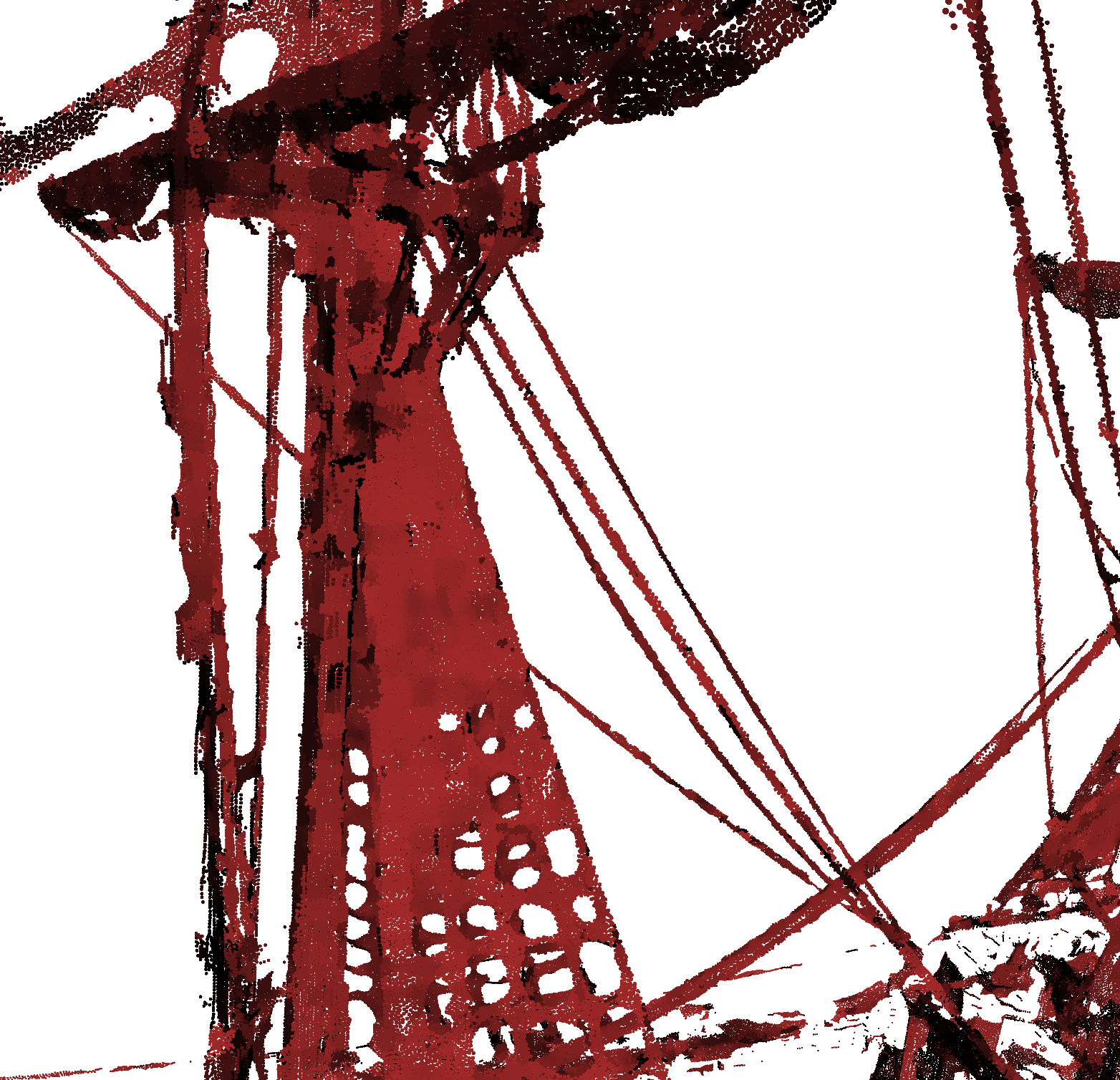}}};
\node[anchor=south west, draw, fill=white] at (0.321\linewidth, 0.002\linewidth) 
{EAR};
\node[anchor=south west,inner sep=0] at (0.32\linewidth, 0.12\linewidth) 
{\fbox{\includegraphics[width=0.12\linewidth]{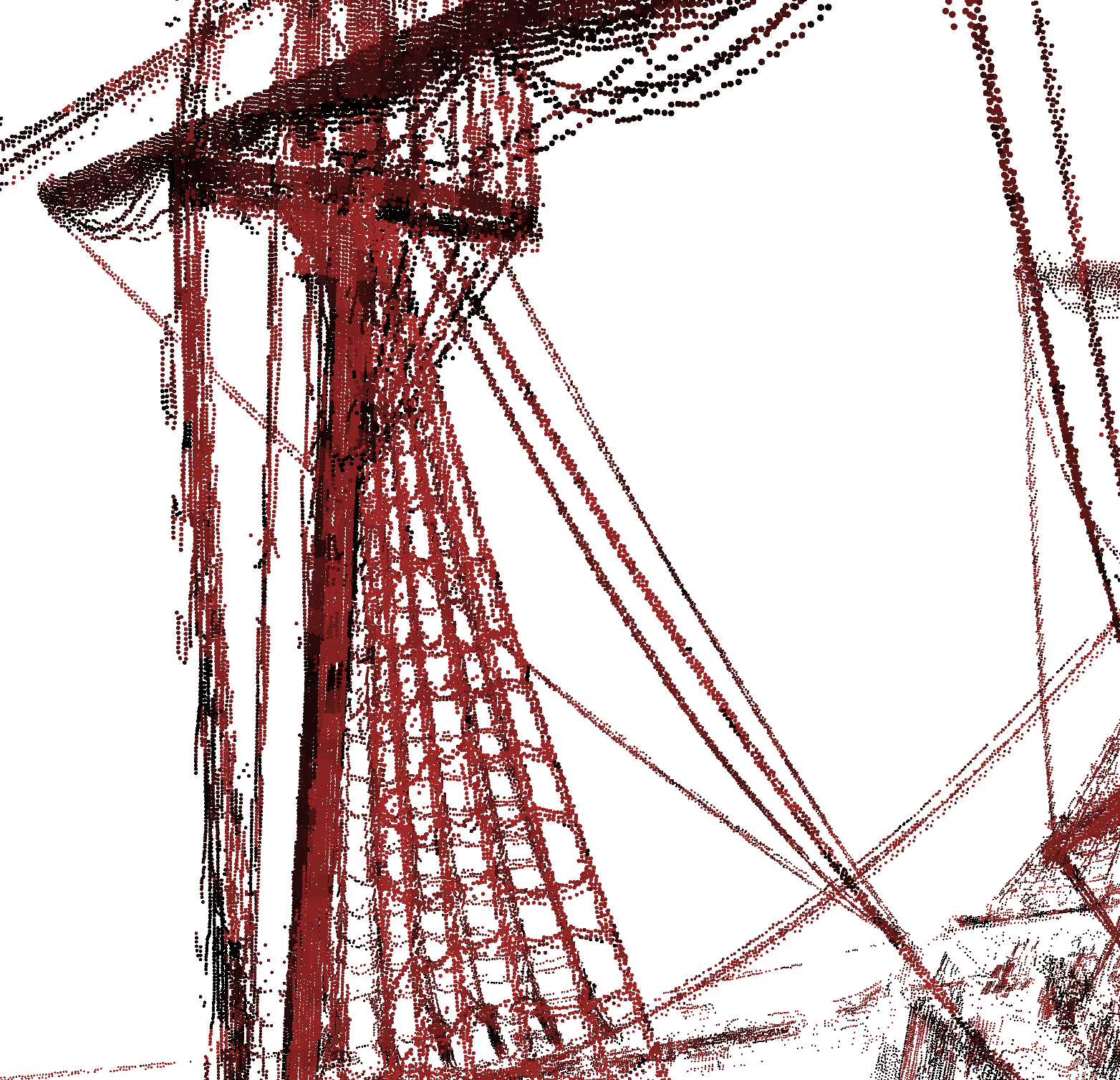}}};
\node[anchor=south west, draw, fill=white] at (0.321\linewidth, 
0.1205\linewidth) {Original};
\node[anchor=south west,inner sep=0] at (0.445\linewidth, 0) 
{\fbox{\includegraphics[width=0.12\linewidth]{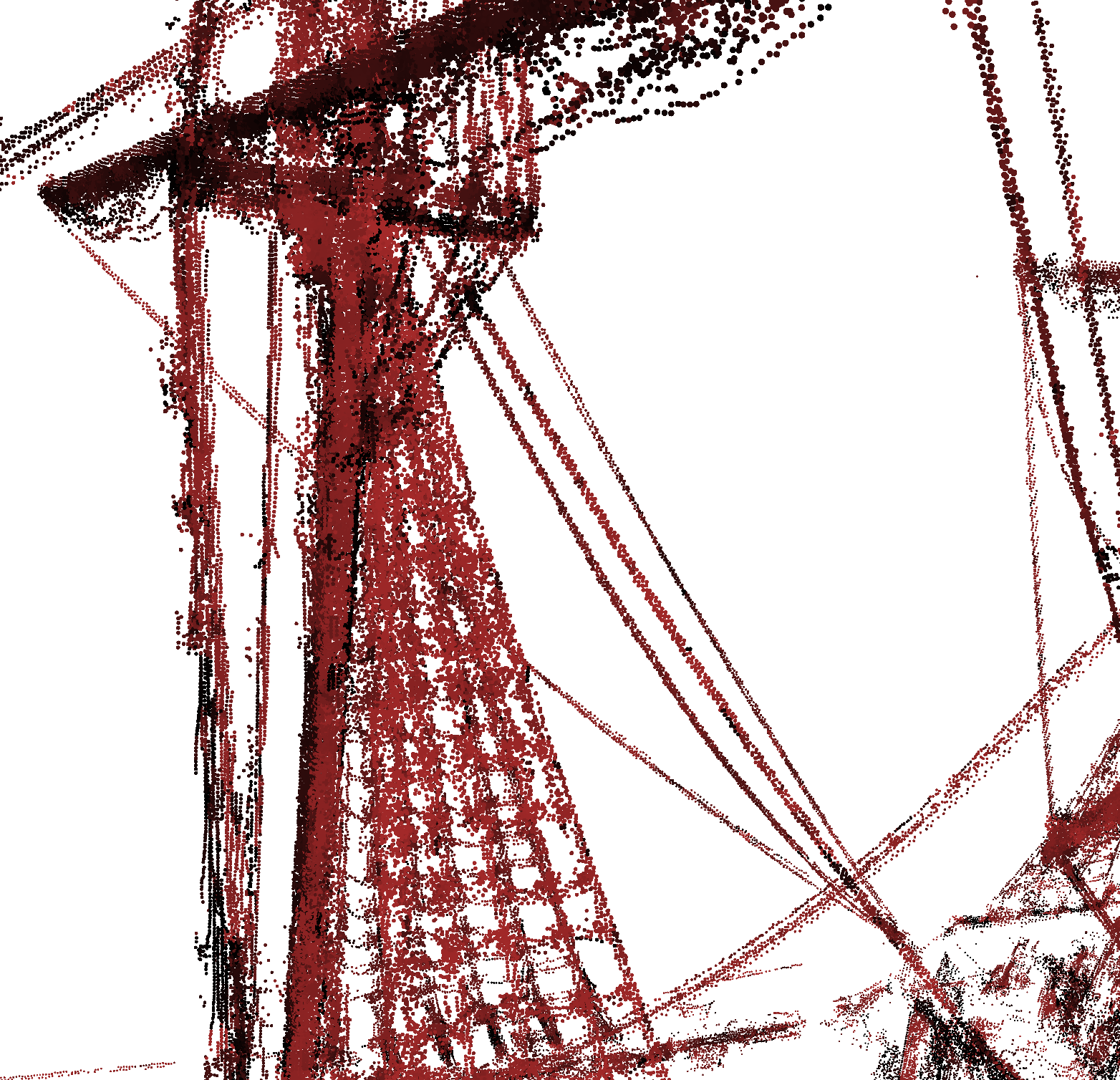}}};
\node[anchor=south west, draw, fill=white] at (0.4451\linewidth, 
0.002\linewidth) {aMLS};
\node[anchor=south west,inner sep=0] at (0.445\linewidth, 0.12\linewidth) 
{\fbox{\includegraphics[width=0.12\linewidth]{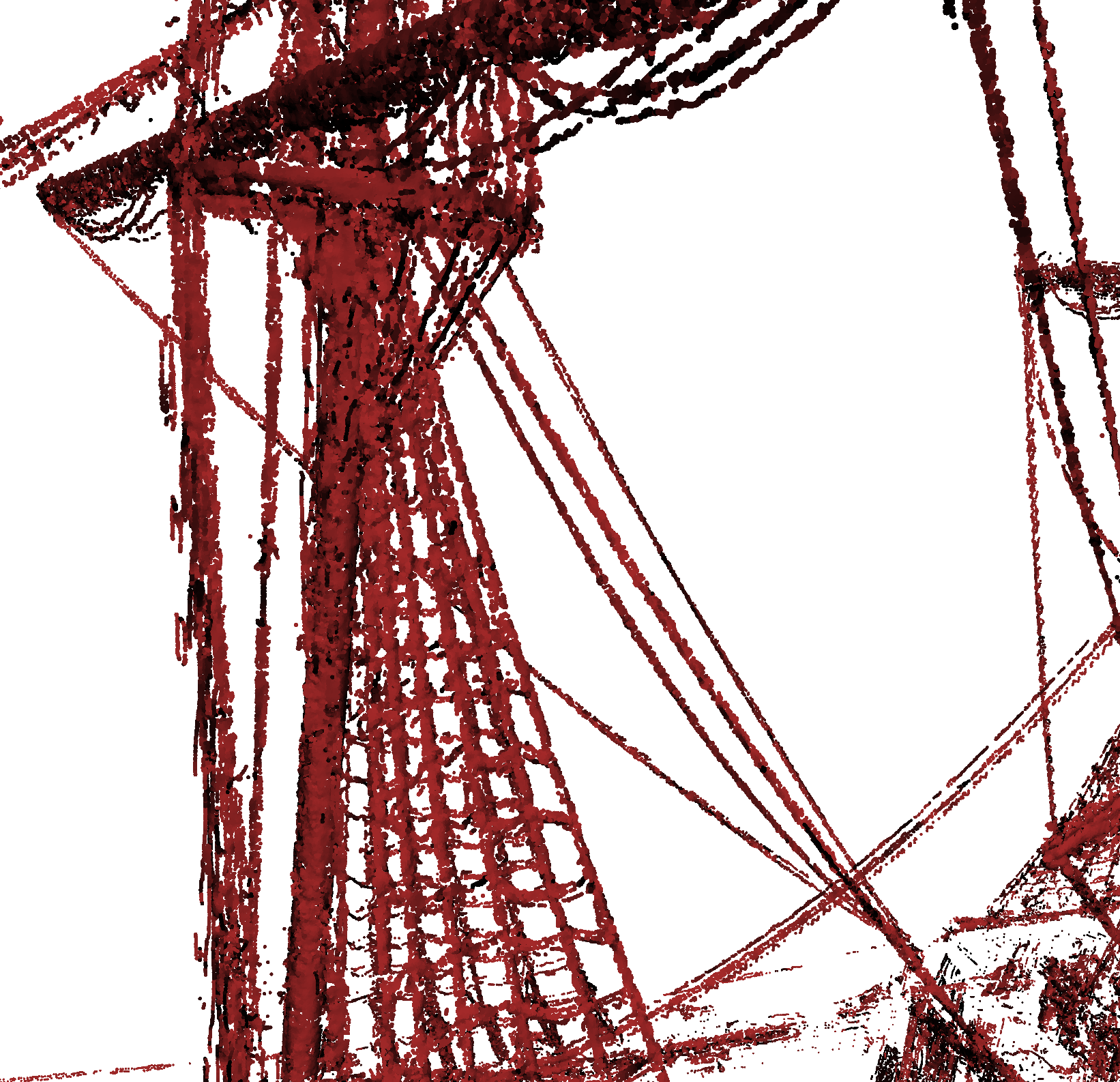}}};
\node[anchor=south west, draw, fill=white] at (0.4451\linewidth, 
0.1205\linewidth) {\textbf{LPF}};

\node[anchor=south west,inner sep=0] at (0.64\linewidth, 0) 
{\fbox{\includegraphics[width=0.12\linewidth]{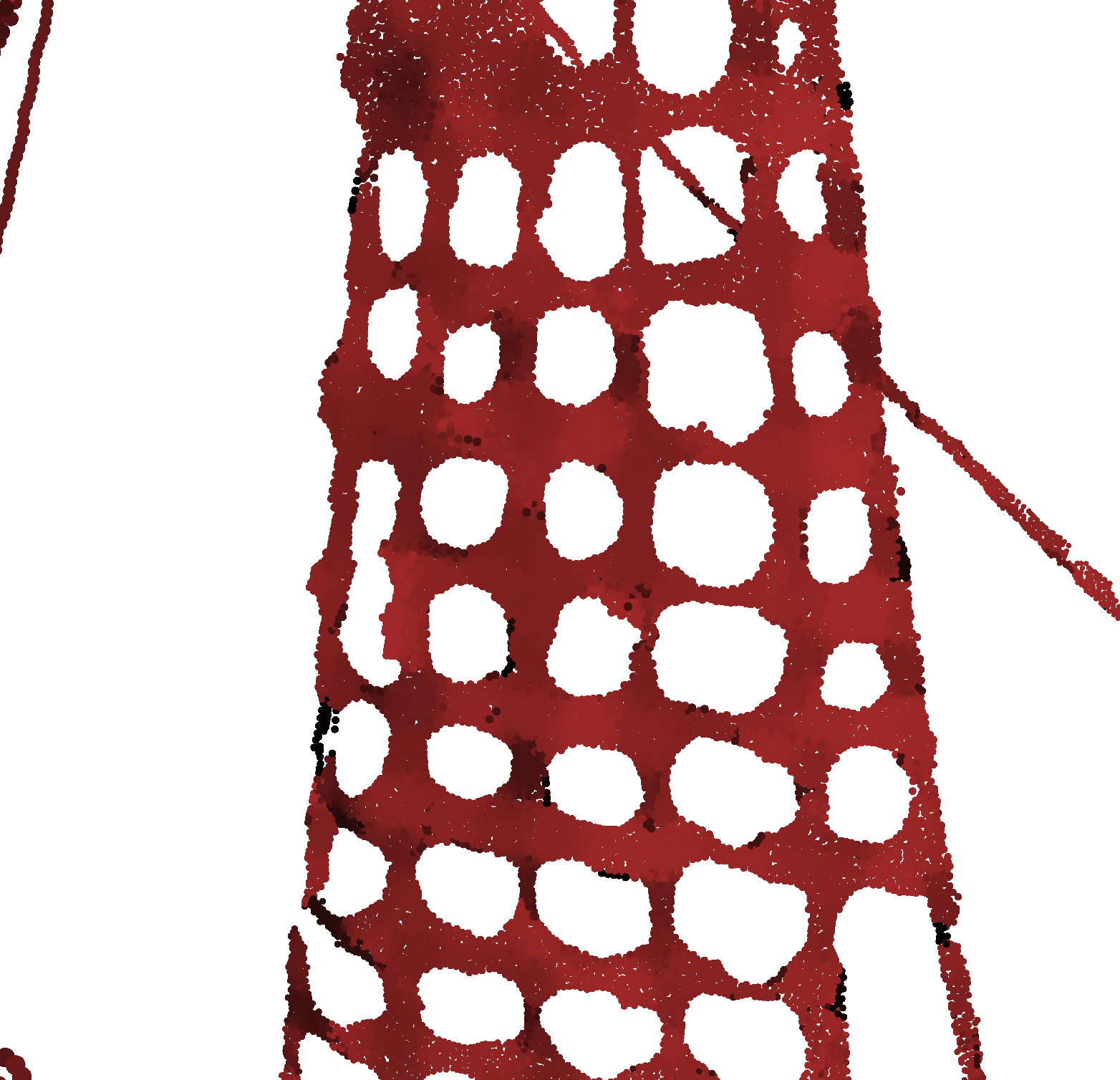}}};
\node[anchor=south west, draw, fill=white] at (0.642\linewidth, 0.002\linewidth) 
{EAR};
\node[anchor=south west,inner sep=0] at (0.64\linewidth, 0.12\linewidth) 
{\fbox{\includegraphics[width=0.12\linewidth]{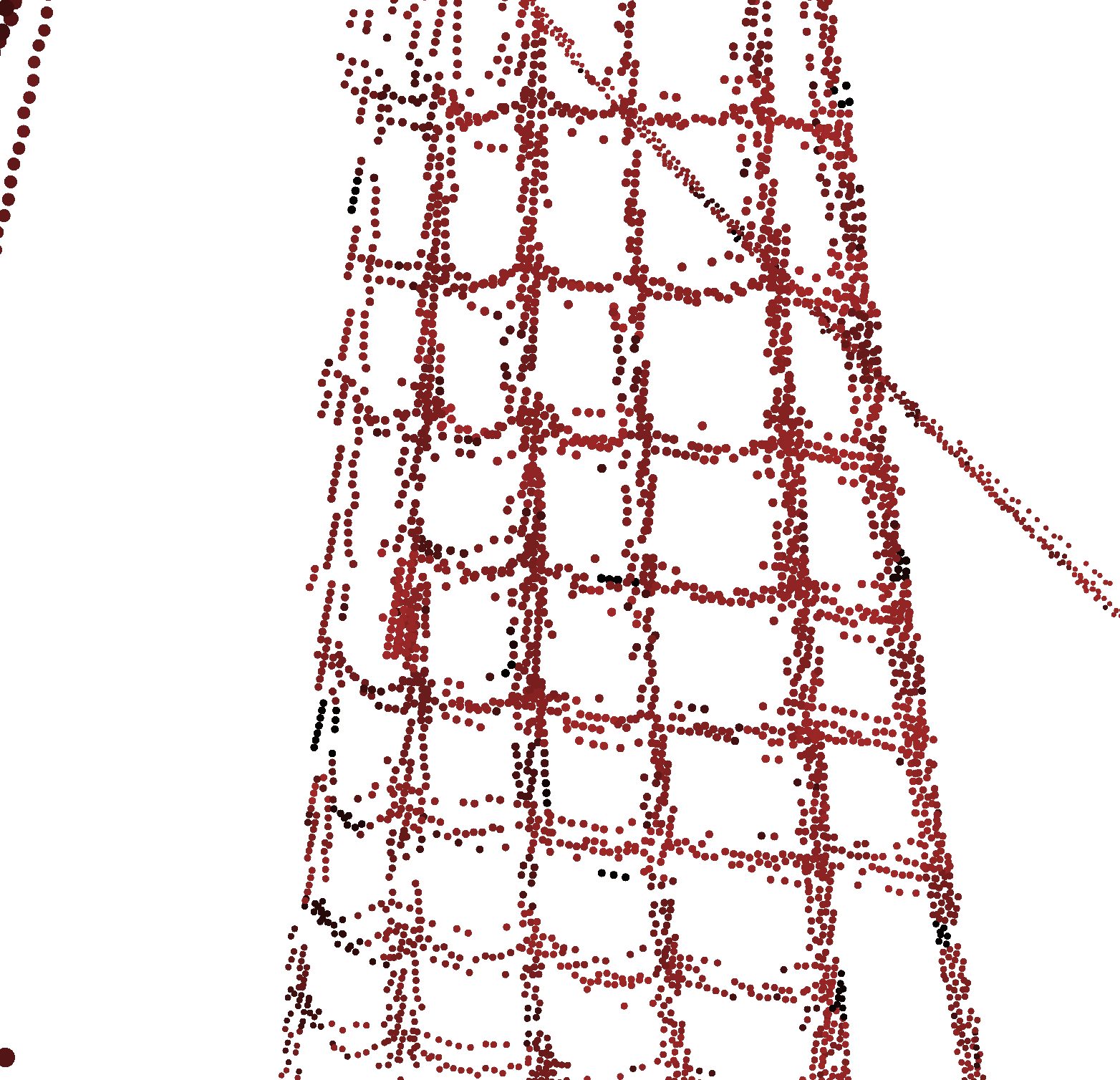}}};
\node[anchor=south west, draw, fill=white] at (0.642\linewidth, 
0.1202\linewidth) {Original};
\node[anchor=south west,inner sep=0] at (0.765\linewidth, 0) 
{\fbox{\includegraphics[width=0.12\linewidth]{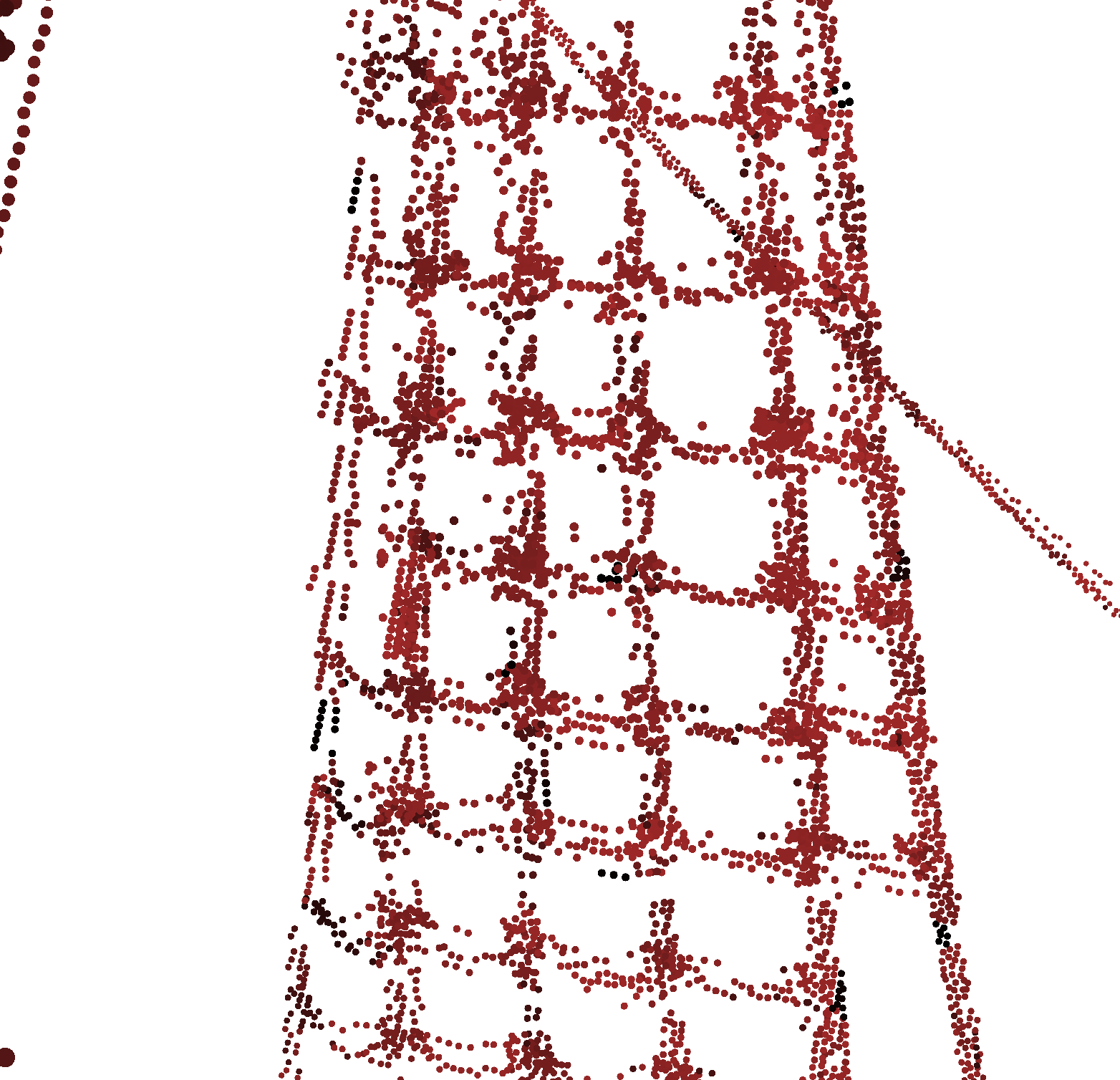}}};
\node[anchor=south west, draw, fill=white] at (0.766\linewidth, 0.002\linewidth) 
{aMLS};
\node[anchor=south west,inner sep=0] at (0.765\linewidth, 0.12\linewidth) 
{\fbox{\includegraphics[width=0.12\linewidth]{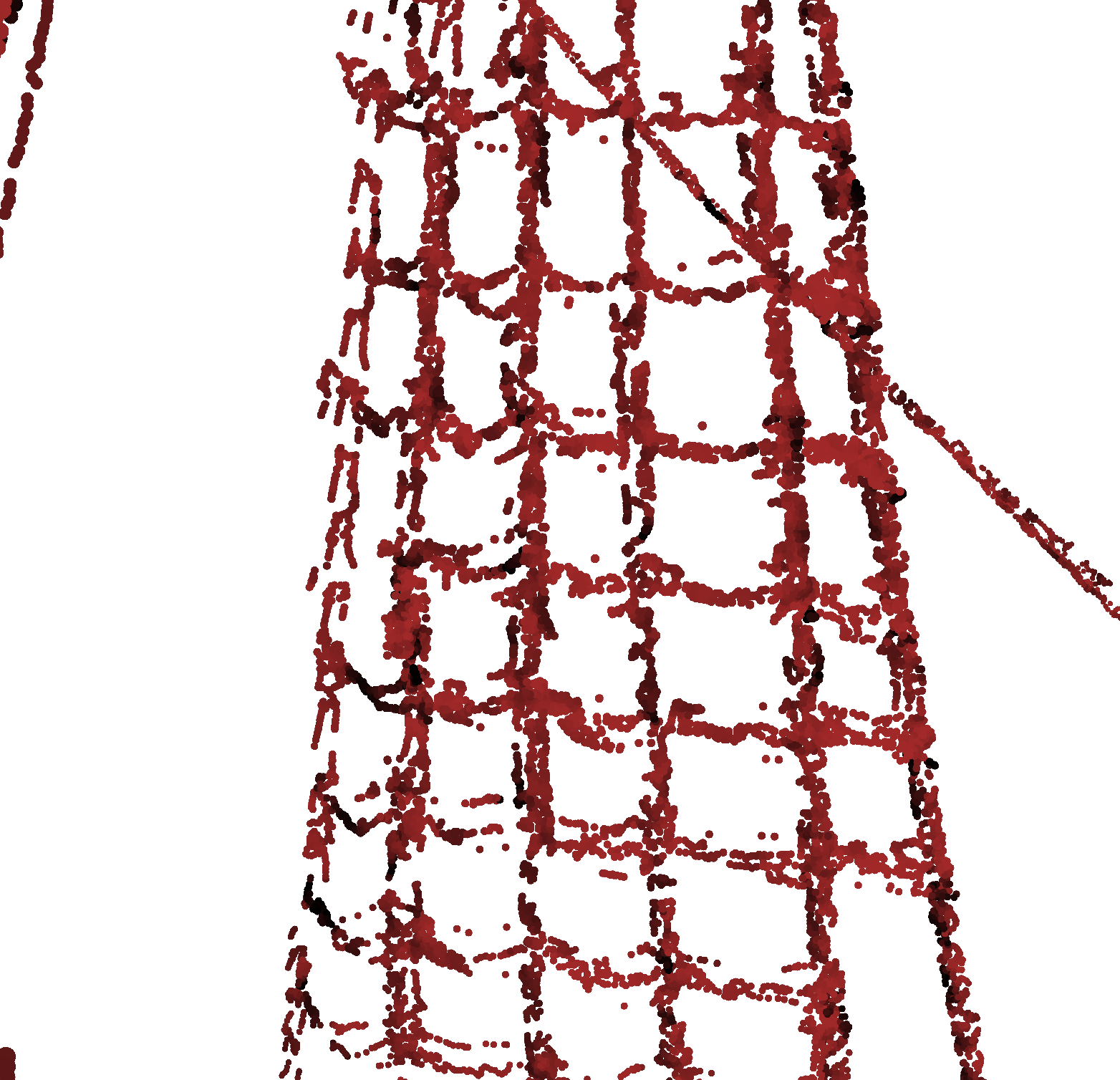}}};
\node[anchor=south west, draw, fill=white] at (0.766\linewidth, 
0.1205\linewidth) {\textbf{LPF}};
\end{tikzpicture}

\caption{Upsampling an ancient ship scanned using LiDaR 
technology. Left: original boat scan (1.1 Million points). Middle and right: comparisons with 
other point set upsampling methods (2.2 Million points): anisotropic MLS and EAR. Our Local Shape Probing analysis 
scheme was able to recover and highlight both curve and surface structures, whereas EAR enlarges curves, and 
    aMLS adds blur. Dataset courtesy of Dorit Borrmann, Jan Elseberg, and Andreas N\"uchter (Jacobs University Bremen). }
\label{fig:ship}
\end{center}
\end{figure*}

\section{Introduction}

Shape analysis is a widely studied topic in Computer Graphics. It is necessary for many different applications such as resampling, denoising, matching, registration.
Furthermore, shapes can be represented in various ways (e.g. point sets provided by 3D scanners or meshes provided by artists) and few methods can address this wide variety of representations. We introduce a statistical method to discover the structures of a given shape by building a dictionary of its local variations yielding a sparse description of the surface. Our method frees itself from the input shape representation format (e.g. mesh or point set) and only needs a probing operator, a tool that associates a point in the ambient space to a point on the local shape. This assumption is very versatile with respect to the shape representation, and can encode manifold surface parts as well as curves. Using the local probing operator, shape variations are represented as Local Probing Fields ($\lpf$) which map instances of a sampling pattern onto the shape. Then, each pattern position and orientation are optimized in order to increase the description efficiency. Finally, by jointly and non-locally analyzing these deformation fields, the shape itself learns its own analysis space, \emph{i.e.} a subspace spanned by a dictionary that best describes it.

\subsection{Related Work}

Exploring images and shapes by looking for structures and repetitions is a fast-developing trend
in Computer Graphics. It can be performed at a larger scale, for discovering shape global
properties through explicit symmetry and structure detection
\cite{symmetry_in_3d_geometry}\cite{discovering_structural_regularity}.
It can also be done at a finer scale to discover and exploit small-scale shape similarities.
This idea, which lies at the heart of our method, is known as non-local analysis and
has been extensively studied in the image processing community.

\noindent\textbf{Non-local analysis.} This principle has been introduced for denoising 2D images
\cite{non_local_algorithm_for_image_denoising}. It processes a pixel using not only its own
neighborhood but also \emph{all} pixels with a \emph{similar} neighborhood, even
if they are distant. Non-local analysis has also been used to analyze images and videos together in \cite{Shechtman07}; 
it has been extended to 3D surfaces for denoising \cite{meshNLmeans}\cite{similarity_based_filtering}, super-resolution \cite{Hamdi2017}, for surface reconstruction
using the famous Point Set Surface framework \cite{non_local_point_set_surfaces}, and surface inpainting
\cite{context_based_surface_completion} by copying similar patches to missing regions.
It has also been exploited in the context of shape resampling and consolidation.
For example, Zheng et al. developed an algorithm for urban scan consolidation using
the repeatability and similarity of urban features \cite{non_local_scan_consolidation}, by means of 
explicit plane fitting and model matching.

\noindent\textbf{Dictionary learning.}
The non-local principle is  strongly related to the dictionary learning
research field: given a set of signals, these methods aim at finding a dictionary as well as a set of coefficients that
best describe the signals.
If those signals really represent similar phenomena,
then few non-zero coefficients will be enough to describe each signal with an
optimized dictionary \cite{Aharon2006}\cite{online_dictionary_learning_for_sparse_coding}.
Dictionary learning can be seen as a way for
the shapes, images or signals to design their own analysis space.
In Computer Graphics, dictionary learning was used for mesh reconstruction
\cite{surf_dico} by considering vertices positions as dictionary atoms over
which the initial point set is decomposed.
Unfortunately, the watertight manifold surface constraint cannot be expressed
as an algebraic problem and needs an ad-hoc construction method.
In this approach, dictionary learning is not used to emphasize similarity but to extract a subset of input points on which a mesh can be constructed, as a piecewise linear approximation of the shape.
Vertices positions are optimized afterwards, but local geometric information is not enhanced as in our case.
Moreover, there is no known efficient way to extend this approach for resampling or denoising, the two applications presented in this paper.
Dictionary learning has also been used for designing shape descriptors in a supervised way~\cite{Litman2014}.
Another dictionary learning application is compression \cite{digne_eg2014} by representing point sets as local height maps
around selected anchors and decomposing them on a dictionary.
But once again the local model is suited for manifold surfaces without boundaries: curves and boundaries will be dilated during decompression.
A complete survey for sparse representations in geometry processing can be found in \cite{Xu2015}.
Representing a point using sparse coefficients on a basis can also be performed by considering only the coordinates (e.g. differential coordinates \cite{diff_coord} or
Laplacian coordinates \cite{laplacian_surf_editing}) which are efficient for surface editing, but are not sampling invariant.

Our work shares some similarity with the manifold reconstruction thread of work \cite{Wang2016}, where local charts are explicitly reconstructed from input meshes to model the manifold surface by optimizing for the compatibility of the representations. Yet our goals and requirements are quite different: we do not require the manifold assumption and will optimize for local representations to better enhance the similarity.

\noindent\textbf{Shape matching and retrieval.}
Surface local representations by descriptors are widely used for registration and matching, such as the Scale-Invariant Feature Transform (SIFT) \cite{lowe2004sift} in the image processing field. Subsequently, many 3D local descriptors have been proposed such as spin images \cite{spin},  SHOT \cite{Tombari2010}, shape contexts \cite{shape_contexts} adapted to 3D \cite{Kokkinos2012},  Mesh-HoG \cite{mesh-hog}, Heat-Kernel Signatures \cite{concise_ms_signature} or descriptors based on Graph Wavelets \cite{Li2016}. These descriptors have been successfully used for 3D scan registration and shape retrieval \cite{shapegoogle}\cite{mattausch14eg}. The shape correspondence problem has also been addressed by Deep Learning approaches \cite{Wei15}\cite{Masci15}\cite{Boscaini16} when the shapes are represented as meshes. However those approaches require large datasets to train the network to be later used on the test data.

\noindent{\textbf{Shape resampling and denoising.}
Many other feature preserving resampling methods have
been investigated, even if they do not explicitly take similarity into account. For
example Huang et al. \cite{Huang2009}\cite{edge_aware_point_set_resampling} proposed
an algorithm to detect sharp features, and resample shapes starting from surface
parts that are distant from the edges and resample gradually
toward the edges in order to better recover them.
This method, called Edge-Aware Resampling (EAR), is built upon the \emph{Locally
Optimal Projection} \cite{lop} and its weighted variant the Weighted Locally Optimal Projection \cite{wlop}, a relaxation method to resample shapes.
Other successful shape resampling techniques build on the Moving Least Squares
approach \cite{pss}\cite{defining_pss}, to resample a shape using local fitted models.
Several improvements to better preserve sharp features have been proposed either
relying \emph{e.g.} on outlier-robust statistics \cite{rmls}, local fitting of
algebraic spheres \cite{algebraic_pss} or outlier-robust kernel regression
\cite{featurepss}.
Consolidation can also be tackled from a more global perspective, for example using a shape skeleton to complete a shape~\cite{deep_points}.
Our method will on the contrary focus on the statistical analysis of local properties.}
Denoising methods have also been introduced for meshes and point sets. \cite{Wang2014} proposed to decouple noise from surface by extracting a smooth surface and denoising the residual (as in \cite{similarity_based_filtering}) using $\ell^1$ decomposition. This efficient denoising technique however relies on Laplacian decomposition, which is computationally intractable for large point sets.
Similarly \cite{l0denoising} proposes to denoise a point set by first denoising the normals based on the $\ell^0$ norm and then denoising the points positions using the estimated normals. This type of denoising is particularly well suited for piecewise planar surfaces. Such a two-step strategy was also adapted in a Total Variation framework \cite{Zhang15}. Other approaches include explicit feature detection before denoising \cite{Lu16}, or taking advantage of both facet and vertex normals in the mesh case \cite{Wei15b}.

\begin{figure*}[ht]
\centering
\includegraphics[width=0.325\linewidth]{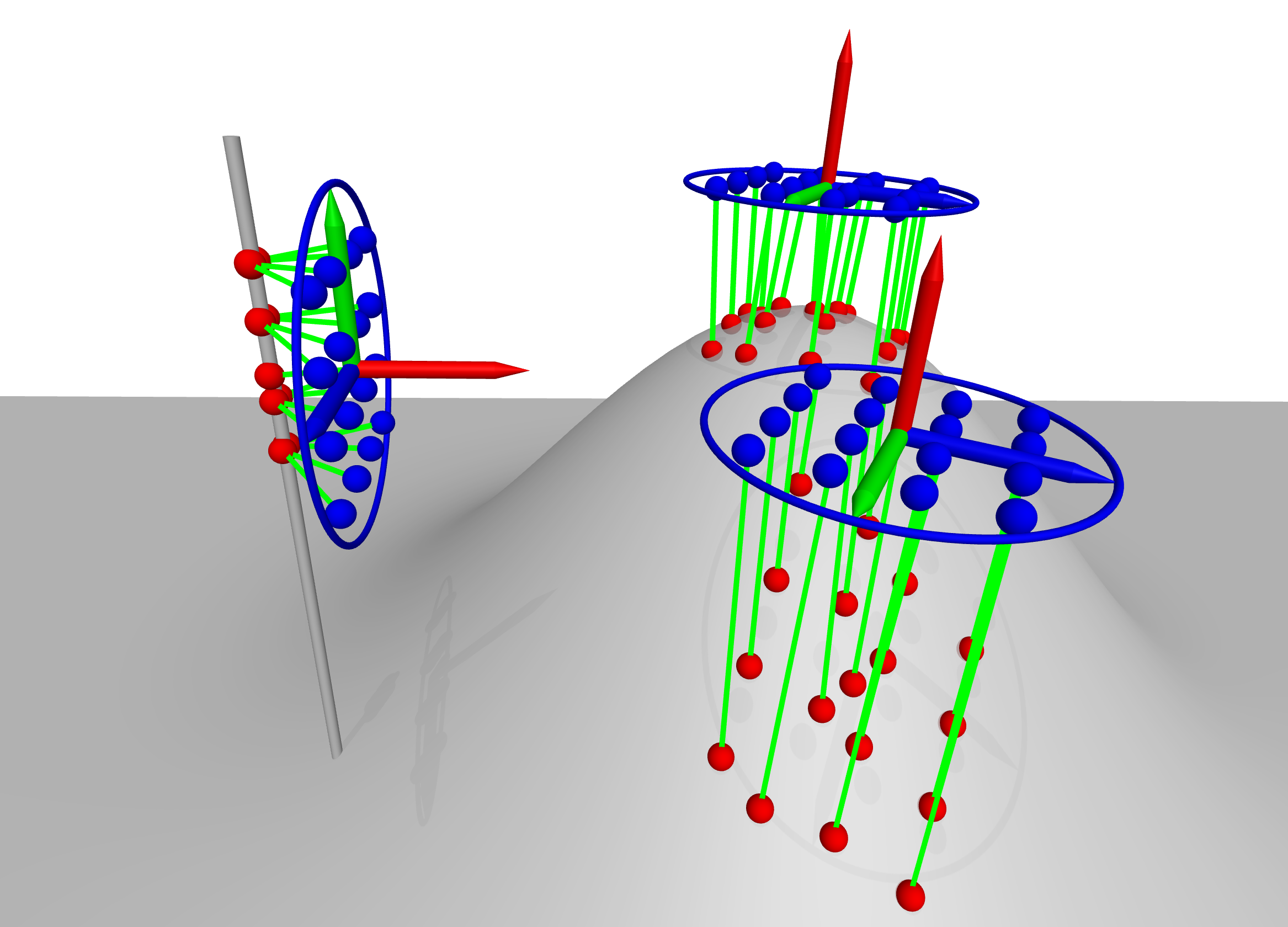}
\hfill
\includegraphics[width=0.325\linewidth]{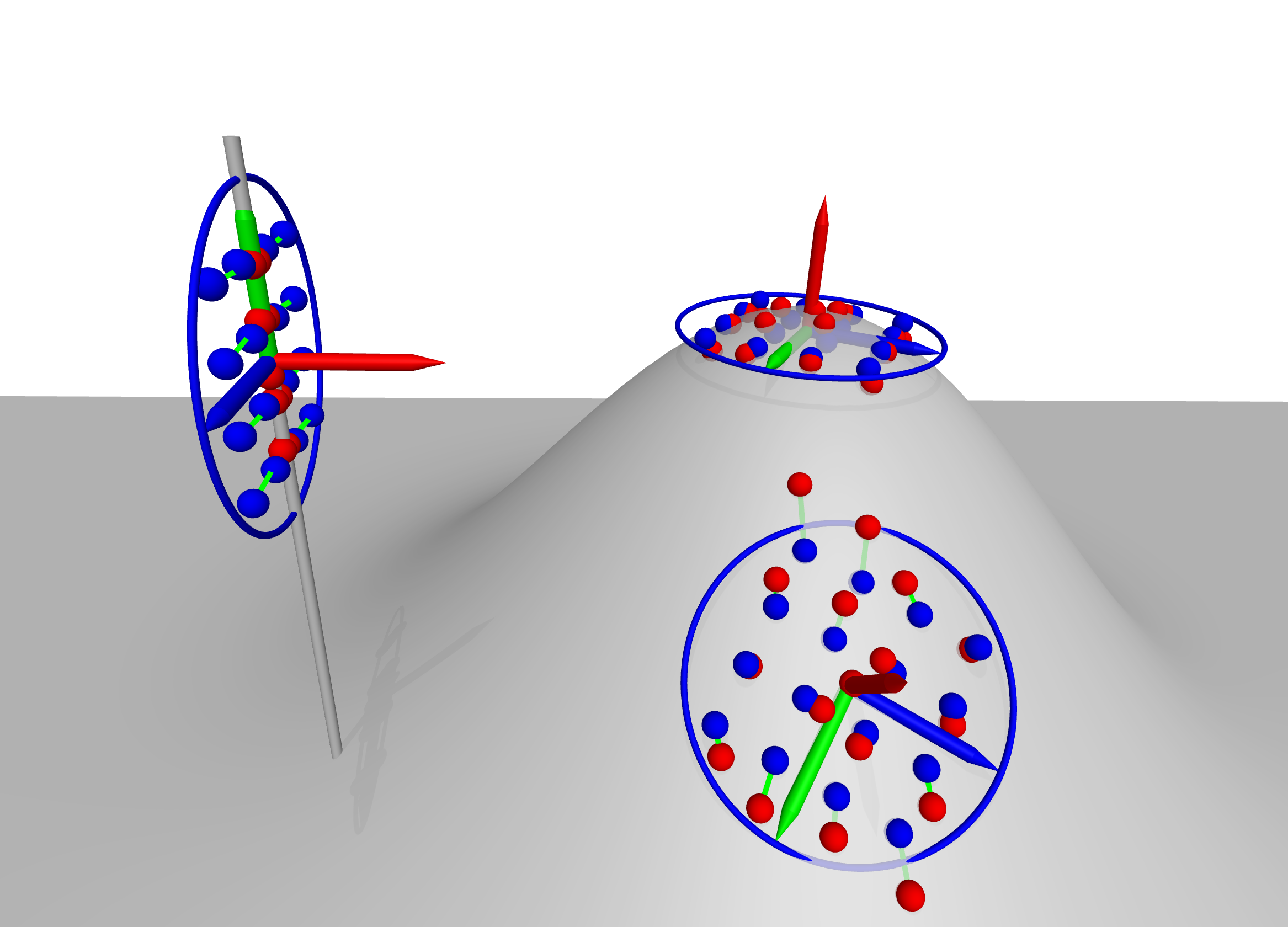}
\hfill
\includegraphics[width=0.325\linewidth]{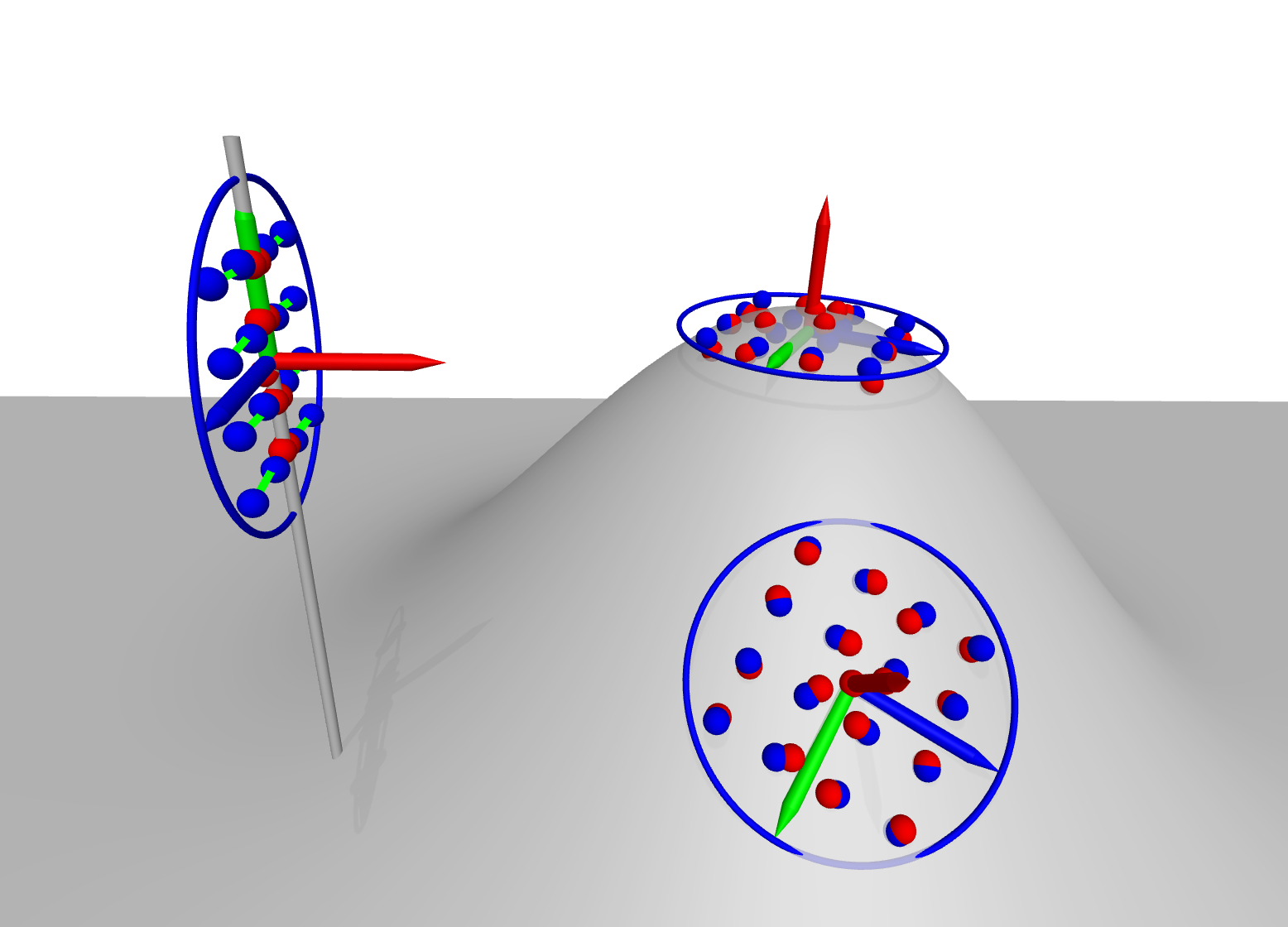}

\caption{Three $\lpf$s in the vicinity of a shape. $\lpf$s can be initialized from any position and
orientation of the pattern. Left: initial position. Middle: after position and rotation
optimization. Right : after
projection on the shape.}
\label{fig:lpf}
\end{figure*}

\subsection{Contributions}

We propose a shape analysis framework that reveals the shape similarities but also its local dimensions.
To do so we introduce a way to represent the shape by a set of deformation fields that represent the
deformation of a pattern onto the local shape regardless of its local dimensionality (e.g. 2-manifolds, curves, boundaries).
We call these representations Local Probing Fields ($\lpf$).
We exploit the idea that if the patterns are positioned optimally all over the shape,
the deformations fields can be compared with each other in a joint statistical learning process.
The analysis space is changed: instead of studying shape variations in the traditional $\bb R^3$
space,
we study them in the space of deformation fields.
A joint analysis of these deformation fields allows to extract a dictionary over which the shape
can be decomposed sparsely, thereby enhancing the structures and similarities of the shape.

The main contributions of this paper are the following:
\begin{itemize}
 \item A local descriptor based on a probing operator that can handle shapes with heterogeneous intrinsic dimensionality.
 \item A novel algorithm to discover local similarities in a shape.
 \item The construction of a geometrically relevant shape dictionary.
\end{itemize}

\section{Local Probing Fields}\label{sec:lpf}

Our framework consists in the construction, analysis and exploitation of deformation fields that
are defined locally in the ambient space.

\subsection{Definitions}
\begin{figure}
\centering
\includegraphics[width=0.4\linewidth]{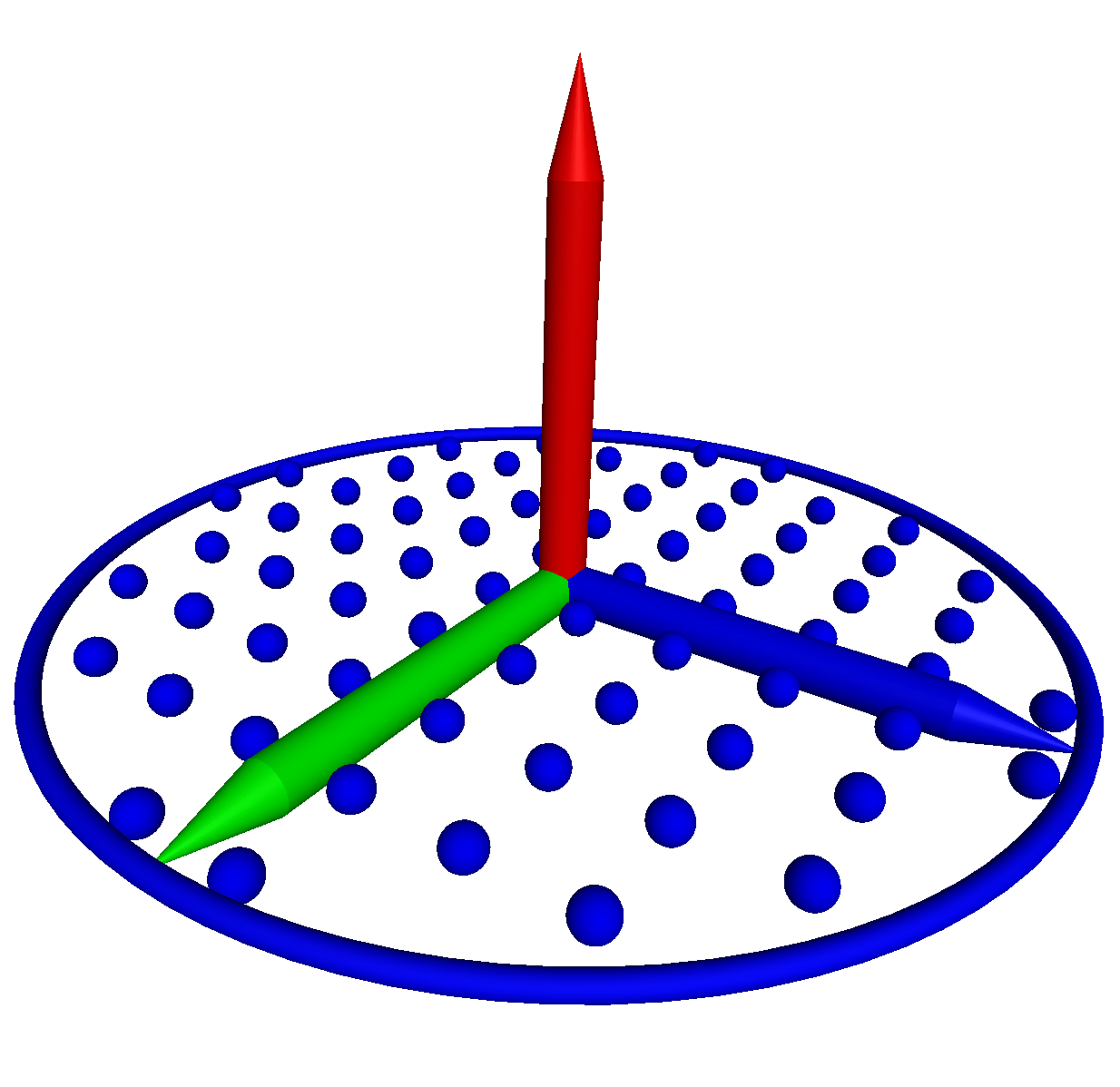}
\hspace{0.1\linewidth}
\includegraphics[width=0.4\linewidth]{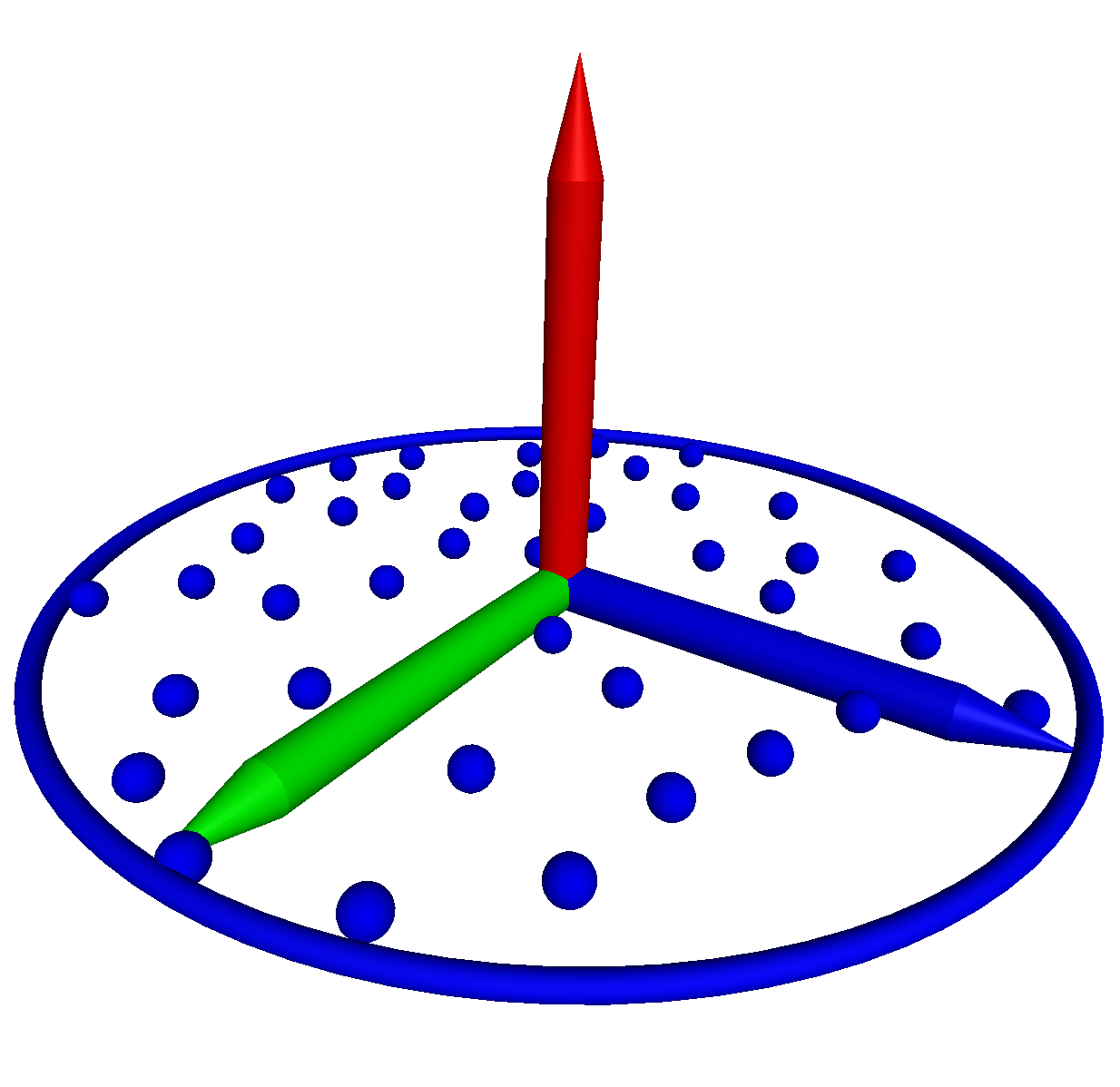}
 \caption{Pattern examples: points regularly or randomly sampled on
a planar disk.}
\label{fig:generic_patterns}
\end{figure}

The principal ingredient for building a $\lpf$ is a \emph{Probing Operator} $\cl P$. In its most
general form, it is defined as an operator that, to each point $x$ in the ambient space assigns a
point of the shape near $x$. We will see in next section how a specific probing operator can
be designed.
A $\lpf$ also requires a sampling pattern, a set of points,
centered around a seed point $s$ and oriented according to an orthogonal frame $(\tt_1,\tt_2,\n)$.
It can be made of points sampled on a surface, as shown on figure \ref{fig:generic_patterns}.
These points are expressed as offset vectors $(\u_{i})_{i=0\cdots M-1}$ from the seed to the points.
There is no strong constraint on the dimensionality of the pattern, even if it seems reasonable to
choose a dimension at least equal to the largest dimension of the input structures. Moreover, the pattern is also free from any regularity contraint. Finally, each $\lpf$ accounts for only a part of 
the shape depending on its initial location. This shape area will be refered to as \emph{target area}.

Let us now consider a pattern centered at a point $s$ on the shape $\cl S$, and aligned
with the local orthogonal frame $(\tt_1,\tt_2,\n)$. Using the probing operator $\cl P$, each point of the pattern is assigned to a point of the target area. Then, the \emph{Local Probing Field centered at $s$} is the displacement of each point of the pattern to its image under $\cl P$. Let us define:
\begin{equation}
\lpf(s) = \{\v_{i}\}_{i \in 0 \cdots M-1}
\end{equation}
where $\v_i$ is a $\bb R^3$ vector such that $\v_i = \cl P(s+\u_i) - (s+\u_i)$.
Hence, a $\lpf$ is a vector field encoding the deformation of the pattern
points onto the shape, without any smoothness requirement or model for
the deformation field.
Thus, instead of having a prior model, we will encode the shape as a set of
local transformations and solely work on these.

\begin{figure*}[ht]
\begin{center}
\includegraphics[width=0.3\linewidth]{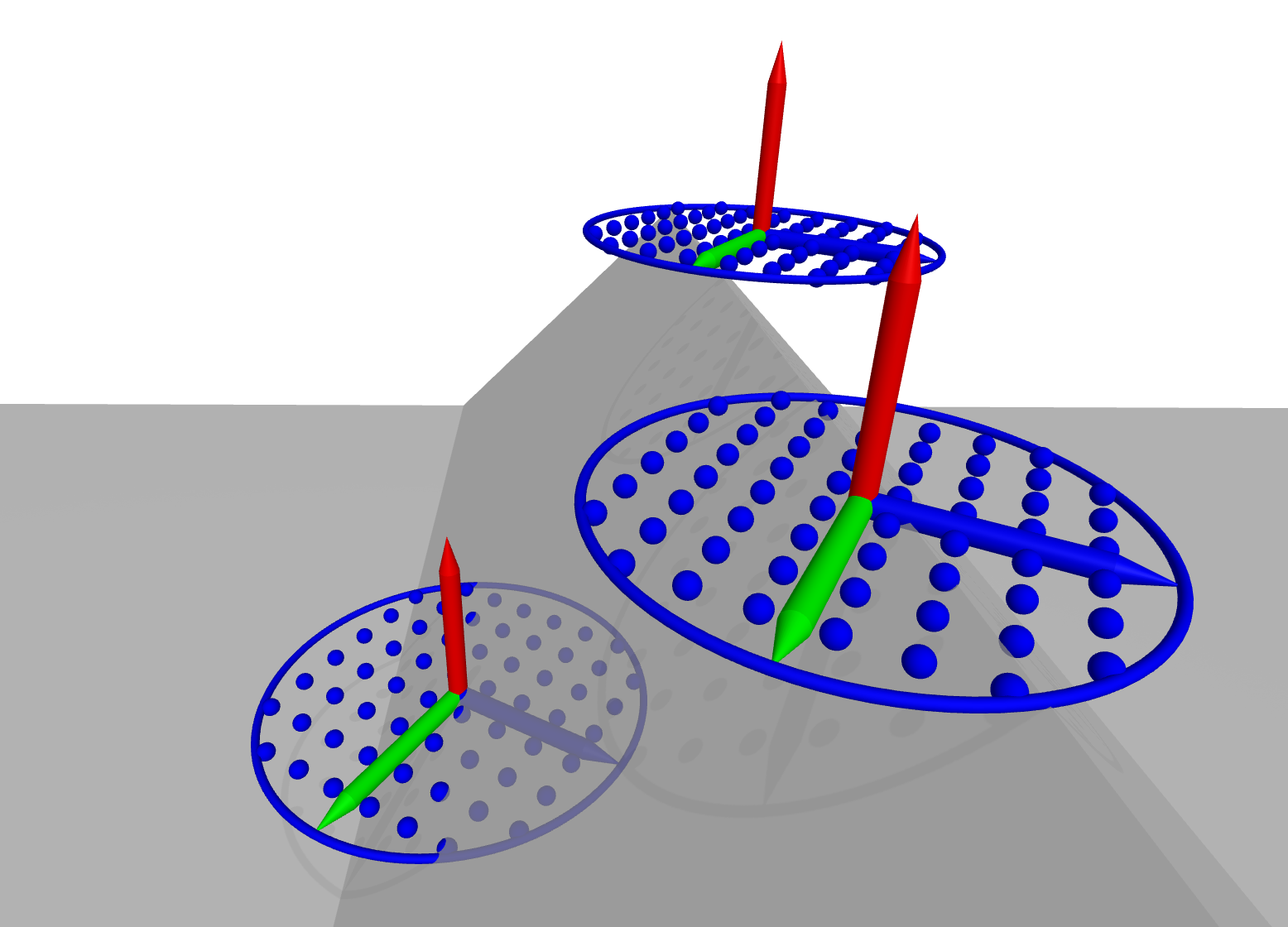}
\hfill
\includegraphics[width=0.3\linewidth]{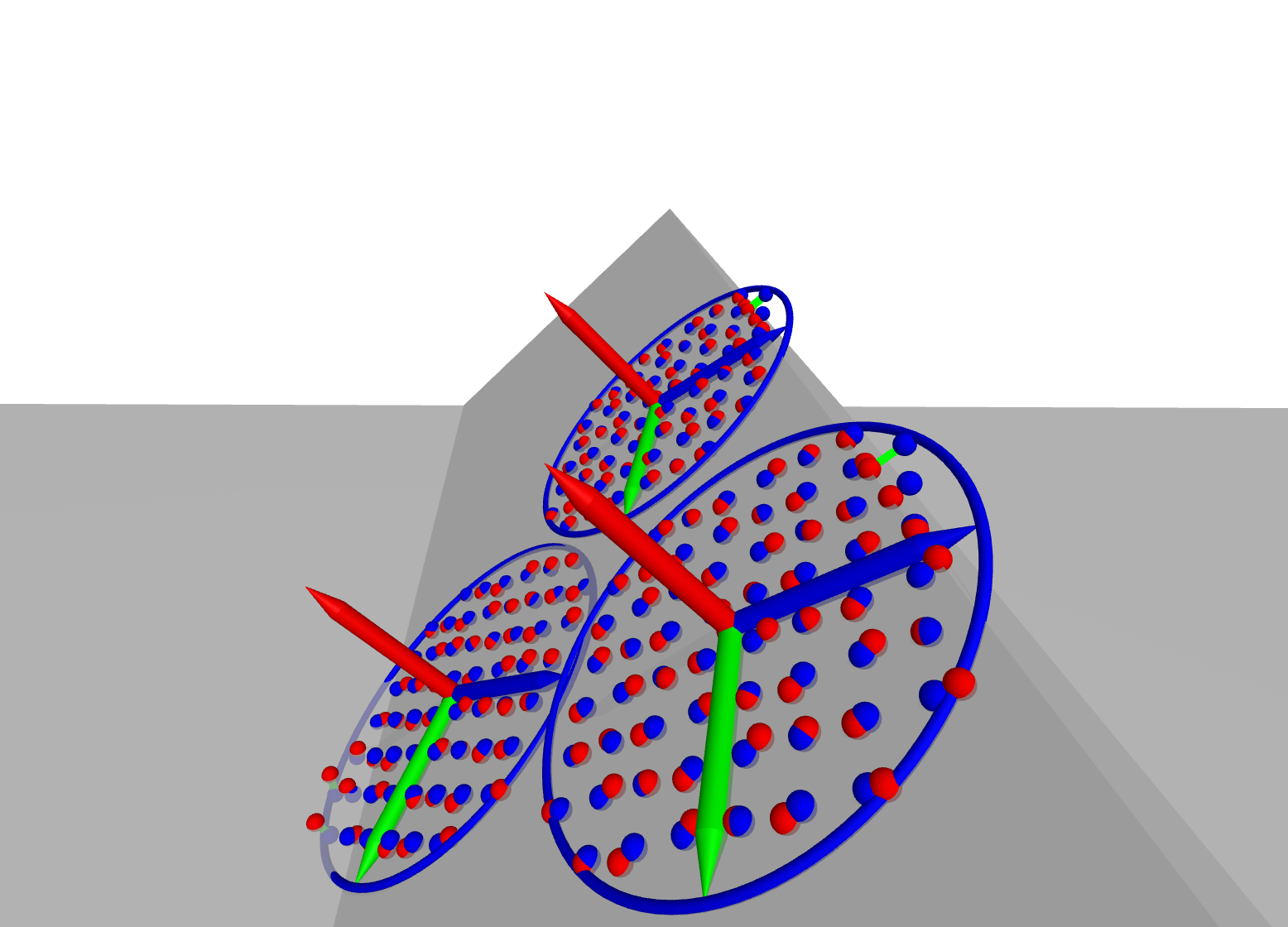}
\hfill
\includegraphics[width=0.3\linewidth]{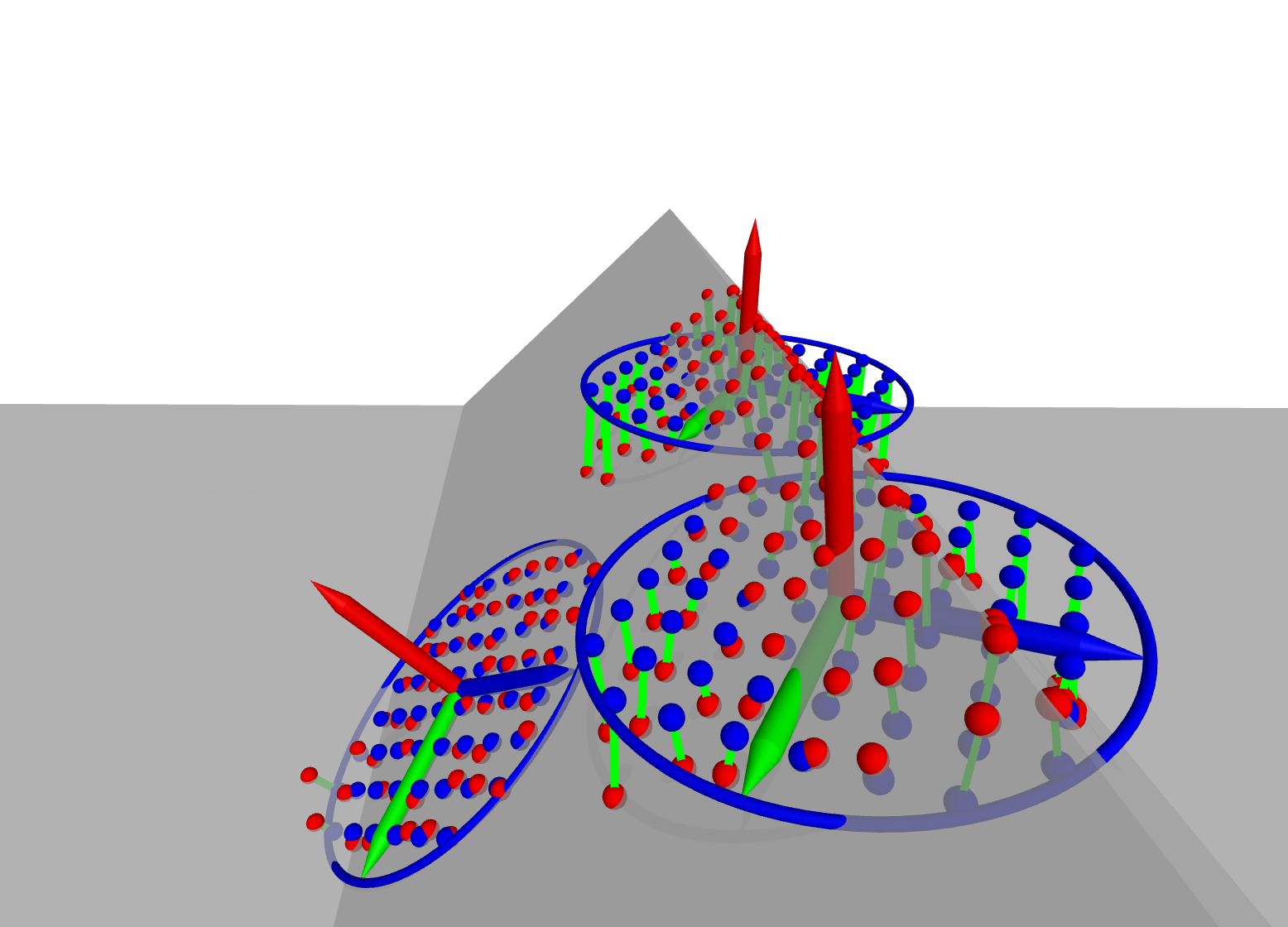}
\end{center}
\caption{Capturing sharp features. Left: initial $\lpf$s positions. Middle: optimization with the nearest neighbor projection causing $\lpf$s to move away from sharp features. Right: optimization using our probing operator, $\lpf$s remain close to their initial target area}
\label{fig:lpf_capture}
\end{figure*}

\subsection{As-Orthogonal-As-Possible LPFs}
\label{sec:AOAP}

Given a pattern and a local probing operator, a $\lpf$ is entirely
determined by its orientation and position. The challenge is to ensure that the
resulting representation is efficient enough for the further joint analysis to
reveal the shape similarities.

In the case of manifold surfaces, height maps come to mind as a natural and efficient way to
represent local surface variations, as height maps contain only geometric information.
In our far more general case, we cannot assume that the shape is a manifold, since we want to infer
its dimensionality by the learning process. Ideally the $\lpf$ should be similar to a height map near a locally manifold surface, but
should also handle other cases such as boundaries and curves.
Therefore we propose to render the $\lpf$ as-close-as-possible to a height map,
by maximizing the
orthogonality of the $\v_i$ vectors with respect to the pattern support plane $(\tt_1,\tt_2)$.
Intuitively, when the surface is manifold, the plane parameterized by $(\tt_1,\tt_2)$ should be close
to the regression plane.
With this goal in mind, several probing operators can be devised, depending on the
input description of the shape. One can use a simple orthogonal projection on the nearest triangle
if the shape is represented by a triangular mesh or a Newton-Raphson projection operator if the
shape is represented by an implicit function. In case of point sets, the possible projections range
from nearest point projection to variants of the Moving Least Squares (e.g.
\cite{pss}\cite{algebraic_pss}).

We use a different probing operator aiming at creating as-orthogonal-as-possible $\v_i$s:
each point $p$ of the pattern is assigned to the point $q$ of the target area whose
projection $q^\prime$ on the pattern plane is the closest to $p$. As shown on figure
\ref{fig:lpf}, this probing operator captures well both curves (as would a nearest
point projector) and manifold surface parts (as would a height map computation).

Moreover, $\lpf$s are generically defined with an arbitrary orientation.
We can then optimize each $\lpf$ position and orientation to better fit its target area.
Unfortunately, using $\lpf$s orthogonality as the sole criterion during this optimization would
result in an ill-defined problem. Therefore we propose to minimize instead the norms of the $\v_i$s and
alternate the two following steps for each $\lpf$ independently:
\begin{itemize}
 \item Compute $\v_i = \cl P(s+\u_i) - (s+\u_i), \forall i\in 1\cdots M$. Due to our choice
of $\cl P$, $\v_i$s are close to orthogonal to the pattern.
 \item Find the rotation and translation of the local frame and pattern minimizing the
squared norms of the $\v_i$s: $\min \sum_i \|\v_i\|^2$.
\end{itemize}

If the chosen probing operator is the nearest point projector, then the
method described above is the well known Iterative Closest Point (ICP) registration \cite{icp}
between the pattern and the shape. Using the as-orthogonal-as-possible probing operator yields a variant
of ICP. An interesting side effect of this approach is that, after optimization, $\lpf$s remain close to
their initial target area, whether they are located on geometric features or not. On the other hand,
nearest-neighbor-based optimization tends to move $\lpf$s away from sharp features, as shown in figure
\ref{fig:lpf_capture}.

\section{LPF-based Shape Analysis} \label{sec:analysis}

Local Probing Fields capture local variations of the shape. These variations can be geometrical and
topological. If the set of $\lpf$s covers the whole shape, it is possible to analyze them
jointly and thus learn the shape similarities. This joint learning process is partially based
on dictionary learning.
Describing the shape in the space of deformation fields, changes \textit{de facto} the space where
the shape analysis is performed.

\subsection{Initial LPF positions}

A set of $\lpf$s is built at arbitrary locations around the shape, with arbitrary frame orientation.
The only constraint on this initial $\lpf$ set is that the set of target areas should provide
a possibly overlapping covering of the shape. For example, a pattern can be initially positioned on
a sampling of the shape
(e.g. each vertex in the case of a mesh or a point set) with a random orientation. They can also be
aligned with principal directions if they are defined, which is not the case for curves.
To illustrate the efficiency of our framework, we distribute $\lpf$s on the shape using a Poisson
sampling process and use random initial orientation.
These positions and orientations are further improved using the minimization
defined in section \ref{sec:AOAP}.

In practice, the target area of a given $\lpf$ is set as the intersection of a sphere centered at $s$
with the shape. The radius of this sphere is set to be slightly larger than the pattern radius
in order to relax the position of the LPF with respect to the target area while
still ensuring that the information is well represented. In our experiments, we chose $s = 1.1r$.
With this definition, $\lpf$s should be located near the shape to have non-empty target areas.
As a consequence, although positions and orientations of the $\lpf$s evolve during the analysis, the
target area of each $\lpf$ remains unchanged.

\begin{figure*}[ht]
\begin{minipage}[c]{.34\linewidth}
\centering
\includegraphics[width=\linewidth]{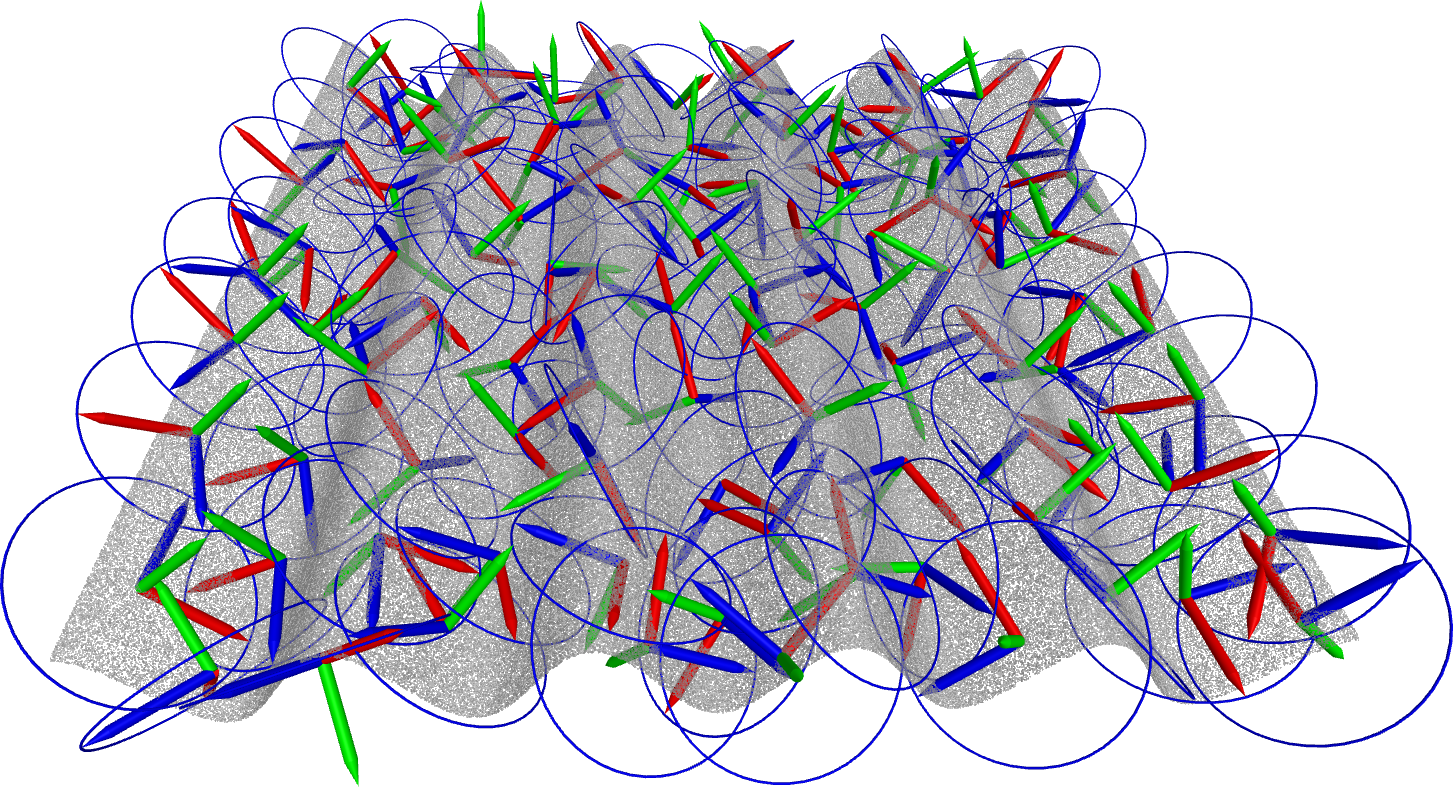}
\includegraphics[width=0.7\linewidth]{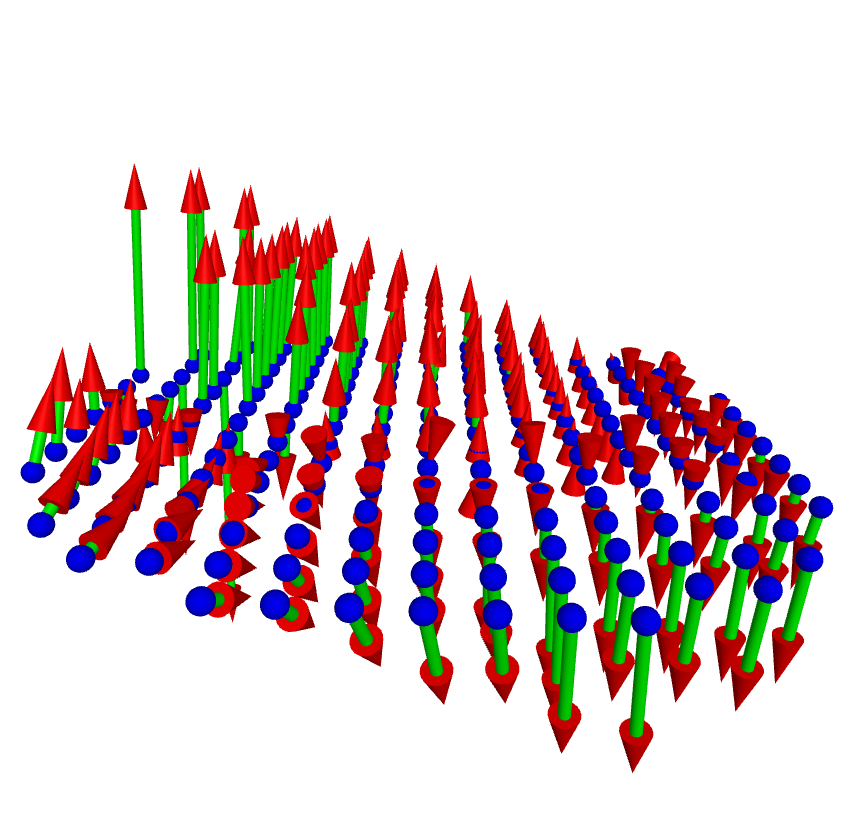}
\end{minipage}
\hfill
\begin{minipage}[c]{.32\linewidth}
\centering
\includegraphics[width=\linewidth]{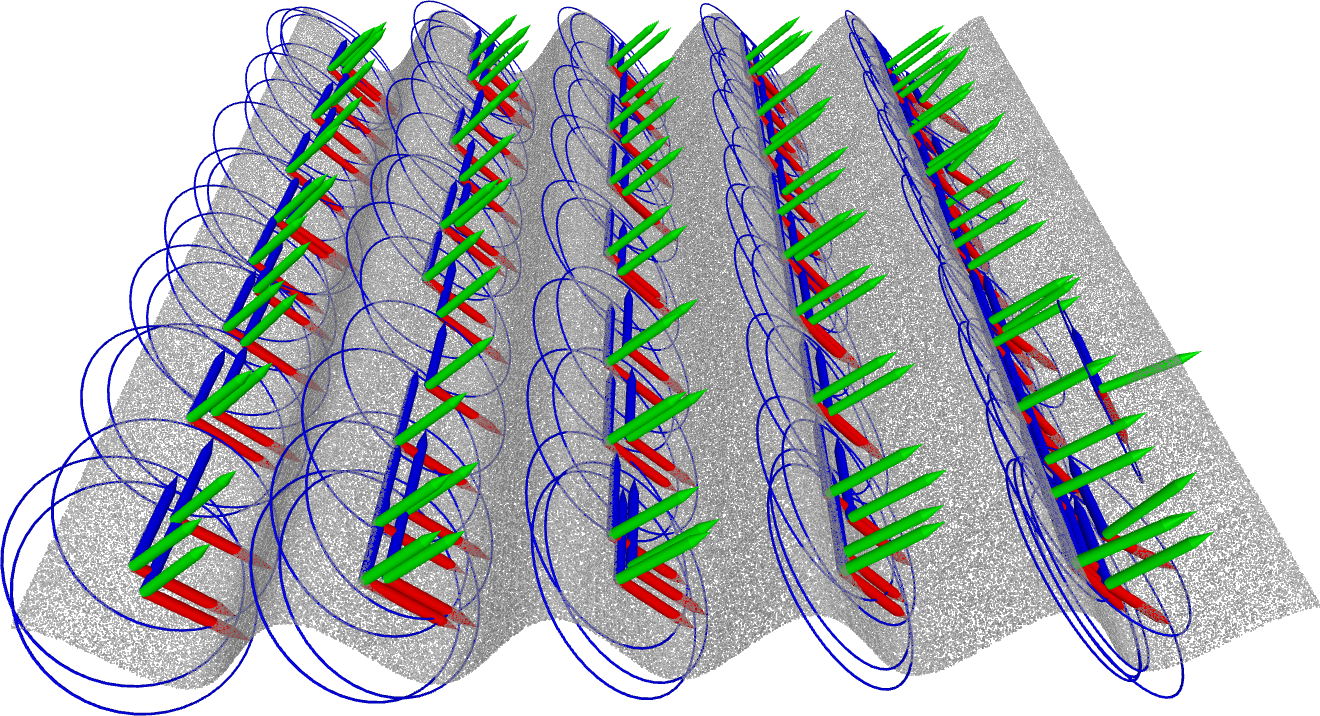}
\includegraphics[width=0.7\linewidth]{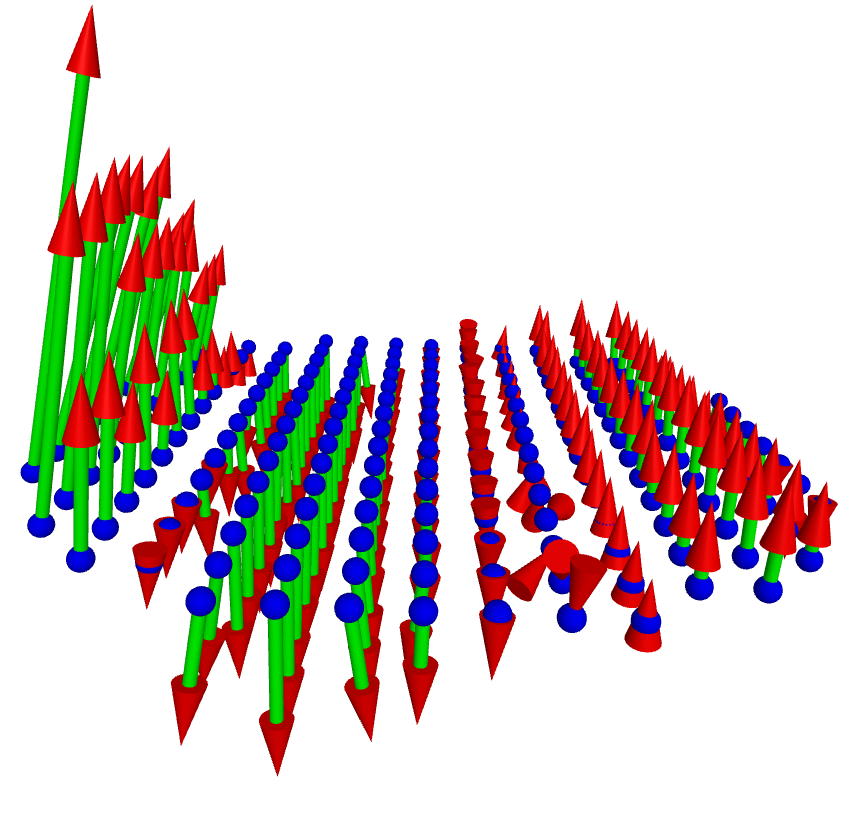}
\end{minipage}
\hfill
\begin{minipage}[c]{.32\linewidth}
\centering
\includegraphics[width=\linewidth]{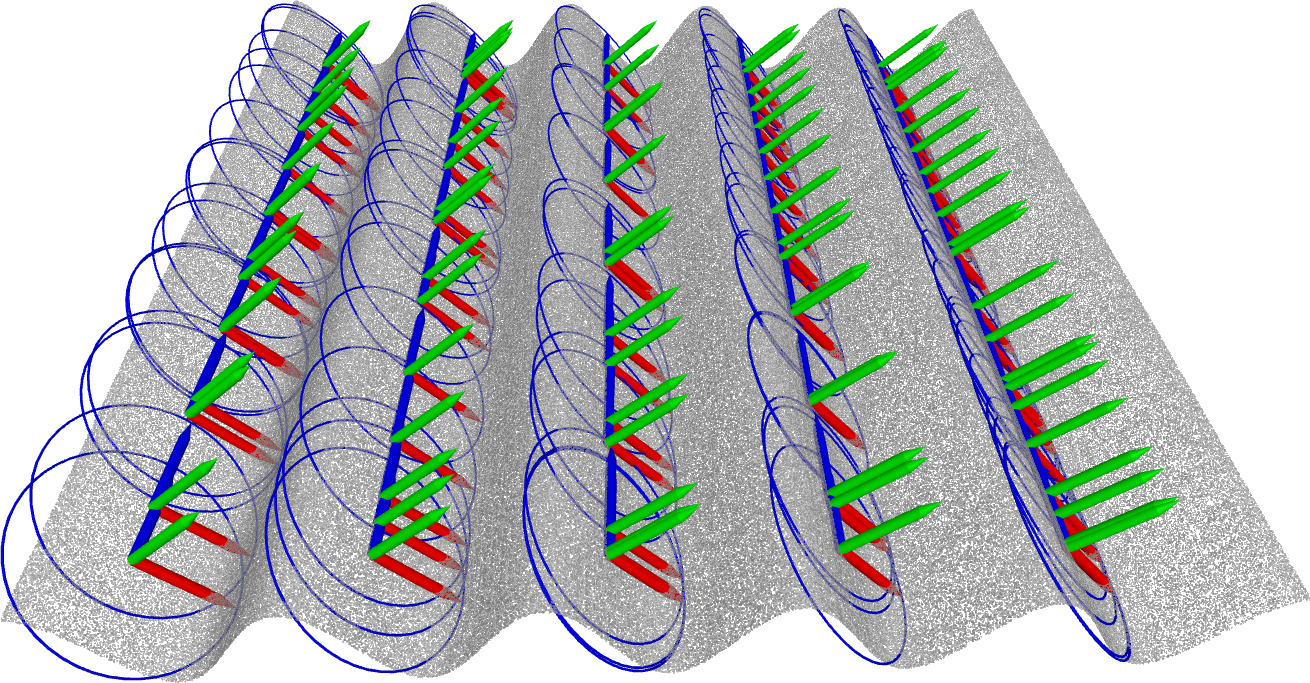}
\includegraphics[width=0.7\linewidth]{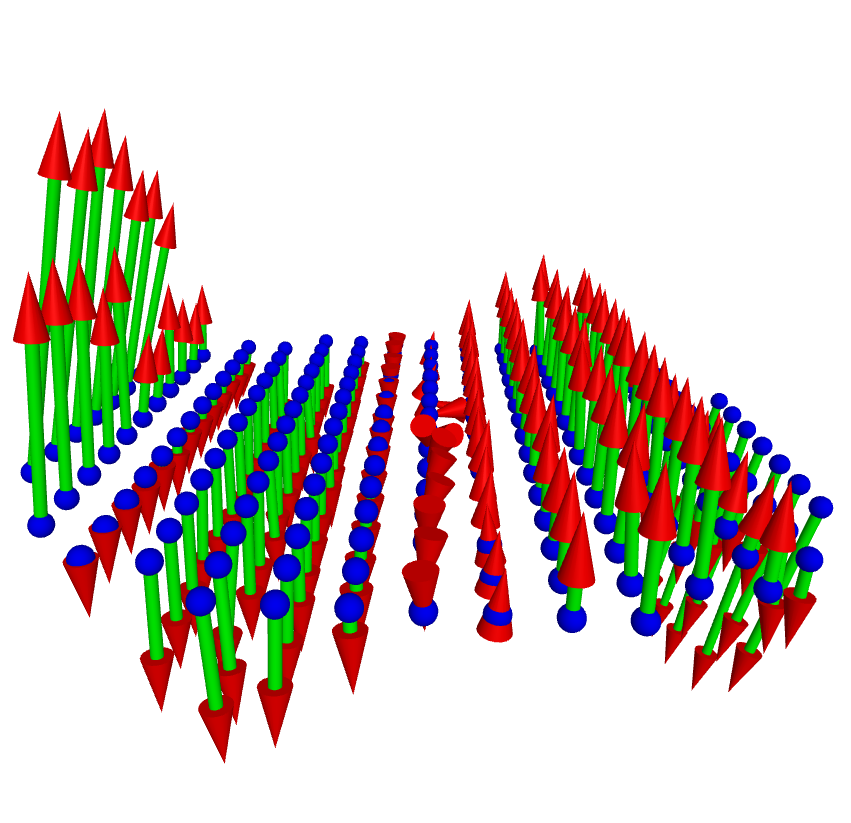}
\end{minipage}
\caption{Optimization of the $\lpf$s on a toy example. From left to right : iteration 1, 10, 99. Top : $\lpf$s tend to align in order to represent similar areas. Bottom : single-atom dictionary, which converges to a smooth sinusoidal deformation. The shape covering constraint has been relaxed for that experiment.}
 \label{fig:sin}
\end{figure*}

Let us notice that if the sampling density of the pattern is below the sampling density of the shape, the $\lpf$ representation is conservative. In that case, the shape can be exactly reconstructed from its set of $\lpf$s.

\subsection{Joint Analysis Overview}

Our shape analysis process computes:
\begin{itemize}
 \item A dictionary of $d$ atoms that best describes the $\lpf$s, and therefore the shape.
 \item A sparse set of coefficients to decompose each $\lpf$ on the dictionary.
 The sparsity constraint enforces a consolidation of the information.
 \item $\lpf$ positions and orientations that best enforce a consistent representation,
 their target areas being fixed.
\end{itemize}

 Let $N$ be the total number of $\lpf$s.
 Writing the $\lpf$s as vectors $V_j \in \bb R^{3M}$, $j \in 0 \cdots N-1$, the
problem can be stated as follows:
\begin{align}
\begin{split}
\min_{V, D,\alpha} & \sum_{j=0}^{N-1} \|V_j - D\alpha_j\|_2^2 +\lambda \|\alpha\|_1\\
\text{ s.t. } & V_j \text{ represents a $\lpf$ with fixed target area}\\
              & \{V_j\}_{j \in0 \cdots N-1} \text{ cover the whole shape}\\
              & D\in \bb R^{3M\times d}\text{, }V_j \in\bb R^{3M} \\
\end{split}
\label{eq:energy}
\end{align}

Previous works on sparsity have mostly focused on minimizing the representation error: given a set
of $N$ signals $V_j$, the aim is to find the best dictionary to represent the set of signals in a
sparse manner (few non-zero coefficients). In our approach, we tackle a more complex problem since
both the signals and their representations are optimized while subject to very
strict constraints.
The dictionary learned on the initial set of $\lpf$s is strongly dependent on
initial positions and orientations of the patterns. As a consequence, two similar
parts of the shape might be described by non-similar $\lpf$s because of different initial poses, whereas
their difference could be reduced after a change in position and orientation.
The goal of the energy minimization is thus to improve dictionary learning via enhancement of the
$\lpf$s similarity.

Our algorithm iterates the following steps:
\begin{enumerate}
 \item \textbf{Dictionary learning:} Solve for $D,\alpha$ minimizing
 $\sum_{j=0}^{N-1}\|V_j - D\alpha_j\|_2^2+\lambda \|\alpha_j\|_1$.
 \item \textbf{$\lpf$ Pose optimization:} For each $\lpf$ $V_j$,
 solve for the translation $\tt$ and rotation $R$ minimizing the representation error and update the $V_j$, seed and frames accordingly.
 \item \textbf{$\lpf$ Update:} Update each $\lpf$ $V_j$ using the probing operator.
\end{enumerate}

Figure \ref{fig:sin} shows how our algorithm captures the self-similarities of a synthetic shape.
Starting with points sampled randomly on a sinusoidal surface, we set $100$
initial $\lpf$s with random position and orientation on the shape. We then iterate our
$\lpf$ joint analysis method.
The result shows that the $\lpf$s align at positions that represent similar parts of the shape, and their optimized orientations are consistent.

\subsection{Sparse Coding For Dictionary learning \label{sec:dic_optim}}
The first optimization aims at learning a good dictionary over which the signals will be
\emph{sparsely} decomposed. Let us write the $\lpf$s as vectors $V_j\in \bb R^{3M}$ for $j\in 0 \cdots N-1$. Then the representation optimization corresponds to the Least Absolute Shrinkage and
Selection Operator (LASSO) problem, which is defined as the minimization of:
\begin{equation}
l(\alpha, D) = \sum_{j=0}^{N-1} \|V_j - D\alpha_j\|_2^2
+\lambda \|\alpha_j\|_1
\label{eq:dico_learning}
\end{equation}
with $D\in \bb R^{3M\times d}$ and $\alpha_j\in \bb R^{d}$.
As stated in \cite{online_dictionary_learning_for_sparse_coding} and
\cite{online_learning_for_matrix_factor}, the $\ell^1$ norm is empirically known to provide sparse
solutions while improving the speed of dictionary learning compared to the so-called $\ell^0$ norm
(number of non-zero elements in a matrix). The dictionary is initialized with $d$ elements drawn
randomly from the set $(V_j)_{j\in 0\cdots N-1}$. Afterwards, the algorithm alternates between
the two following steps:
\begin{itemize}
 \item Sparse coding step to compute $\alpha_j$, $j \in 0 \cdots N-1$, using the Least Angle
Regression (LARS) algorithm \cite{lars}.
 \item Dictionary update step, performed by using the previous dictionary as a warm restart to
minimize the objective function.
\end{itemize}

The steps are iterated until convergence of the representation error is reached. In our setting,
$10$ iterations are sufficient for the error to converge. The $\lambda$
parameter controls the
sparsity of the solution: large values will favor very sparse solutions while
small values
will yield dense solutions. In our tests, we set $\lambda=0.2$ for dictionaries with a large number of atoms, and
$\lambda=0.05$ for small dictionaries, to allow enough non-zero coefficients for shape representation.

The dictionary is initialized to a random subset of the LPFs. Since our algorithm converges to a local minimum, a different random initialization might lead to different minima. We experimentally observed the stability with respect to the random dictionary initialization, by running the shape analysis algorithm $50$ times with the exact same parameters, starting with a different initial random dictionary and measuring the representation error (\ref{eq:dico_learning}) divided by $N$. This experiment yielded an average error of $3.2$ and a standard deviation of $0.007$, thus showing that our algorithm is rather stable to the random dictionary initialization. Although the error converges to similar values, LPF Analysis is unfortunately not guaranteed to converge towards the same dictionary given different initializations (similarly to K-Means or K-SVD). A stable dictionary would be of great interest for applications such as shape matching and retrieval. However, for the applications shown in this paper (resampling and denoising), stability of the dictionary is not crucial, since it is an intermediate representation and not the output of the algorithm.

An alternative for the sparse coding step might be to use Deep Learning and especially auto-encoders but the optimization of position and rotation  to ensure that LPF representation of similar regions are indeed similar does not write in a simple way.

\subsection{Pattern pose optimization}
The previous sparse decomposition steps introduce a representation error. But
reducing this error is still possible by improving each $\lpf$ pose individually,
to find the optimal fit between the shape and each newly modified $\lpf$. This
can
easily be done using a least squares minimization step, similar to what is done during one ICP
iteration. More precisely, one looks for a translation of the $\lpf$ and a rotation of its frame
such that the newly defined vectors $\v_i$ fit better the ones obtained by
dictionary decomposition
$\tilde \v_i$.
After a rotation $R$ of the $\lpf$ frame and translation $\tt$ of its position,
the modified $\v_i$
can be expressed as:

\begin{equation}
\v^\prime_i = R^{-1}\cdot(\u_i+\v_i) - \u_i - \tt.
\end{equation}

The best rotation and translation thus minimize:

\begin{equation}
\sum_{i=0}^{M-1} \|\v^\prime_i - \tilde \v_i\|^2 = \sum_{i=0}^{M-1}
\|R^{-1}\cdot(\u_i+\v_i) - \tt -
(\u_i +\tilde \v_i)\|^2
\end{equation}

which is exactly a least squares minimization for estimating a rigid transform
between two sets of points: $\u_i+\v_i$ and $\u_i+\tilde \v_i$.

Using these optimal rotation and translation for each $\lpf$, the seed position $s$, frame orientation $\tt_1,\tt_2,\n$ and vectors $\v_i$ should be updated accordingly:
\begin{align}
\begin{split}
&s^{\text{\tiny{updated}}} = s - \tt ;\\
&\tt_1^{\text{\tiny{updated}}} = R^{-1} \tt_1 ;\\
&\tt_2^{\text{\tiny{updated}}} = R^{-1} \tt_2 ;\\
&\n^{\text{\tiny{updated}}} = R^{-1}\n ;\\
&\forall i \in{0\cdots M-1},\text{  } \v_i^{\text{\tiny{updated}}} = R^{-1}\cdot(\u_i+\v_i) - \u_i - \tt.
\end{split}
\end{align}
This update maintains the consistency between the shape and the description, position and orientation of the $\lpf$s.

\begin{figure}
\begin{tikzpicture}
\node[anchor=south west,inner sep=0] at (0, 0) {\includegraphics[width=\linewidth]{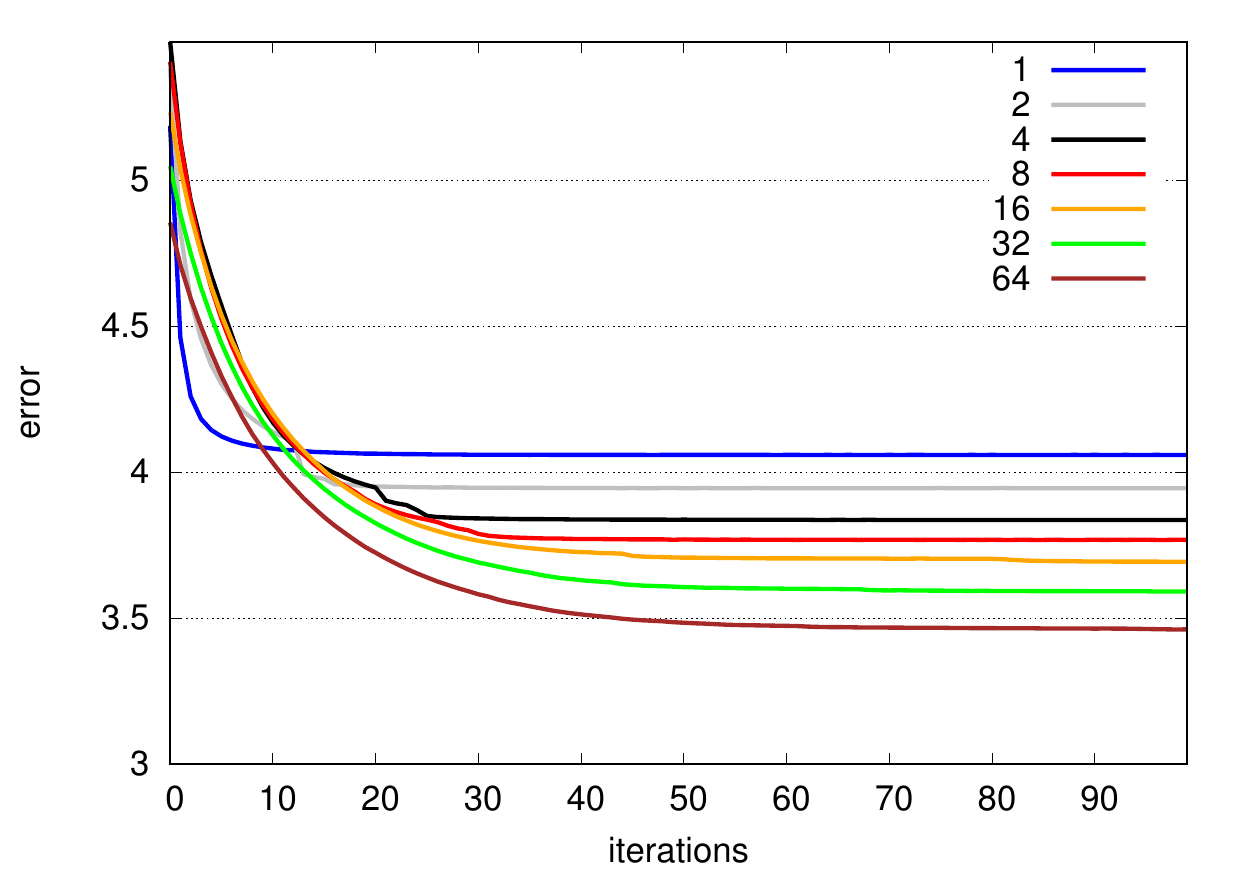}};
\node[anchor=south west,inner sep=0] at (0.4\linewidth, 0.35\linewidth) {\includegraphics[width= 0.2\linewidth]{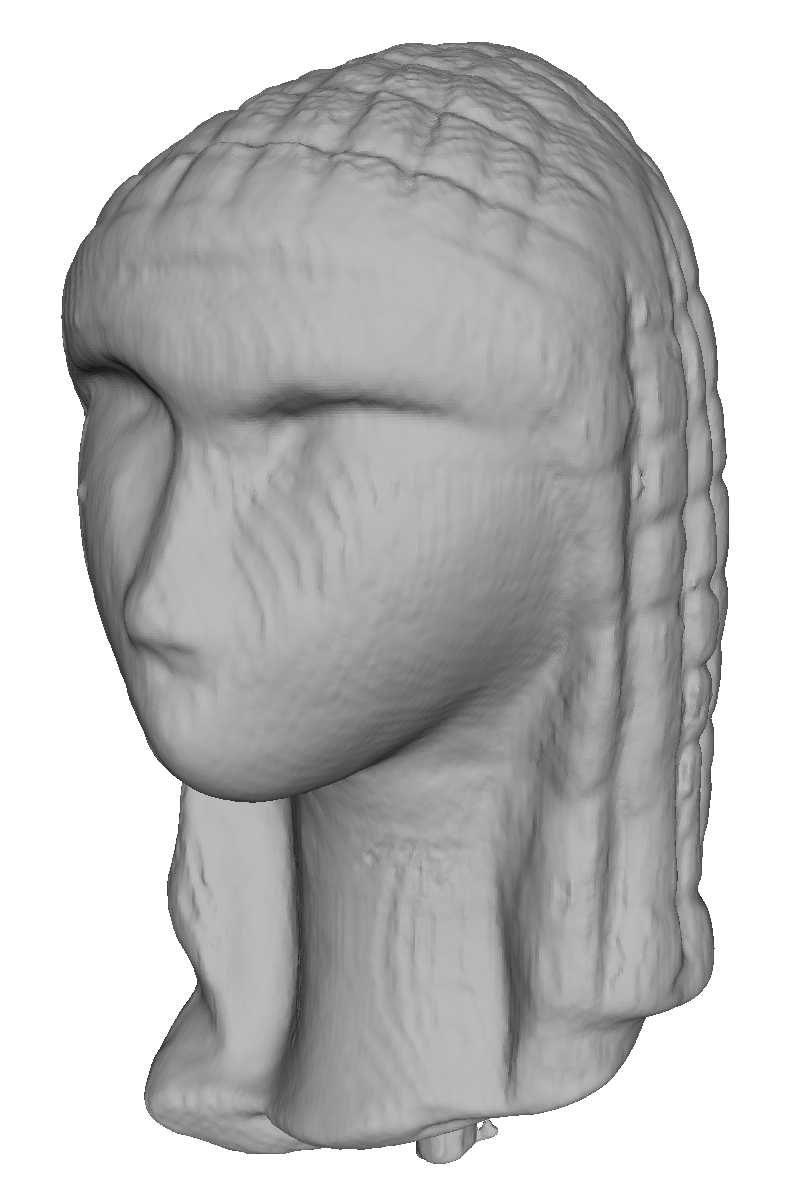}};
\end{tikzpicture}
\caption{Error convergence for the ``Brassempouy'' model, for different dictionary sizes. For visualization purposes, the error is divided by the dictionary size.}
\label{fig:convergence}
\end{figure}

\begin{figure*}[ht]
\centering
\begin{subfigure}[]{}
\includegraphics[width=.18\linewidth]{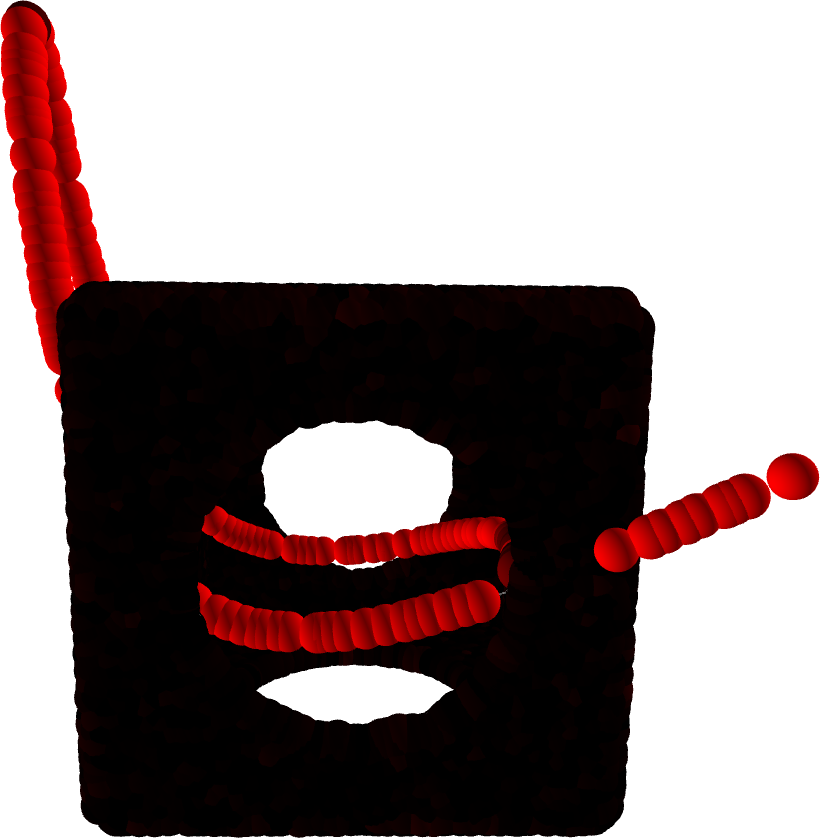}
\end{subfigure}
\begin{subfigure}[]{}
\includegraphics[width=.14\linewidth]{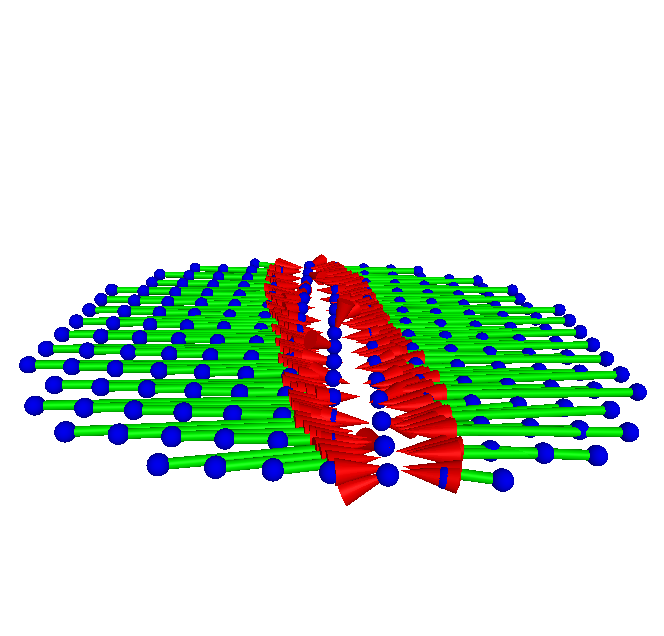}
\end{subfigure}
\begin{subfigure}[]{}
\includegraphics[width=.18\linewidth]{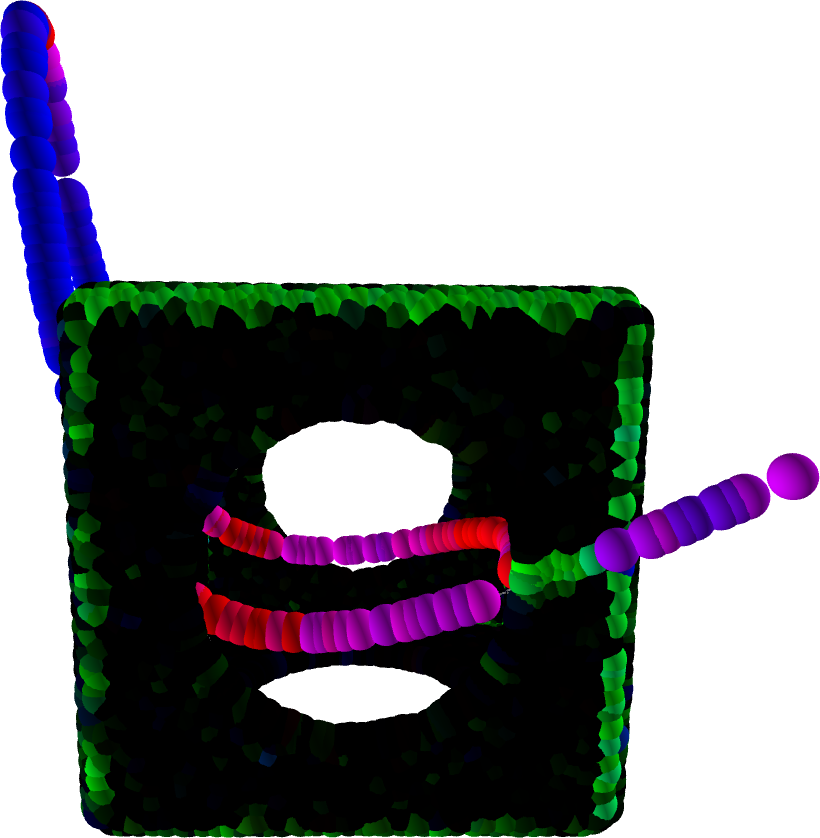}
\end{subfigure}
\begin{subfigure}[]{}
\includegraphics[width=.14\linewidth]{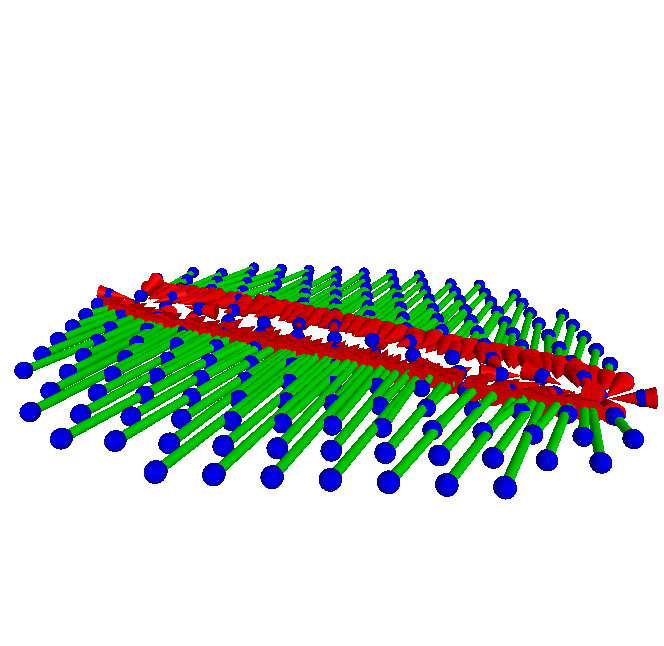}
\end{subfigure}
\begin{subfigure}[]{}
\includegraphics[width=.14\linewidth]{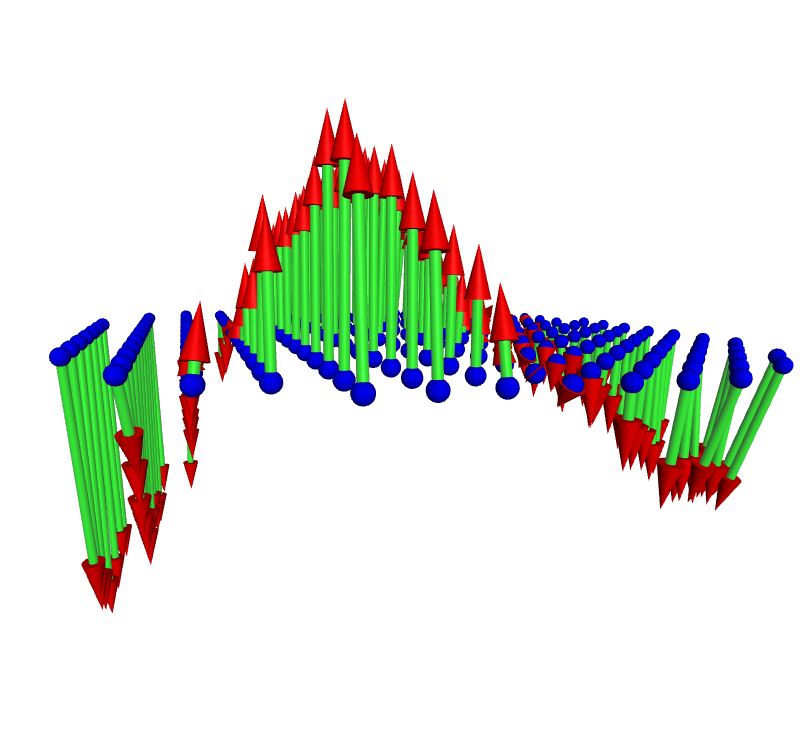}
\end{subfigure}
\begin{subfigure}[]{}
\includegraphics[width=.14\linewidth]{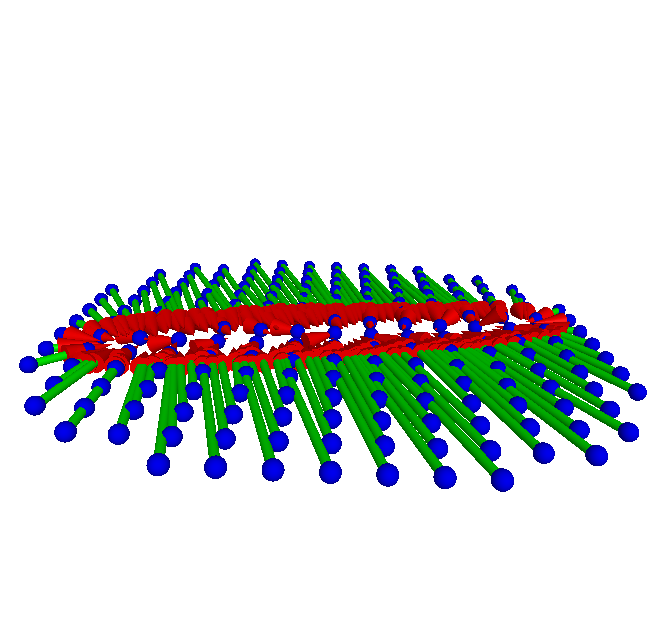}
\end{subfigure}
\caption{Analysis of a cube with a curve (also shown on Fig. \protect\ref{fig:cubecurve}). (a)  : Decomposition using 1 atom, where the amplitude of the decomposition is encoded as the red component. (b) the resulting dictionary atom, which exhibits a shape with intrinsic dimension $1$. This is consistent with the decomposition coefficient representation (a) where only the curve part contains significant coefficients. (c) Decomposition using 3 atoms, each component being encoded respectively using the red, green and blue component. (d), (e) and (f): the three resulting atoms. One atom represents the sharp feature and 2 atoms represent the curve. Notice that the planar parts are decomposed with all coefficients equal to $0$, since our pattern is perfectly planar.}
\label{fig:cubecurve_dict}
\end{figure*}

\subsection{Closing the loop}
After the $\lpf$ pose optimization step, the set of optimized $\lpf$s is better suited for the input
shape analysis. We use once again the probing operator to probe the shape with respect to the
optimized $\lpf$ pose. To avoid $\lpf$s gliding on the shape (with the risk of losing the
shape covering), we restrict the update to its original target area.
Therefore, although each $\lpf$ pose and sparse decomposition is optimized,
it always accounts for the same shape area.
This update can theoretically increase the error but we observed experimentally that the error decreases across the iterations (see Fig. \ref{fig:convergence}).
Then, we are able to repeat all the processing steps until convergence is reached.
Figure \ref{fig:convergence} shows the evolution of the error with varying dictionary sizes on the Brassempouy point set.

Figure \ref{fig:cubecurve_dict} shows the analysis result for a shape representing a cube with a curve when using a dictionary of size $1$ and $3$.
When using 1 atom, the decomposition (a) mostly encodes the curve, while sharp features are undetected.
When using 3 atoms, the decomposition (c) exhibits significant coefficients on sharp edges or on the curve. For both cases, the remaining parts of the shape are represented with small coefficients for all
atoms. (b), (d), (e) and (f) show atoms representing sharp features and curves. These experiments
show that the final dictionary retrieves important features of the shape.

\subsection{Controlling the shape analysis}

Our shape analysis approach is driven by three parameters: the pattern
radius $r$, the number of points $M$ of the pattern, and the number of
dictionary atoms $d$.

The pattern radius $r$  is linked to the shape geometry:
it represents the scale at which the similarities can be exploited to build the
dictionary. On Figure \ref{fig:sphere}, we illustrate the
importance of the radius: with a large radius, the faceted sphere mesh is interpreted as a sphere,
whereas it stays a piecewise linear shape with a small radius.

Once $r$ is fixed, the number of points $M$ in each pattern controls the
accuracy of the analysis. This accuracy can be measured by the average distance
$\tau_s$ between pattern points. Assuming that the points are uniformly sampled
on a disk of radius $r$, each point represents a region of area $\frac{\pi
r^2}{M}$. Thus, an estimation of the distance between points is
$\tau_s = \frac{r}{\sqrt M}$. Hence, once $\tau_s$ is fixed, $M$ can be
computed as: $M=\frac{r^2}{\tau_s^2}$. $M$ can then be adjusted at will, to
perform different sampling scenarios, depending on the input shape: if the
probing operator accuracy is $\tau_p$, setting $\tau_s=\tau_p$ will perform a
$1:1$ sampling scheme. Setting $\tau_s = 0.5 \tau_p$ will double the number of
sampling points, as shown on figure \ref{fig:ship}. In practice, an efficient way of
building a pattern is to generate a regular grid with a step of size
$\tau_s$ and keep only the points that are
included within a radius $r$.

The number of atoms $d$ controls the amount of consolidation. If the number of
$\lpf$s is equal to the number of atoms, then no $\lpf$ learning is performed
since all $\lpf$s can be represented independently, and there is no consolidation.
Conversely, fewer dictionary atoms implies a stronger consolidation.
Therefore the number $d$ of atoms in the dictionary controls the degrees of
freedom in the representation, i.e. the supposed variability of local parts of
the shape.
$d$ can be increased until the error falls under a threshold that is
consistent with the accuracy $\tau_p$ of the probing operator.
Interestingly, when the sampling pattern is planar, all dictionaries naturally allow to represent planar regions as any
atom multiplied by $0$ results in a planar shape.

Finally, let us notice that the analysis framework is entirely driven by the minimization of Equation (\ref{eq:energy}).
First, the dictionary learning minimizes the error
with fixed LPFs. Then, the pose estimation step keeps the dictionary and
coefficients fixed and minimizes the error by aligning the $\lpf$s to match the
dictionary decomposition, therefore, the $\ell^1$ component remains unchanged
while the $\ell^2$ component decreases making the sum of the two components
decrease.

The last recomputation step is not guaranteed to reduce the error, since
the probing operator is used once again to recompute the $\lpf$s. Without
this step, the error defined by Equation (\ref{eq:energy}) would always decrease and convergence
would be guaranteed. In our experiments, we observed that in most cases, this recomputation step
made the error decrease. However, in a few cases the error increased
marginally but that was compensated by the decrease in the two other steps.
Moreover, using the recomputation step yields lower errors and better dictionaries.

\begin{figure*}[ht]
\centering
\subfigure[Original]{\includegraphics[width=0.24\linewidth]{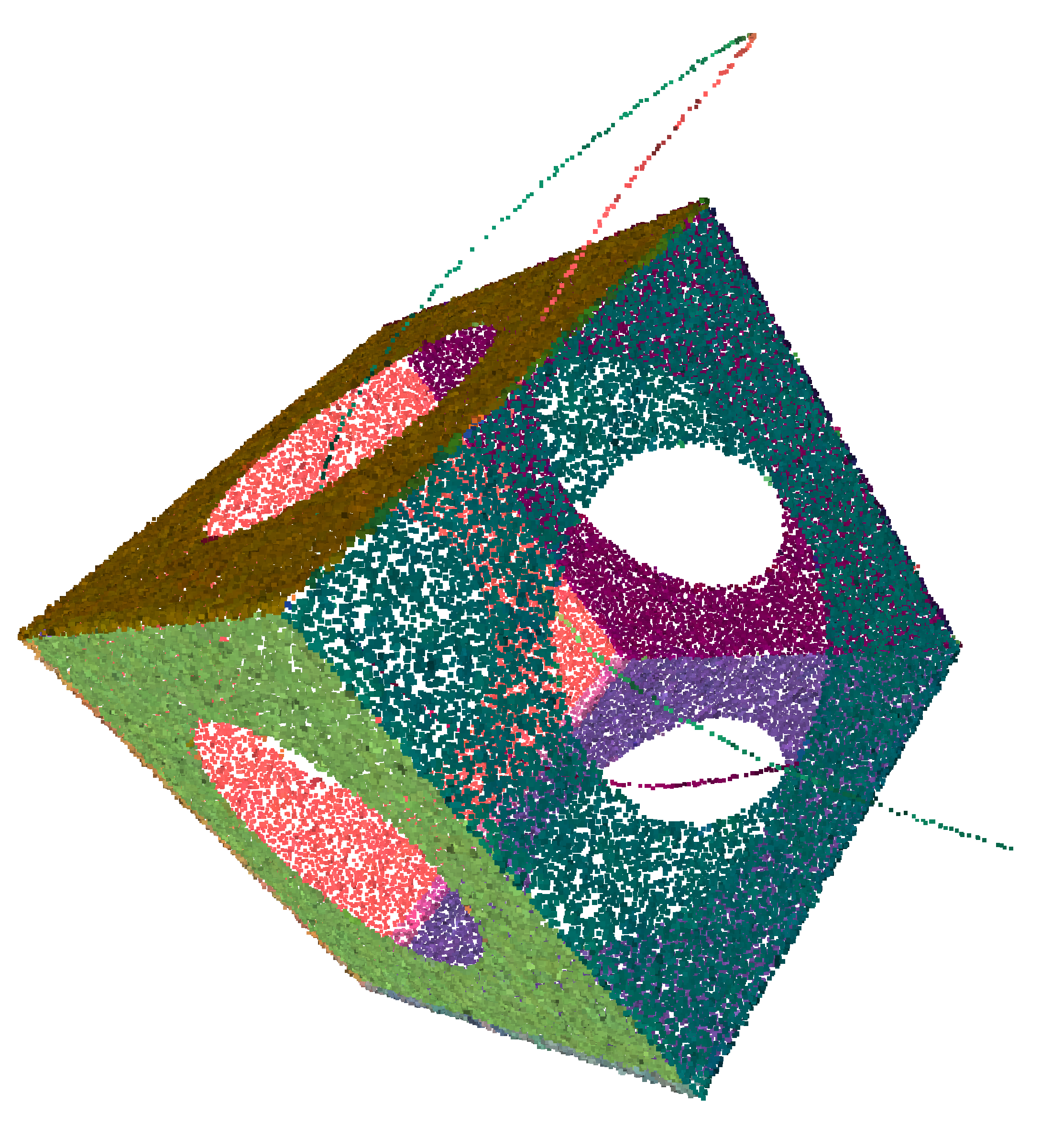}}
\subfigure[MLS]{\includegraphics[width=0.24\linewidth]{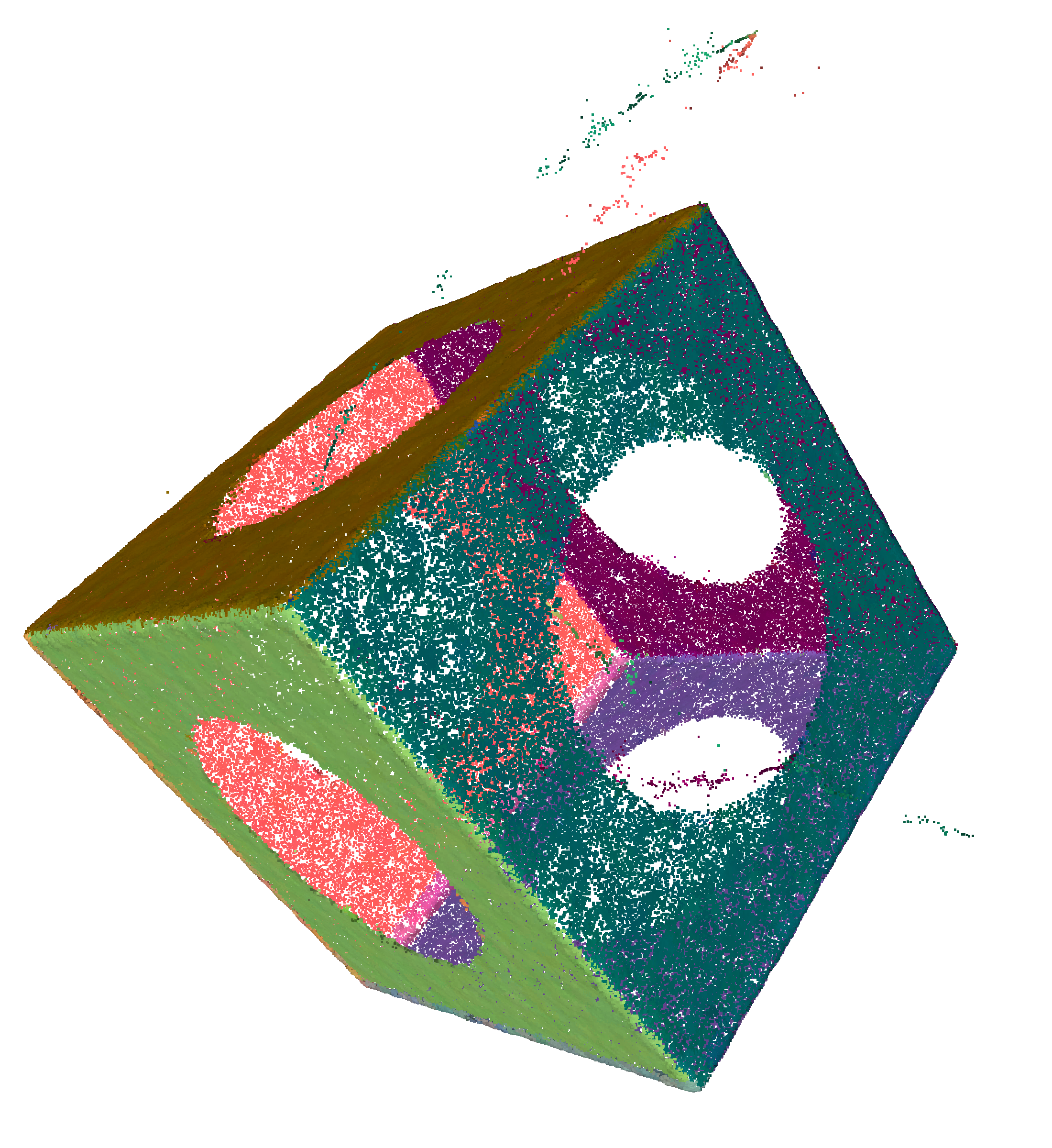}}
\subfigure[EAR]{\includegraphics[width=0.24\linewidth]{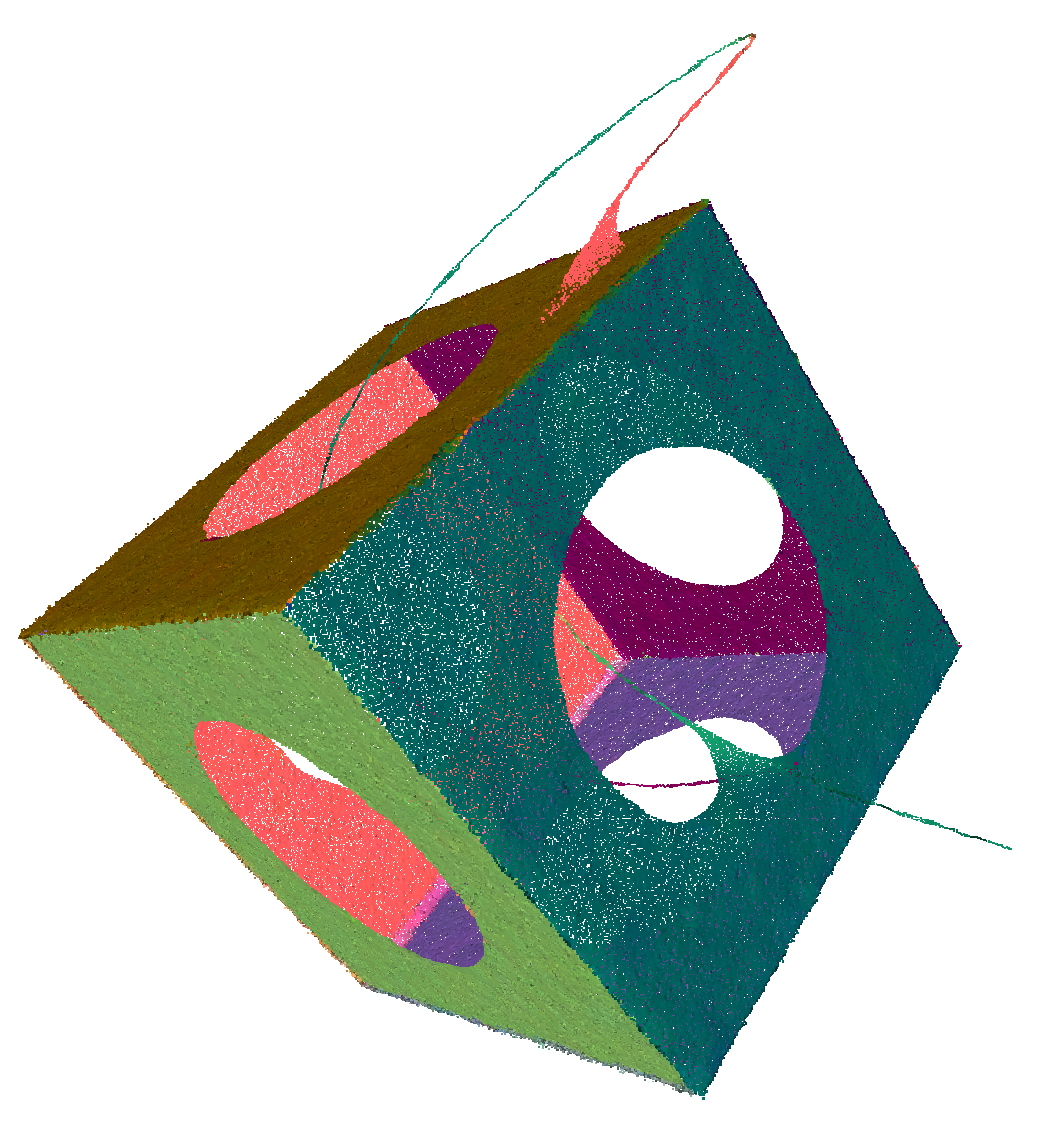}}
\subfigure[\textbf{LPF}]{\includegraphics[width=0.24\linewidth]{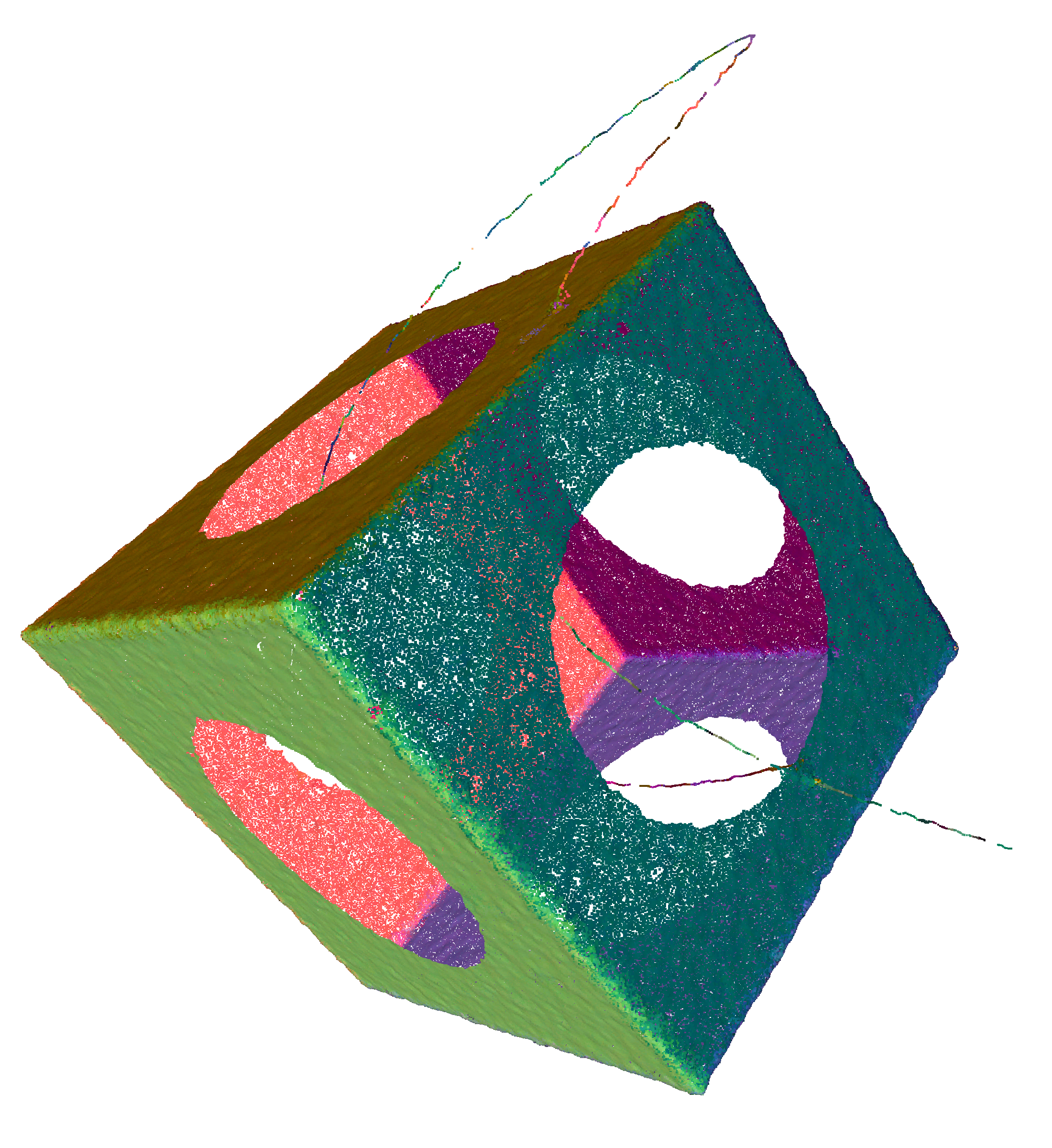}}
\caption{Upsampling a slightly noisy hollow cube intersected by a curve. Original point set
($80K$ points), MLS, EAR, $\lpf$ ($360K$ points). MLS is unstable around thin lines. EAR performs
better on the edges but tends to impose a local surface model, as can be seen near the
intersection of the line and the cube ($r=0.25$, $M=793$, $d=64$, Shape diagonal: $8.8$).}
\label{fig:cubecurve}
\end{figure*}

\section{Applications}\label{sec:applis}

\subsection{Shape Resampling}\label{sec:resampling}

\begin{figure}
\begin{center}
\includegraphics[width=0.5\linewidth]{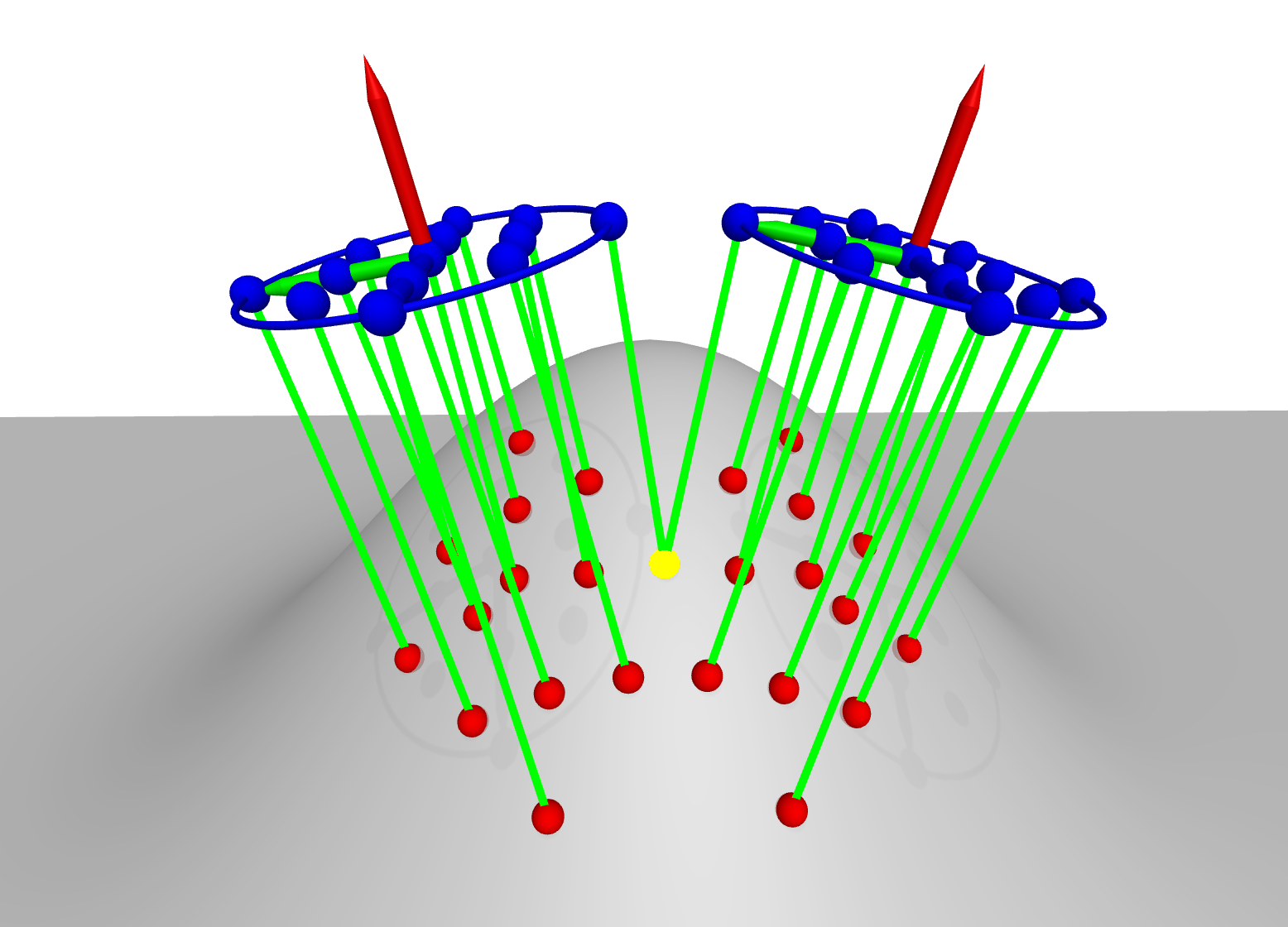}
\end{center}
\caption{$\lpf$ consolidation: during resampling, a point on the shape (yellow
sphere) may be
shared by two different $\lpf$s.}
\label{fig:lpf_consolidation}
\end{figure}

The shape analysis of section \ref{sec:analysis} provides us with optimized
$\lpf$  positions $\tilde s$ and orientations $\tilde \tt_1, \tilde \tt_2, \tilde \n$
as well as a shape dictionary $D$ and the associated decomposition coefficients
$\alpha$.
We propose to use these optimized $\lpf$s to sample points on the shape.

First, the vector fields are recomputed using their decompositions
$\tilde{V}_j=D\alpha_j$.
Each $\lpf$ yields a set of points $p_i$ such that:
\begin{equation}
p_i = \tilde s + (\u_i.x +\tilde \v_i.x)\tilde \tt_1 + (\u_i.y + \tilde
\v_i.y)\tilde \tt_2 +  (\u_i.z
+\tilde \v_i.z)\tilde \n
\end{equation}

where $\u.x,\u.y,\u.z$ are the $x,y,z$ coordinates of vector $\u$ (and similarly for $\v$).
Since the $\lpf$s overlap, the reconstructed information in the overlapping
areas may not coincide exactly.
In these areas, a consensus point distribution, still driven by the error minimization, must be built.
Let us call $q$ an hypothetical best consensus position. This point is located
inside the target areas of several $\lpf$s.
A point $(s^\cl L, \u_i^\cl L, \tilde \v_i^\cl L)$ of a reconstructed $\lpf$
$\cl L$ is said to conflict with a point $q$ if the target area of $\cl L$
includes $q$ and $\|s^\cl L +\u_{i}^\cl L + \tilde \v_{i}^\cl L - q\|\leq
\tau_p^2$.
In practice this means that the $\lpf$ $\cl L$ yields a point close to $q$.
Let us define $\cl A(q)$ the influence zone of $q$ as the set of conflicting $\lpf$ points.
A consensus point $q$ should minimize:
\begin{equation}
\sum_{(s^\cl L, \u_i^\cl L,\tilde \v_i^\cl L) \in \cl
A(q)}\!\!\!\!\!\!\!\!\!\! \| q - s^\cl L -
\u_i^{\cl L} -\tilde \v_i^\cl L\|_2^2.
\end{equation}
This minimization is non-trivial since $\cl A(q)$ varies with $q$.
Fortunately, we can simplify it by fixing the set $\cl A(q)$ which
yields a least squares minimization resulting in:
$q = \frac1{\#\cl A(q)}\sum_{\cl A(q)} s^\cl L+\u_i^\cl L +\tilde\v_i^\cl L$.
In practice this means that when a $\lpf$ proposes a position $q$, the best
consensus point is found inside $\cl A(q)$, and $q$ is moved at this optimal
position.
Afterwards, no point conflicting with $q$ can be added to the resampled point
cloud.

\begin{figure}[t]
 \begin{center}
  \includegraphics[width=0.31\linewidth]{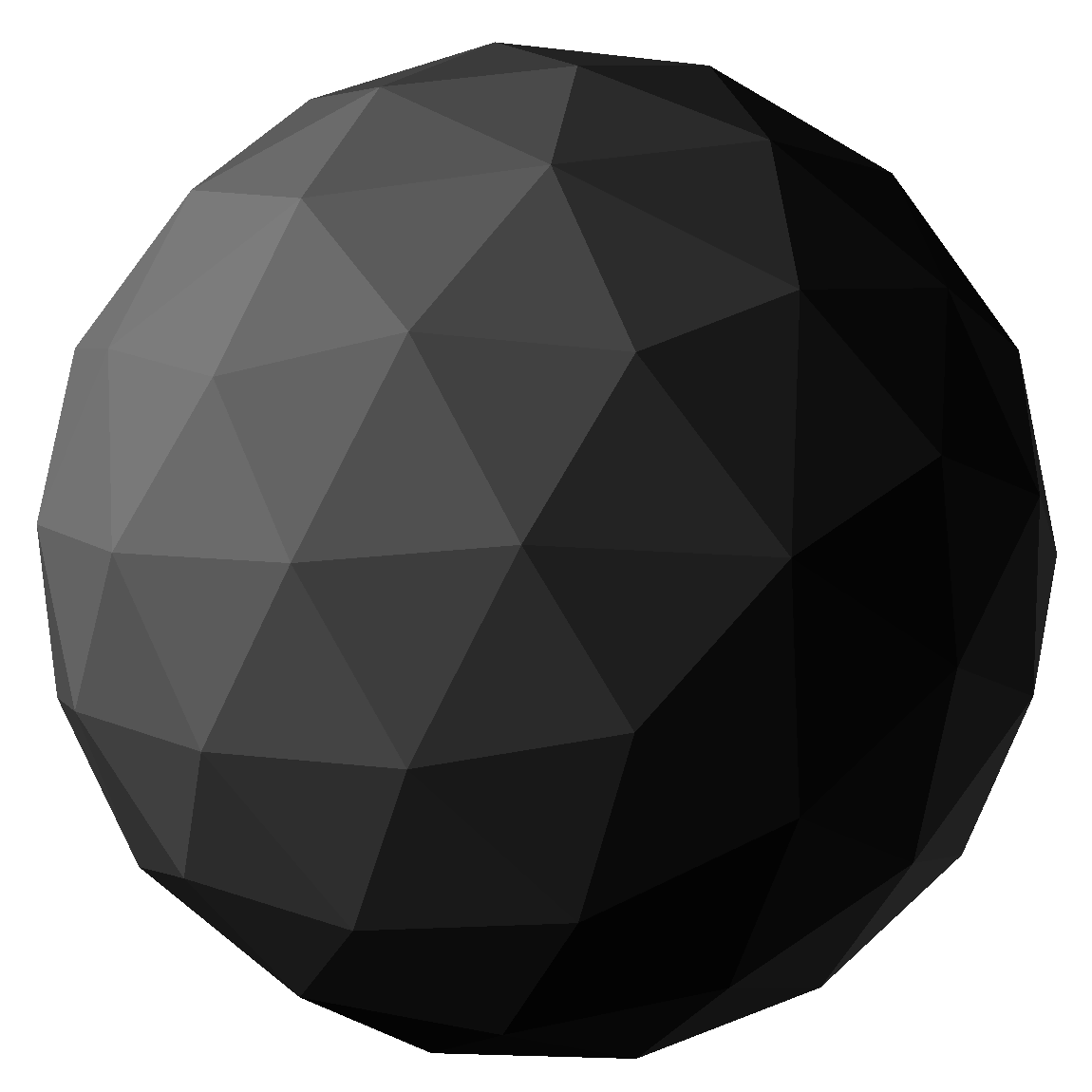}
  \includegraphics[width=0.31\linewidth]{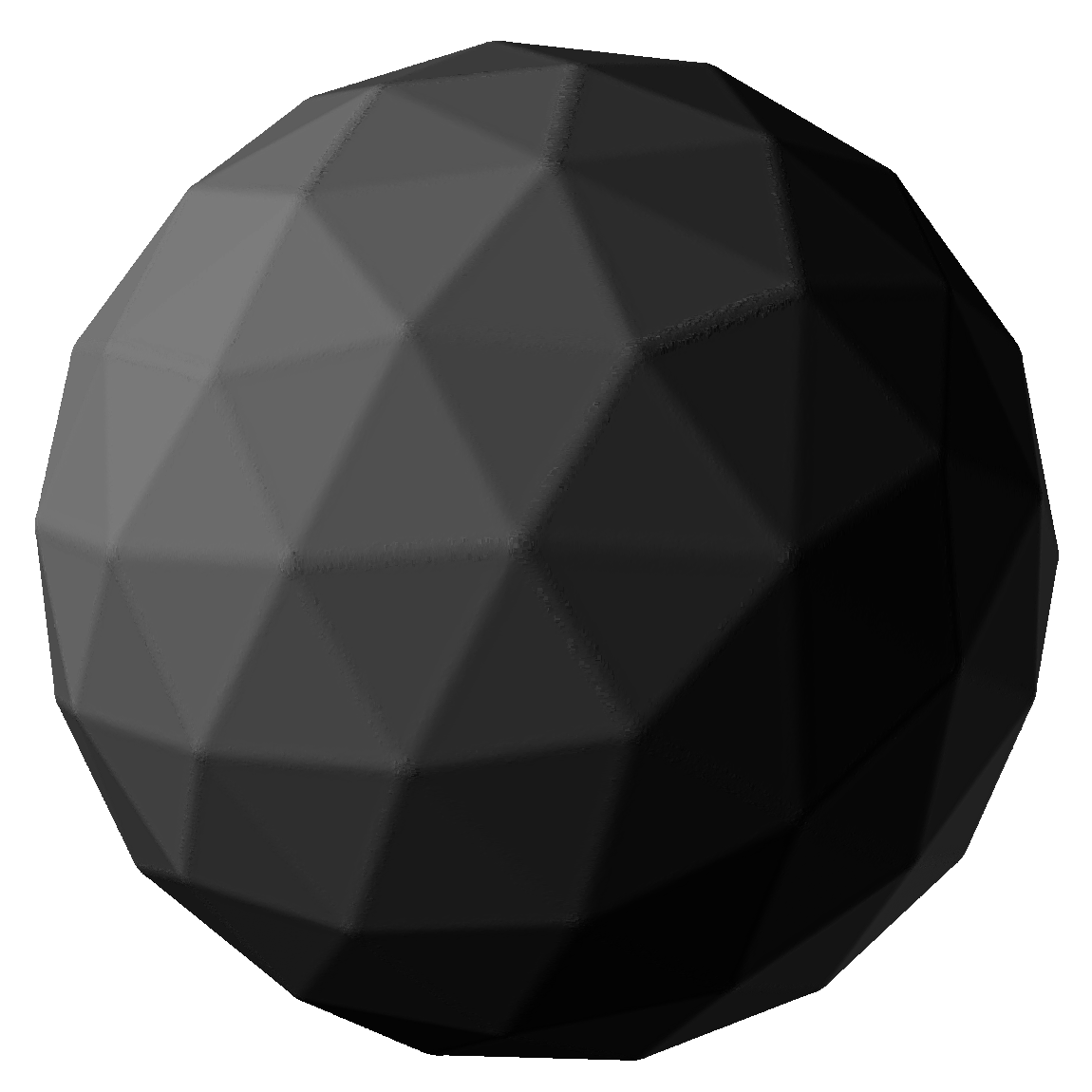}
  \includegraphics[width=0.31\linewidth]{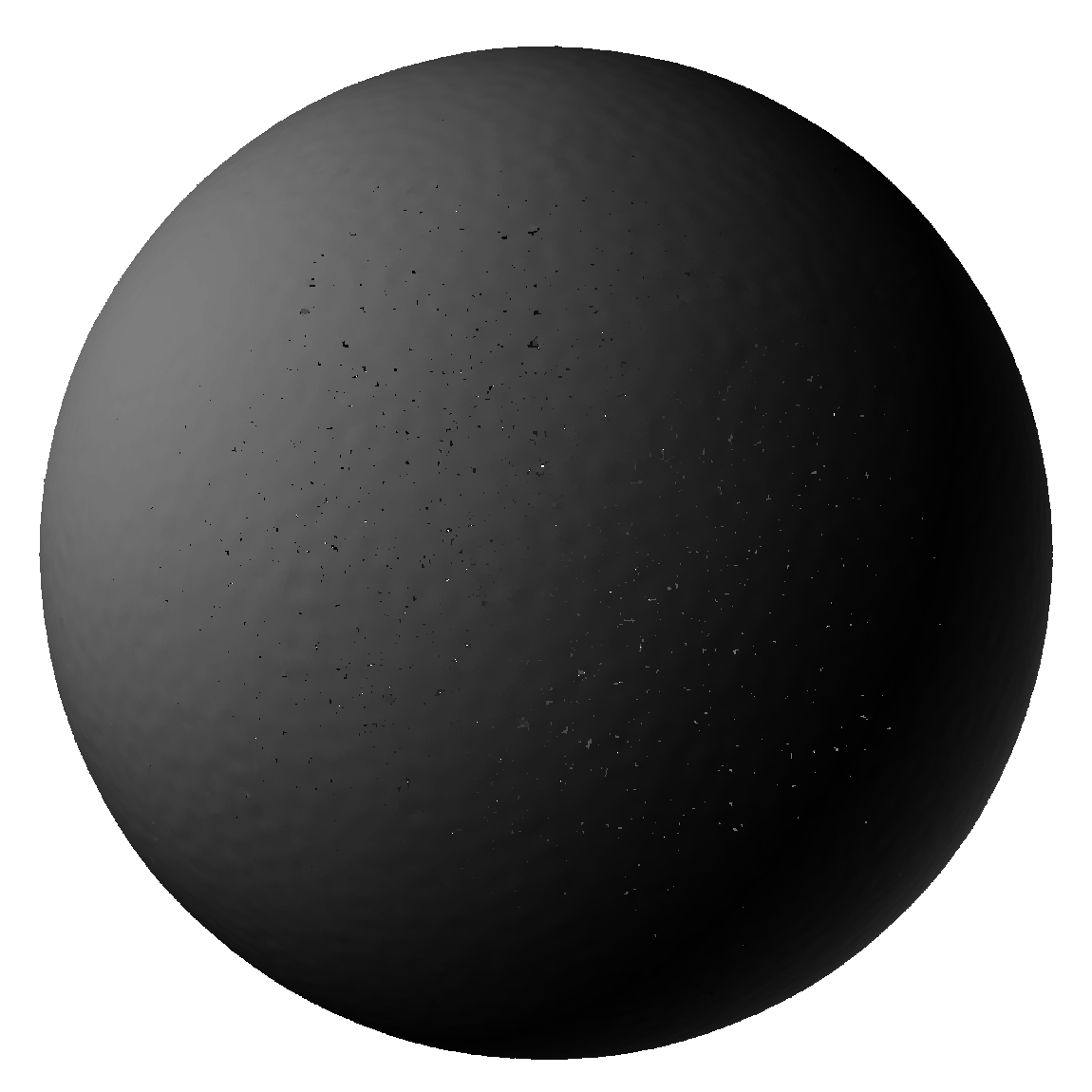}
 \end{center}
 \caption{Influence of the pattern radius $r$: mesh of a sphere with $100$ vertices (left), extracting
the dictionary and resampling from it with a small radius (middle) and a large radius
(right). A larger radius captures large scale similarities. Both dictionaries
contain $16$ atoms. The normals for the point sets are estimated by PCA.}
\label{fig:sphere}
\end{figure}

\begin{figure}
 \begin{center}
  \includegraphics[width=0.48\linewidth]{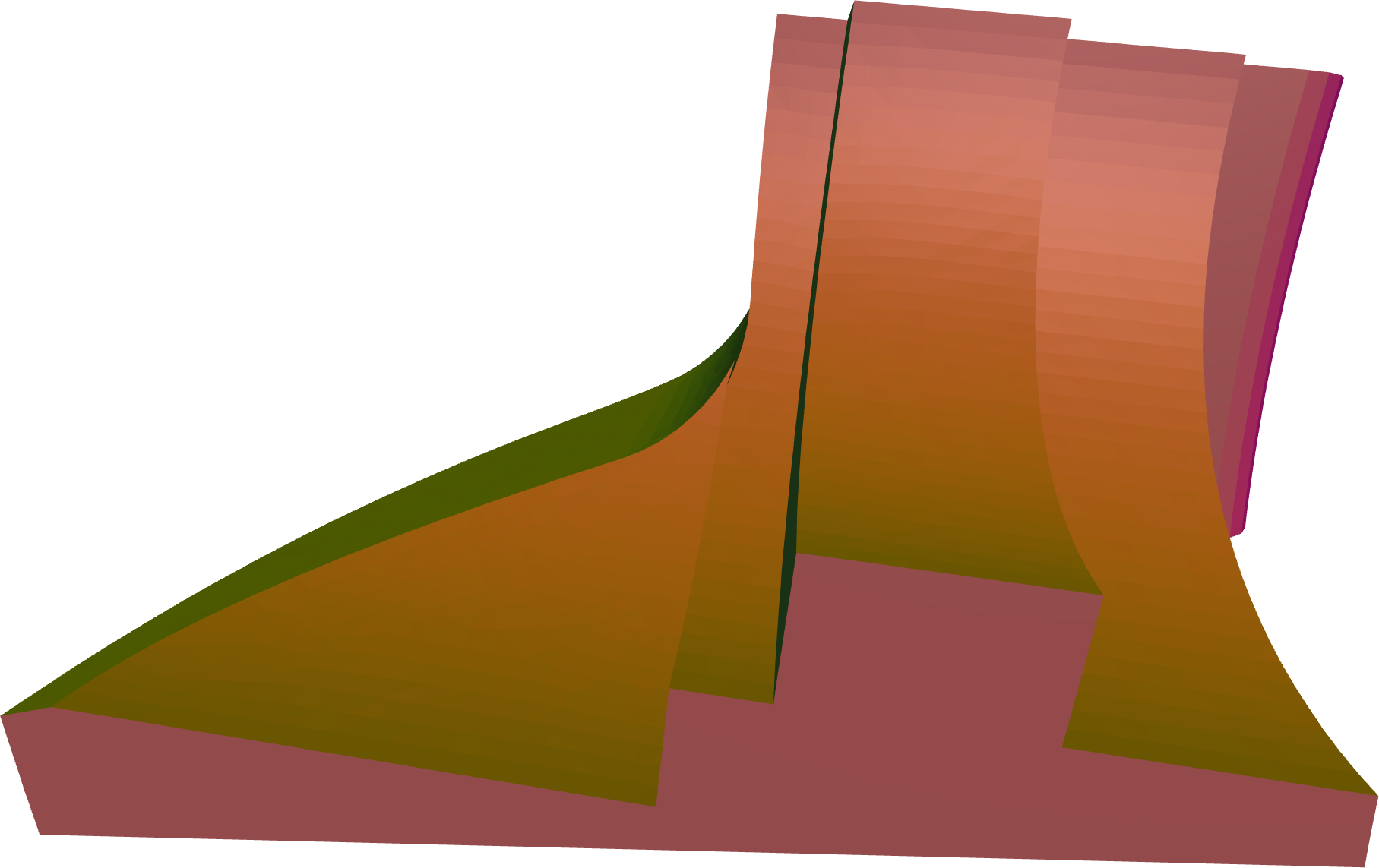}
  \includegraphics[width=0.48\linewidth]{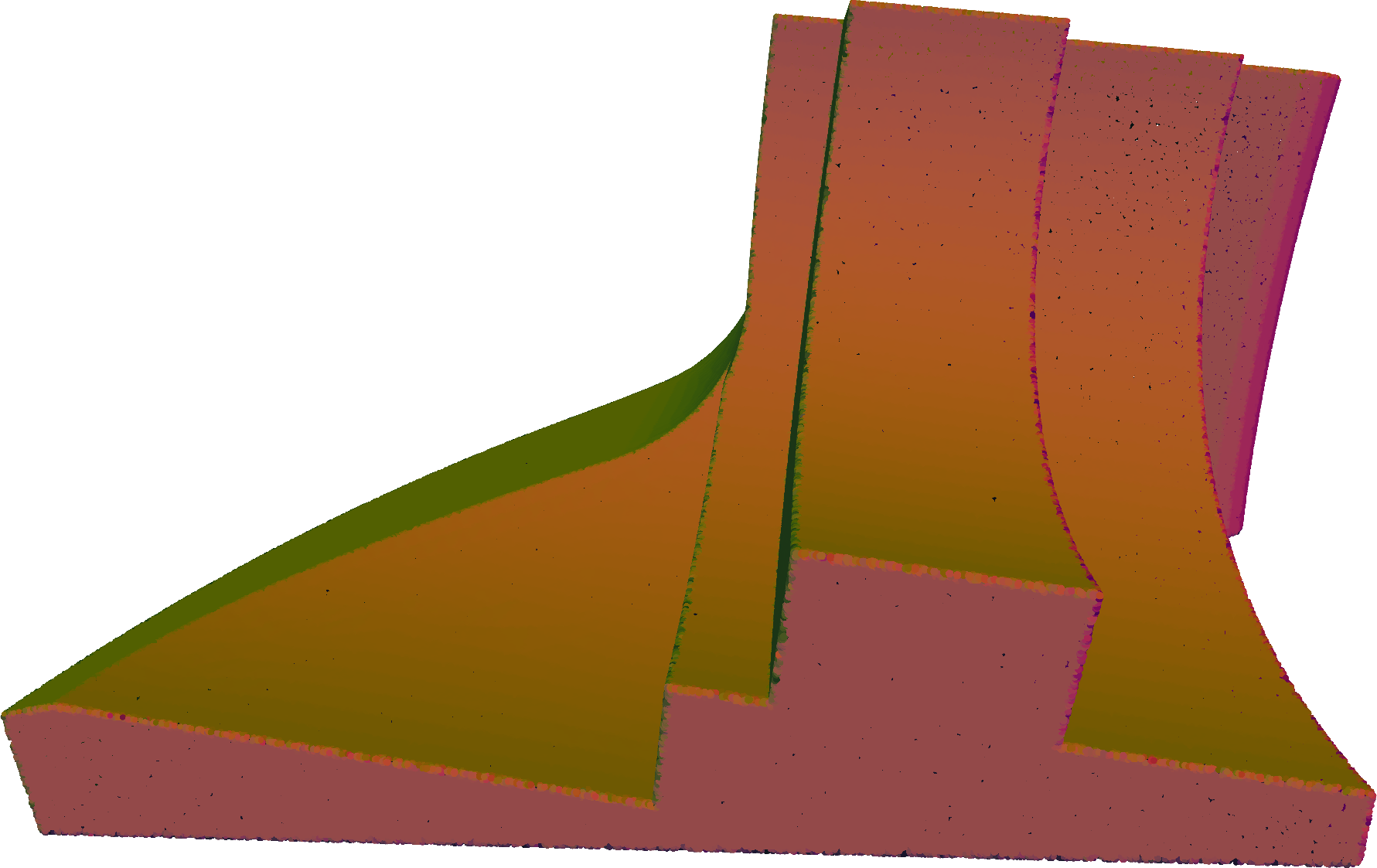}
 \end{center}
\caption{Sampling the Fandisk mesh. Left: original mesh (6k vertices). Right:
our resampling to 1M points, which shows the accuracy of our LPF-based method. Normals are computed by PCA analysis for the point
set ($r=0.2$, $M=793$, $d=16$, Shape diagonal: $7.61$). }
\label{fig:fandisk}
\end{figure}

\begin{figure}

 \centering

\setlength{\fboxsep}{0pt}
\setlength{\fboxrule}{1.5pt}

\begin{tikzpicture}
\node[anchor=south west,inner sep=0] at (0.1\linewidth, 0) {\fbox{\includegraphics[width=0.9\linewidth]{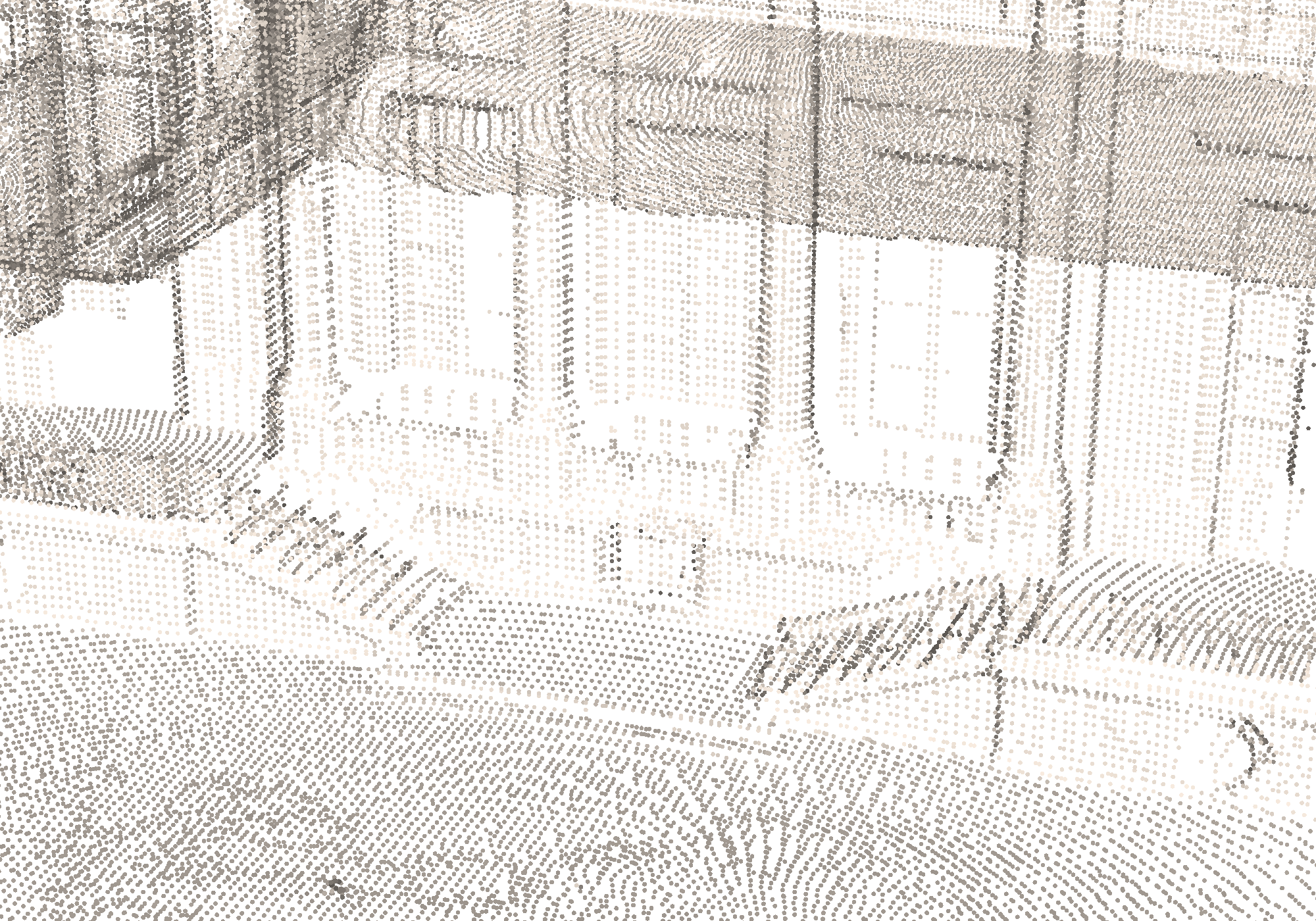}}};
\node[anchor=south west,inner sep=0] at (0,0.4\linewidth ){\fbox{\includegraphics[width=0.25\linewidth]{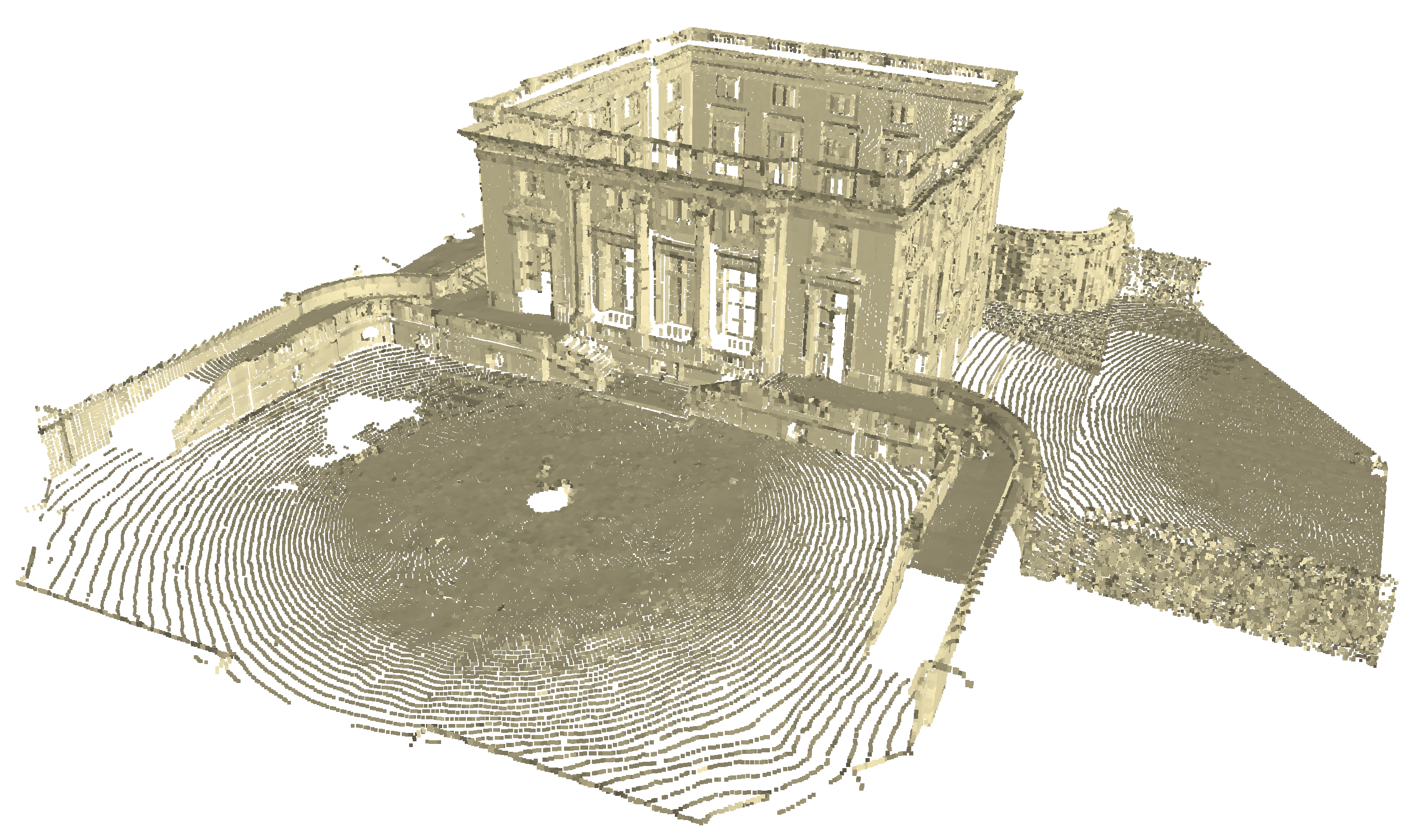}}};
\end{tikzpicture}
\vspace{5mm}

\begin{tikzpicture}
\node[anchor=south west,inner sep=0] at (0.1\linewidth, 0) {\fbox{\includegraphics[width=0.9\linewidth]{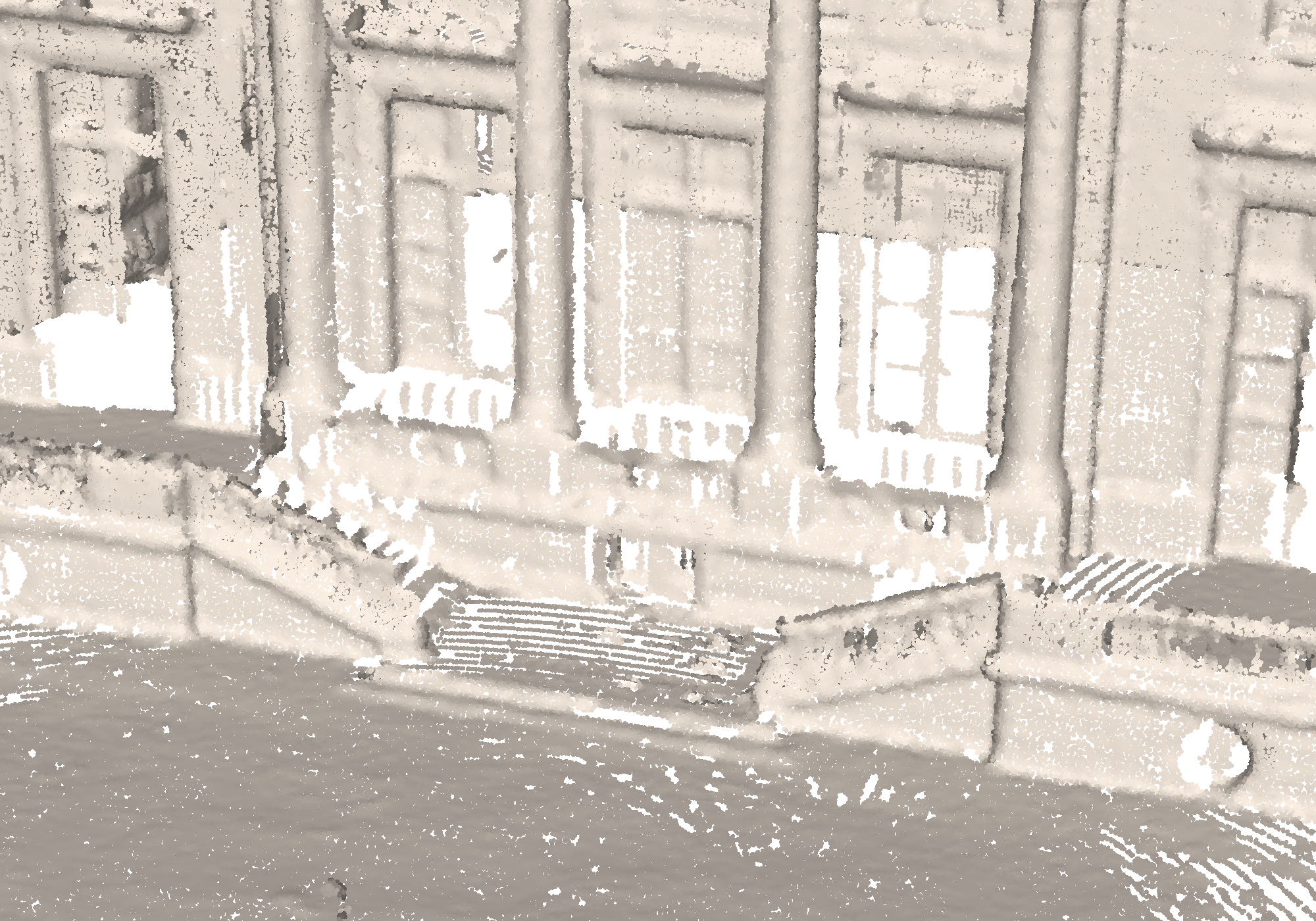}}};
\node[anchor=south west,inner sep=0] at (0, 0.4\linewidth) {\fbox{\includegraphics[width=0.25\linewidth]{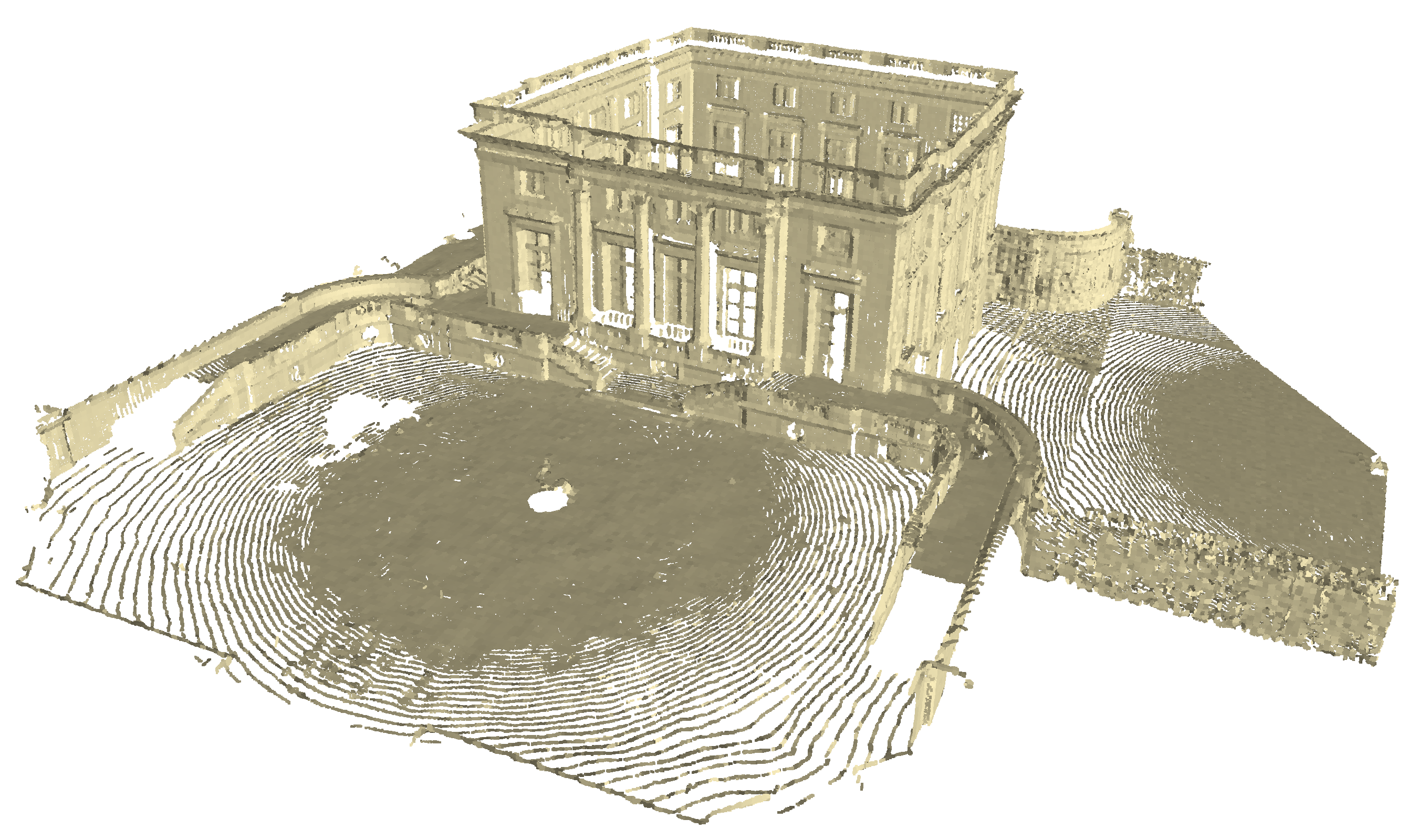}}};
\end{tikzpicture}
 
\caption{Resampling the Trianon point set with close-up views. Top : original with $200K$ points; Bottom: result ($2.2M$ points). Notice how the details are enhanced through the resampling while preserving the point set borders ($r=0.003$, $M=193$, $d=32$, shape diagonal: $1.36$). The Trianon point set is courtesy of CNRS-MAP laboratory.}
\label{fig:trianon}

\end{figure}

\begin{figure}[t]
 \centering

\setlength{\fboxsep}{0pt}
\setlength{\fboxrule}{1.5pt}

\begin{tikzpicture}
\node[anchor=south west,inner sep=0] at (0.1\linewidth, 0) {\fbox{\includegraphics[width=0.9\linewidth]{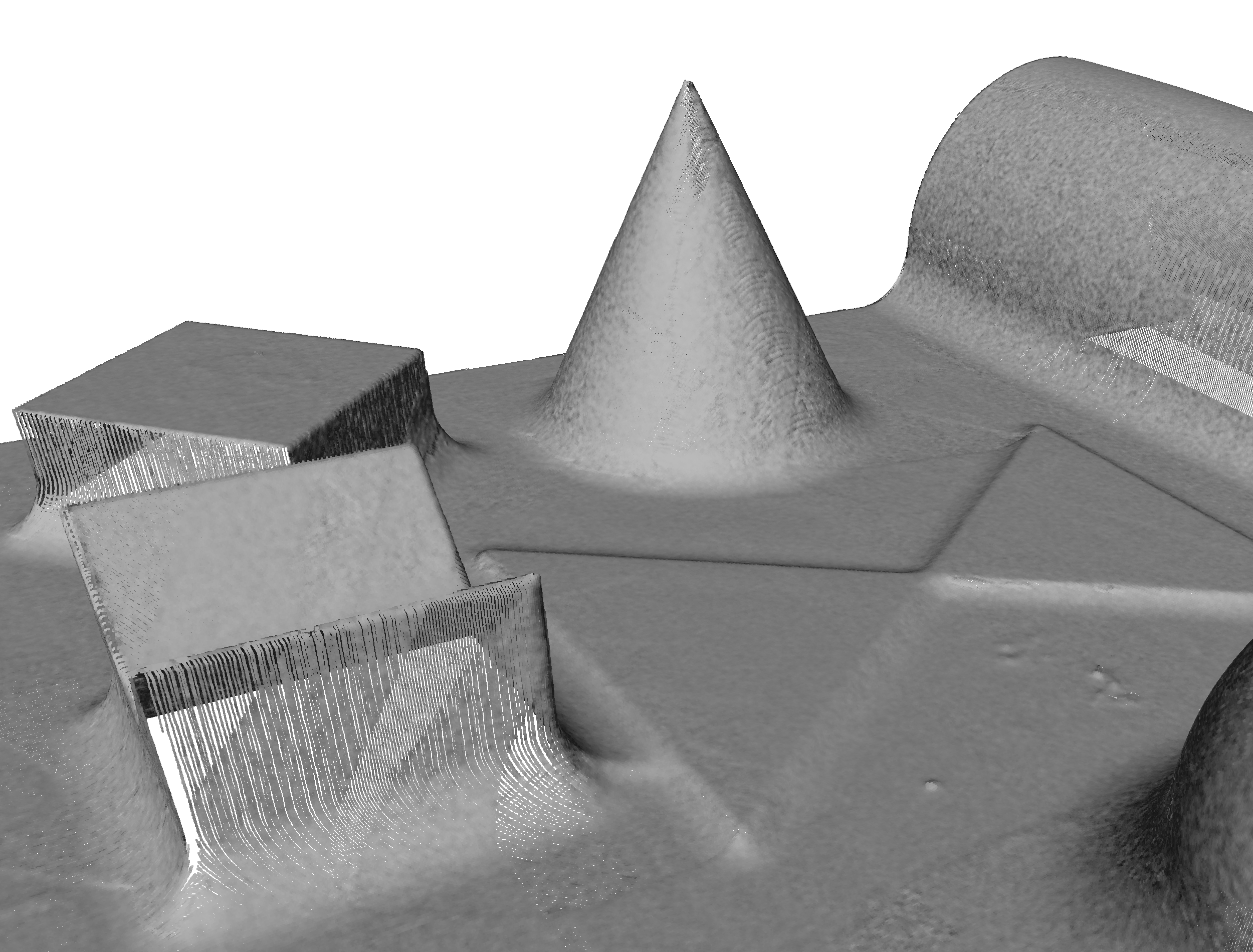}}};
\node[anchor=south west,inner sep=0] at (0, 0.48\linewidth) {\fbox{\includegraphics[width=0.25\linewidth]{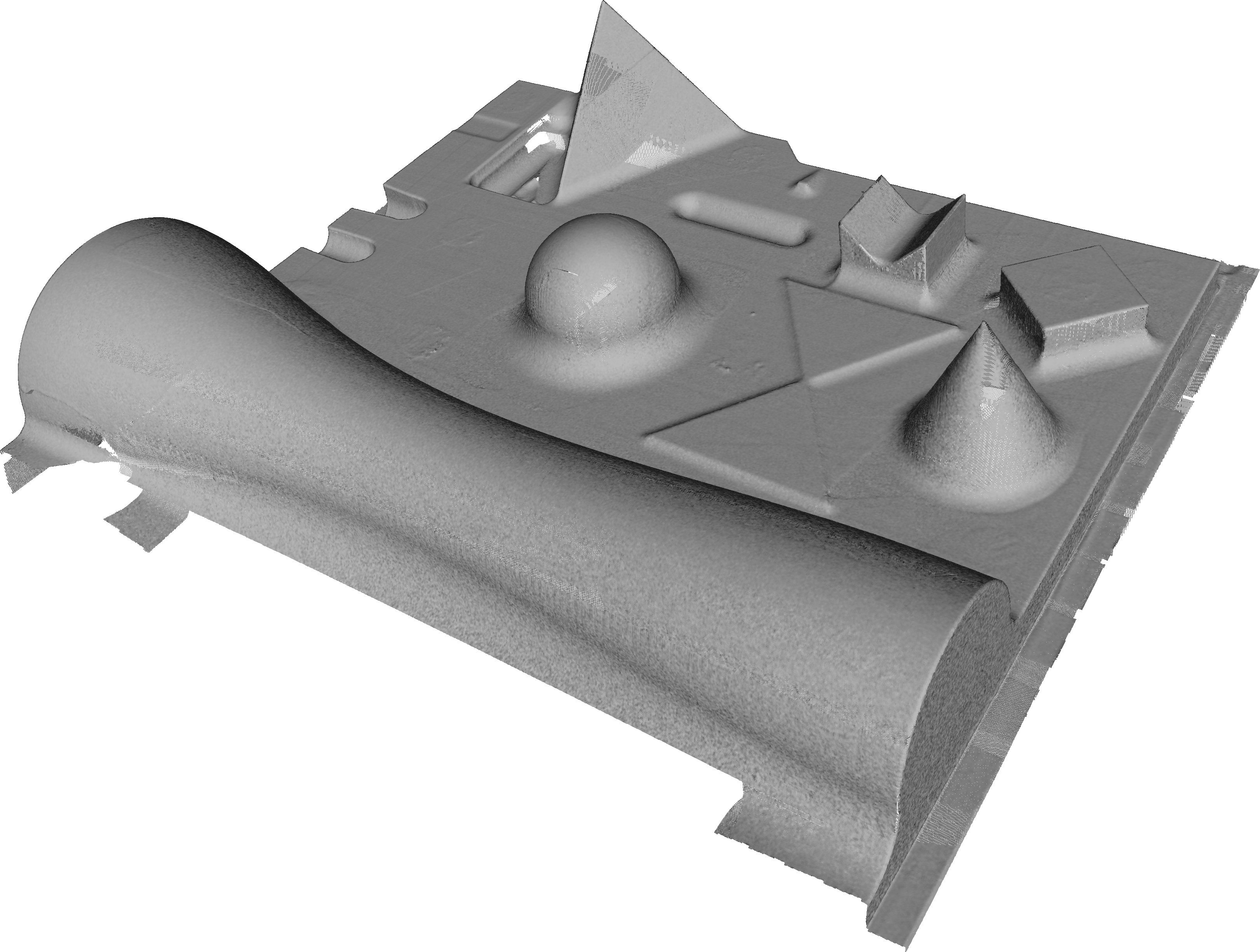}}};
\end{tikzpicture}
\vspace{5mm}

\begin{tikzpicture}
\node[anchor=south west,inner sep=0] at (0.1\linewidth, 0) {\fbox{\includegraphics[width=0.9\linewidth]{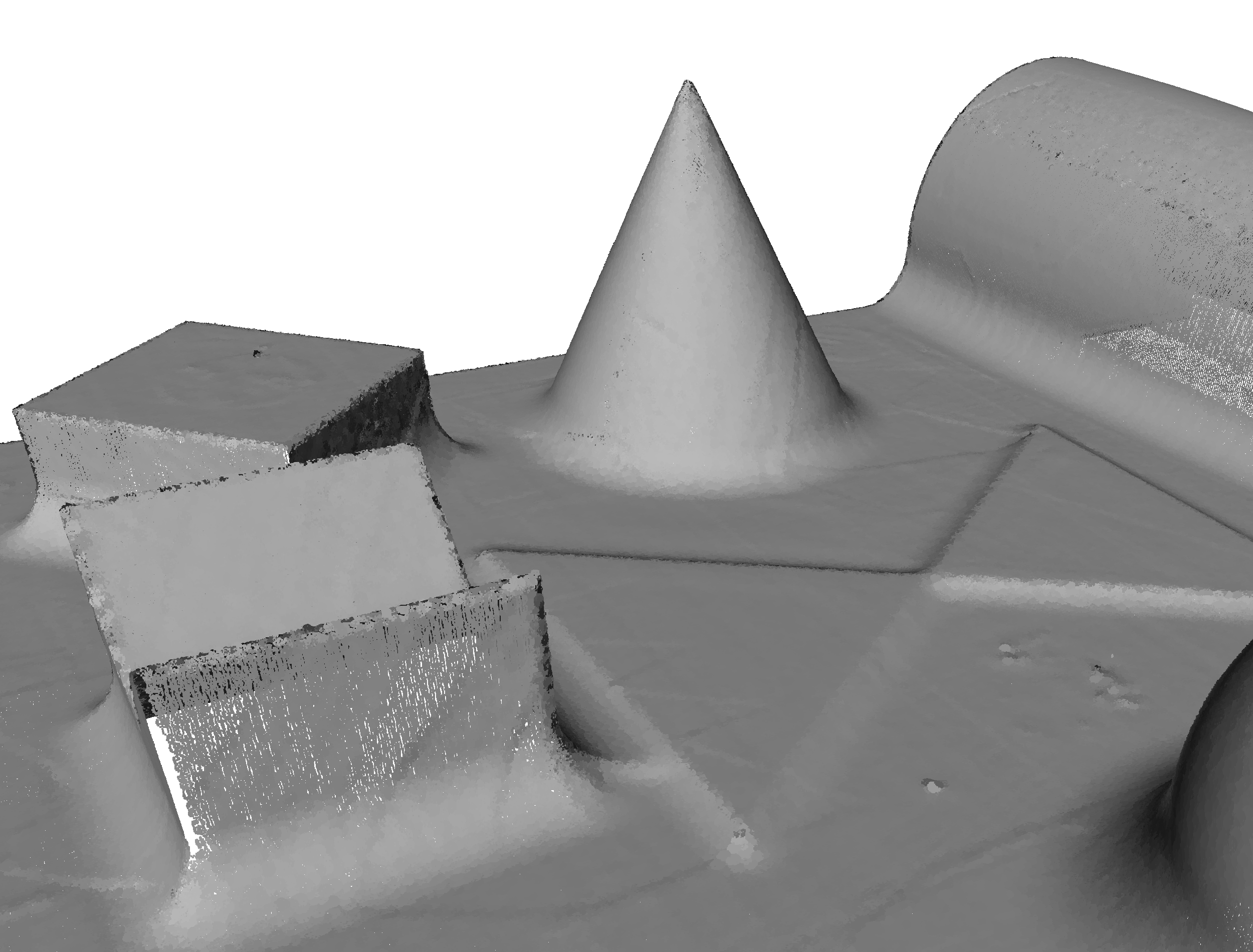}}};
\node[anchor=south west,inner sep=0] at (0, 0.48\linewidth) {\fbox{\includegraphics[width=0.25\linewidth]{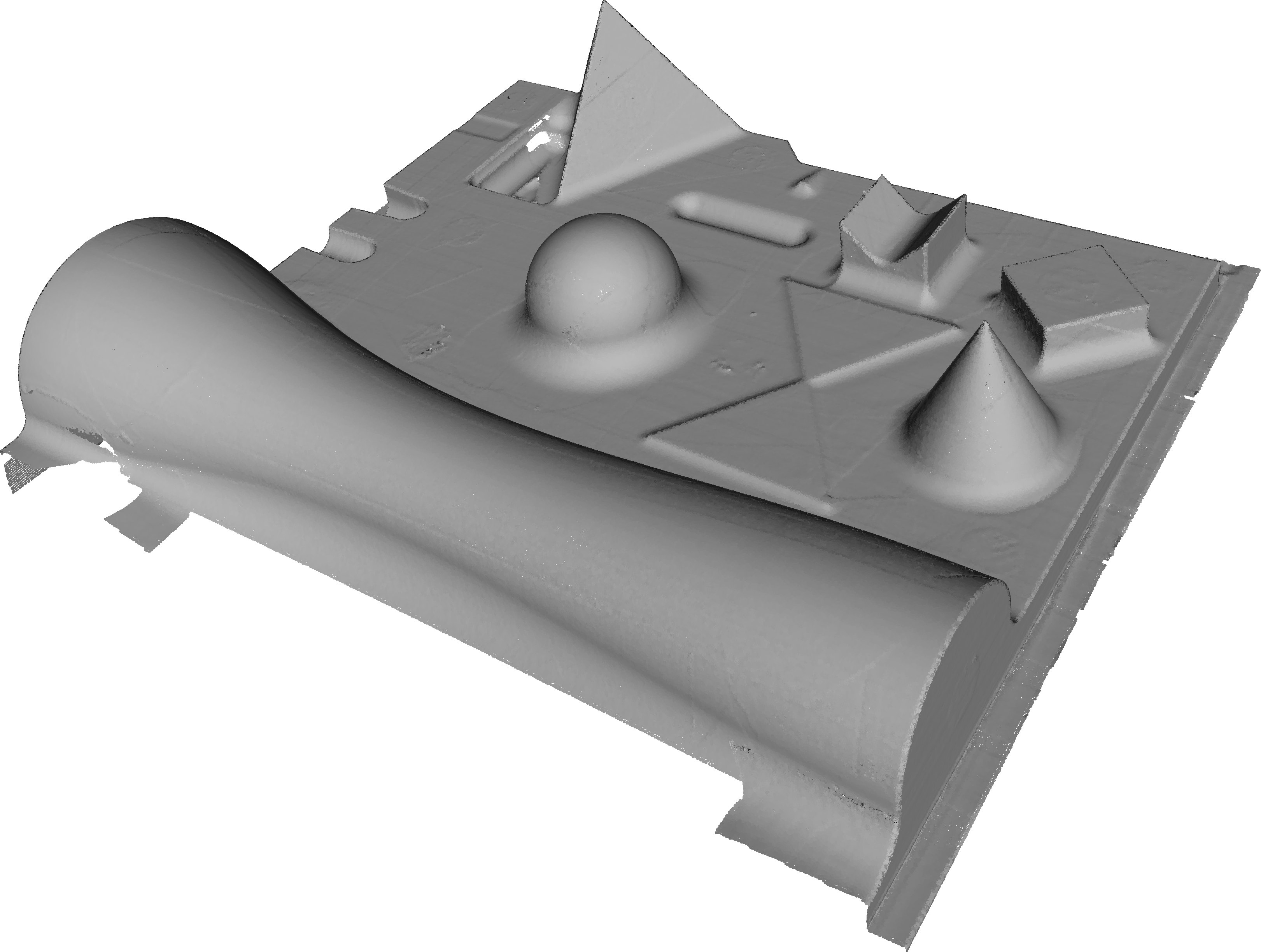}}};
\end{tikzpicture}

\caption{Resampling the mire, by keeping the same number of points ($15M$). Top: original
obtained by laser scanning technology, bottom: $\lpf$ resampling. The noise is well removed, and the
sampling is regularized all over the shape ($r=0.4$, $M=193$, $d=64$, shape diagonal: $285.35$).}
\label{fig:mire}
\end{figure}

\begin{figure}[ht]
\begin{center}
 \includegraphics[width=0.35\linewidth]{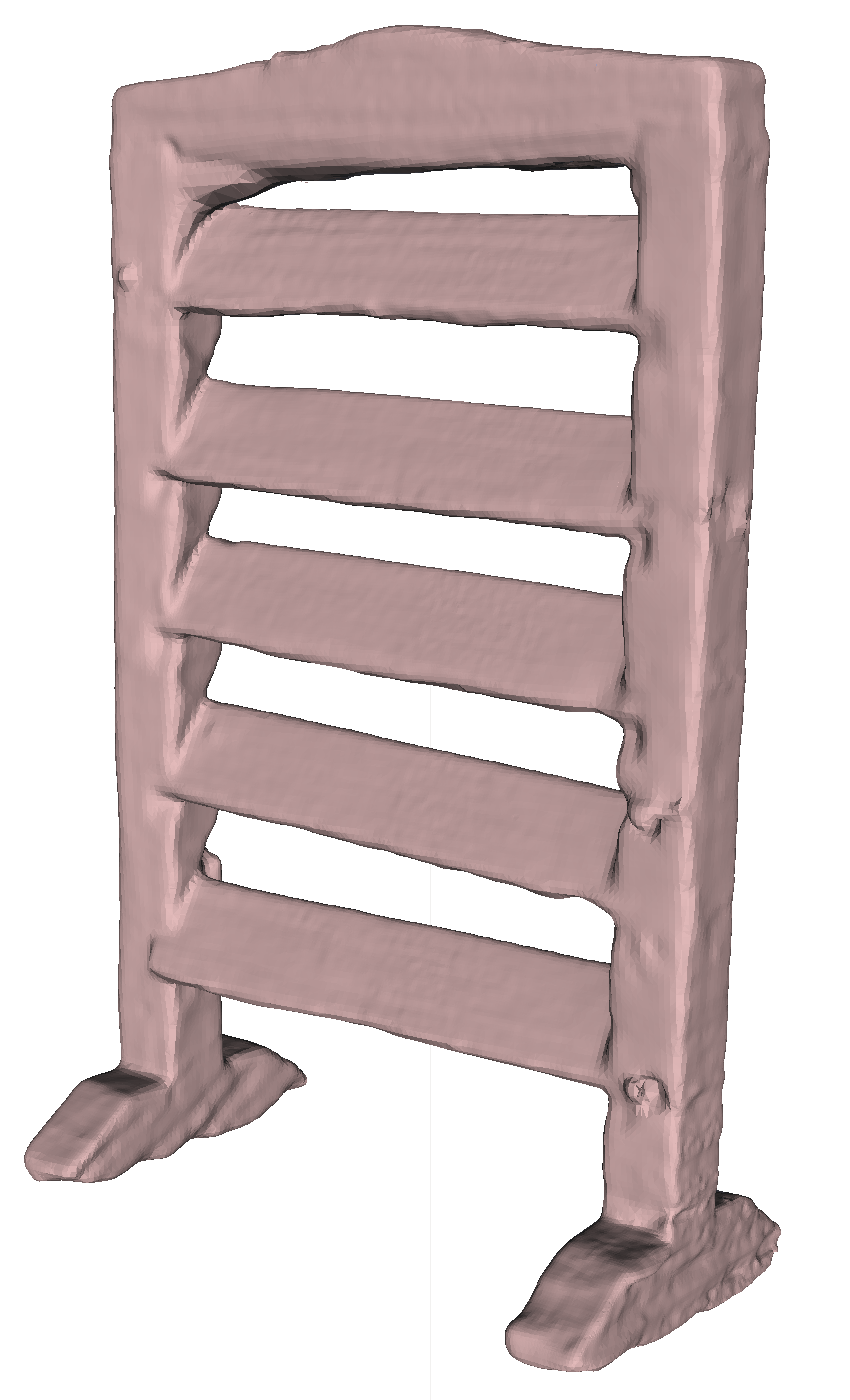}
 \includegraphics[width=0.35\linewidth]{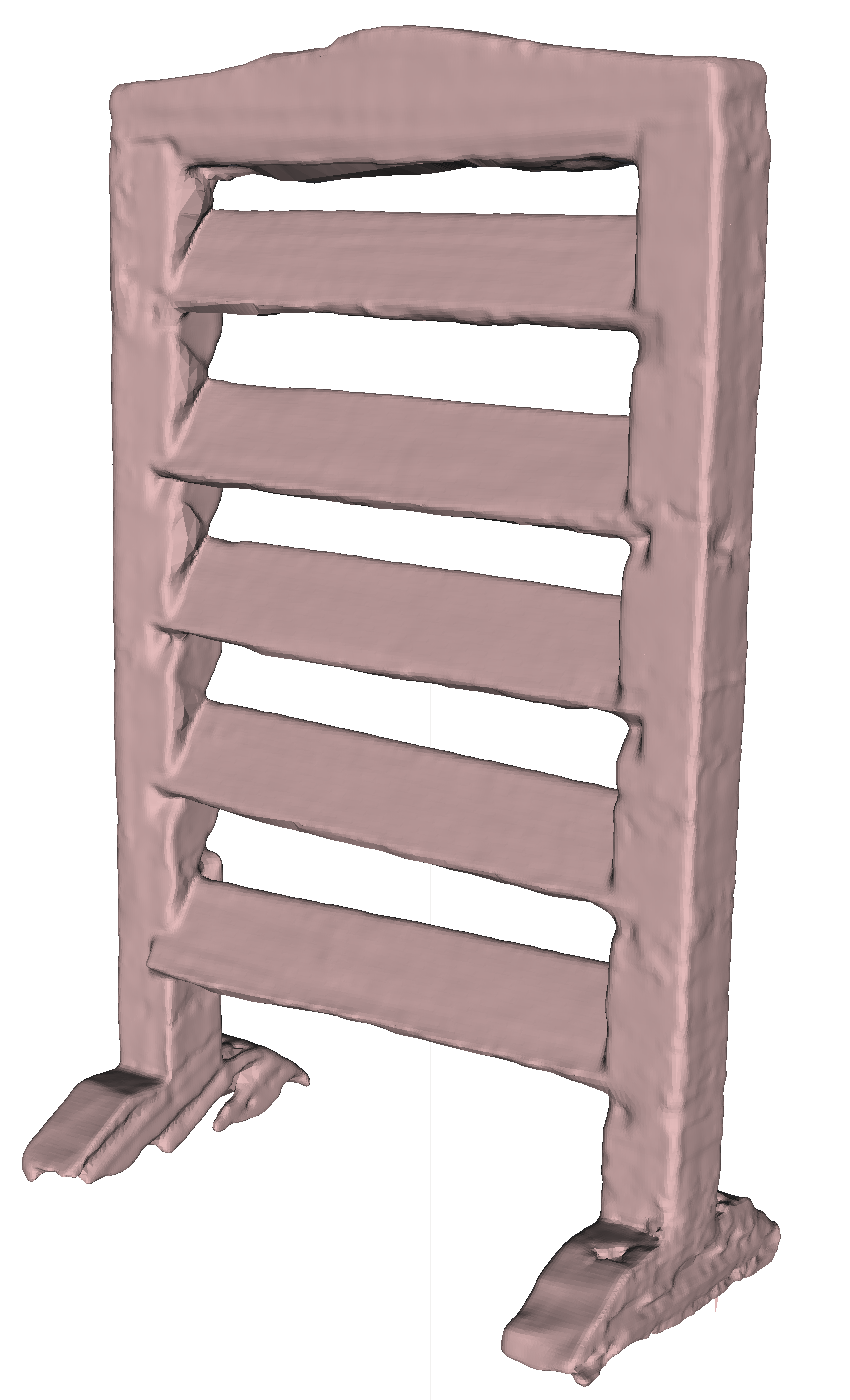}\\
 \setlength{\fboxsep}{0pt}
 \setlength{\fboxrule}{1.5pt}
 \fbox{\includegraphics[width=0.35\linewidth]{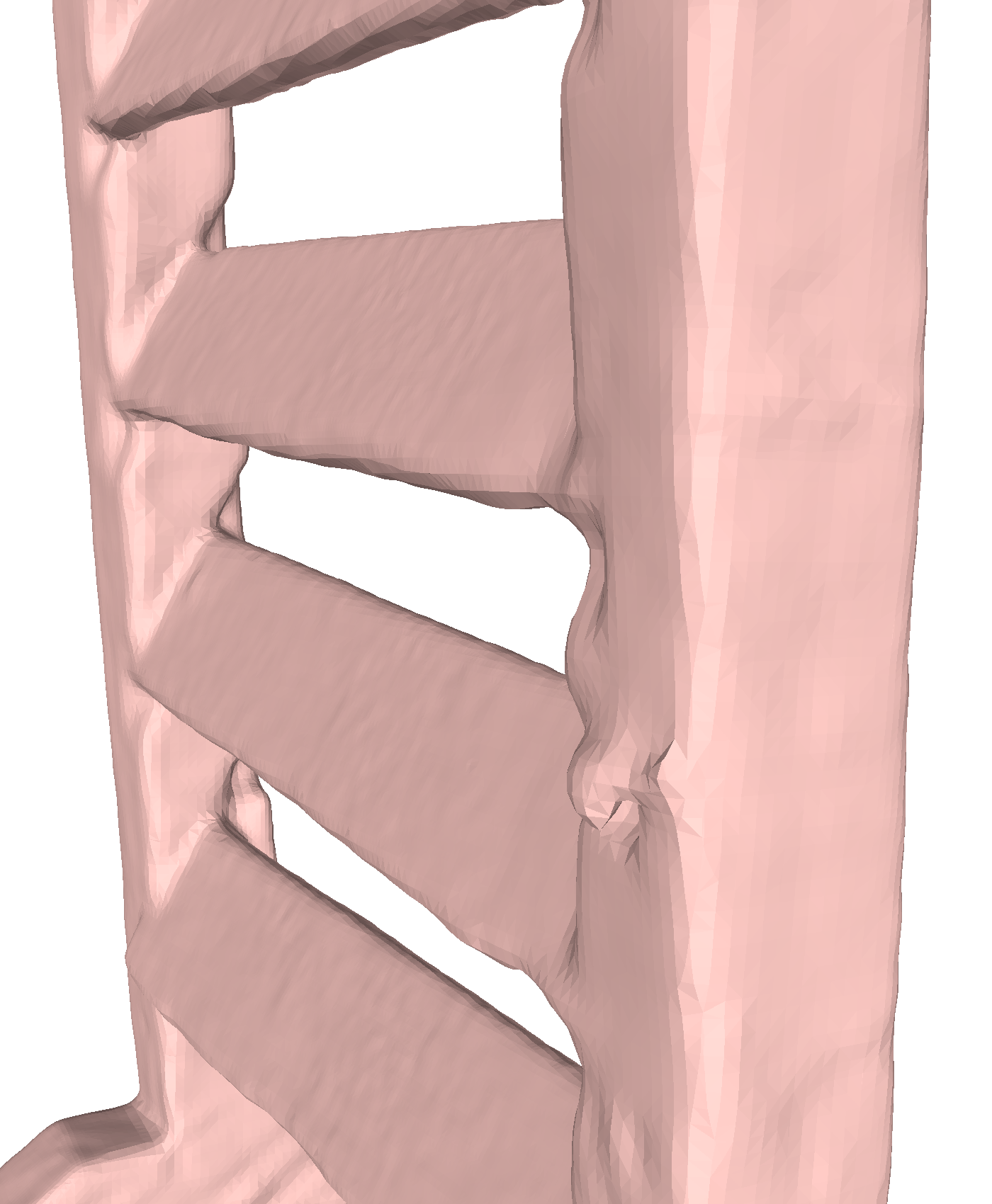}}
 \fbox{\includegraphics[width=0.35\linewidth]{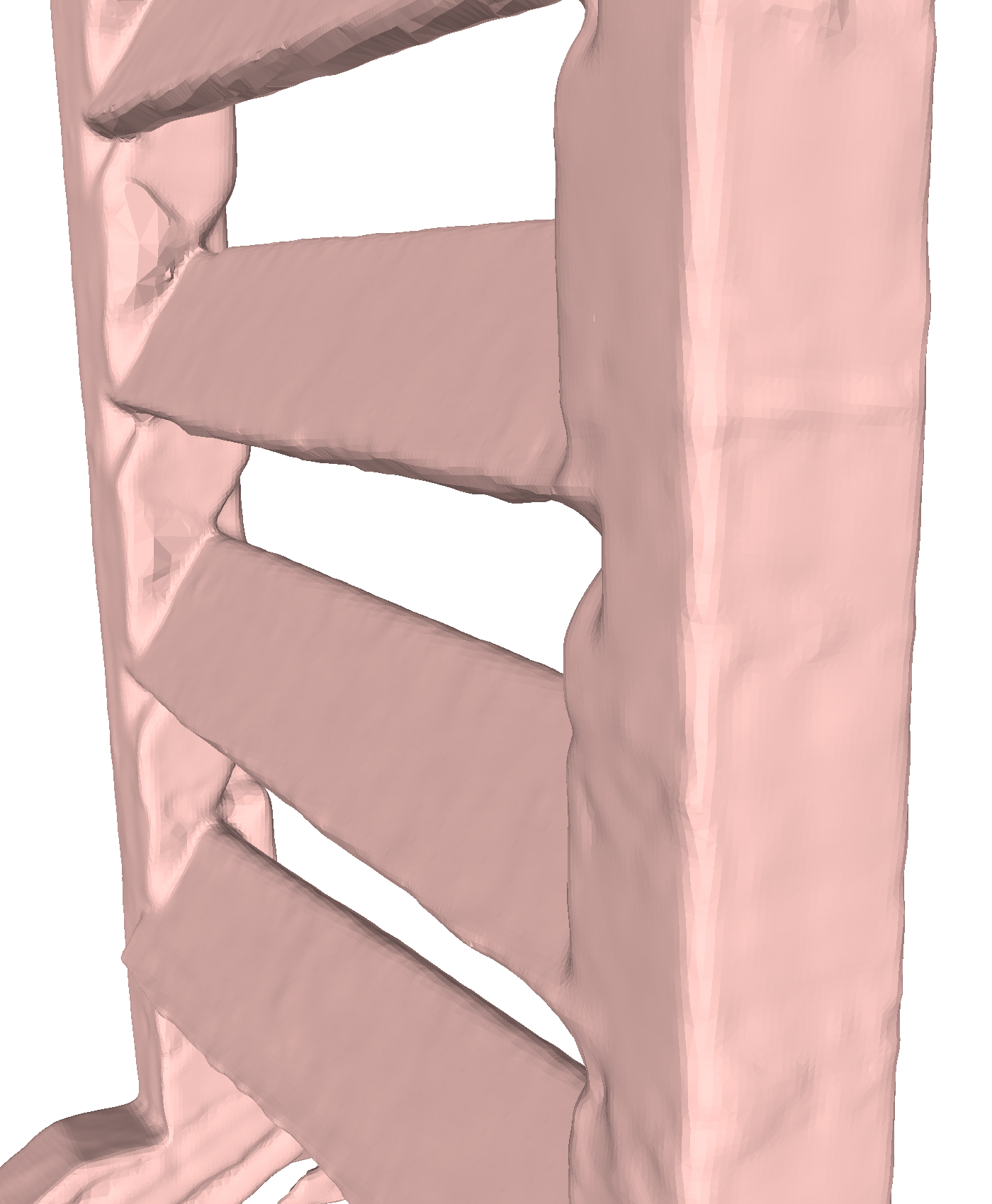}}
\end{center}
 \caption{The shutter model. Original point set obtained by laser scanning
technology (left, $290K$ points),
resampling (right, $2.0M$ points), both reconstructed with Screened Poisson
Reconstruction \protect\cite{screened_poisson} using the same parameters. Notice
the enhancement of the sharp edges using our shape learning process ($r=0.02$, $M=193$, $d=64$, shape diagonal: $2.07$).}
 \label{fig:shutter}
\end{figure}

We experimented our framework on shapes represented as surface meshes (Figures
\ref{fig:sphere}, \ref{fig:fandisk}) and point sets (e.g. Figures \ref{fig:ship},
\ref{fig:shutter}).
On a LiDaR point set (Fig. \ref{fig:trianon}), our $\lpf$ framework yields
a detail-enhancing resampling which preserves well the point set borders.
We first illustrate how the resampling removes the noise in a feature preserving
way (Figures \ref{fig:mire}, \ref{fig:shutter}), in particular for open
surfaces.
On Fig. \ref{fig:shutter}, we show the result of applying the Screened Poisson
Reconstruction \cite{screened_poisson} after resampling the shutter: the sharp
edges and details are much better enhanced.

\begin{figure}
\centering
\subfigure[Original]{\includegraphics[width=0.45\linewidth]{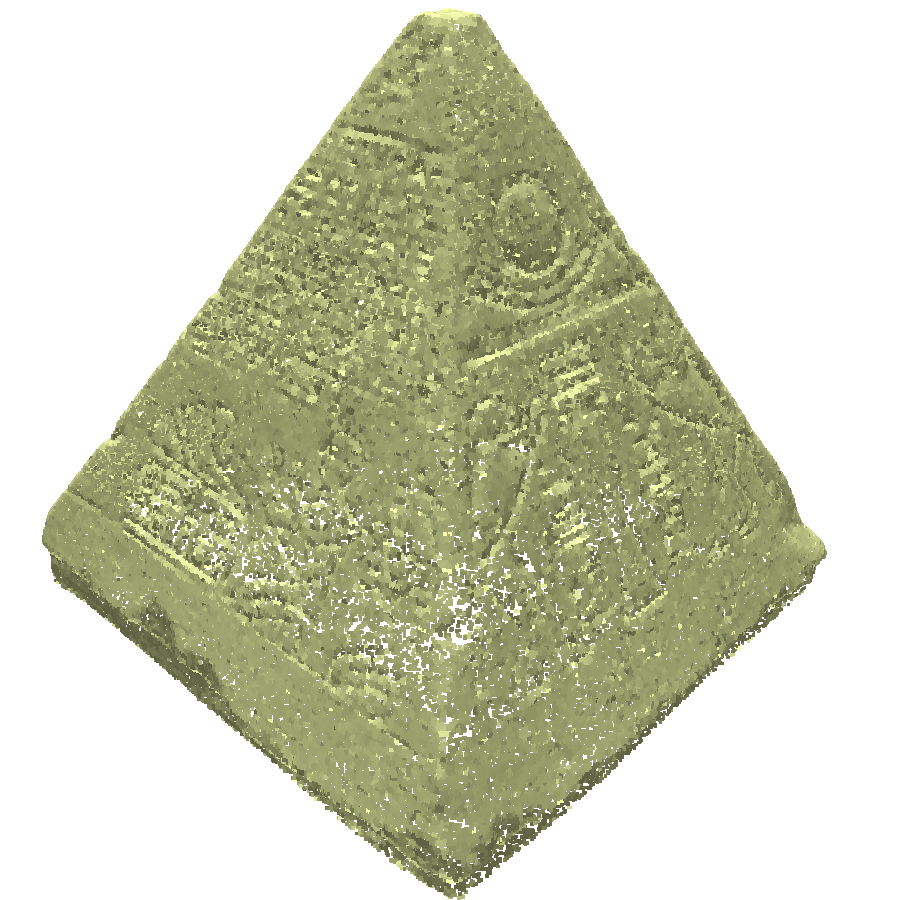}}
\subfigure[MLS]{\includegraphics[width=0.45\linewidth]{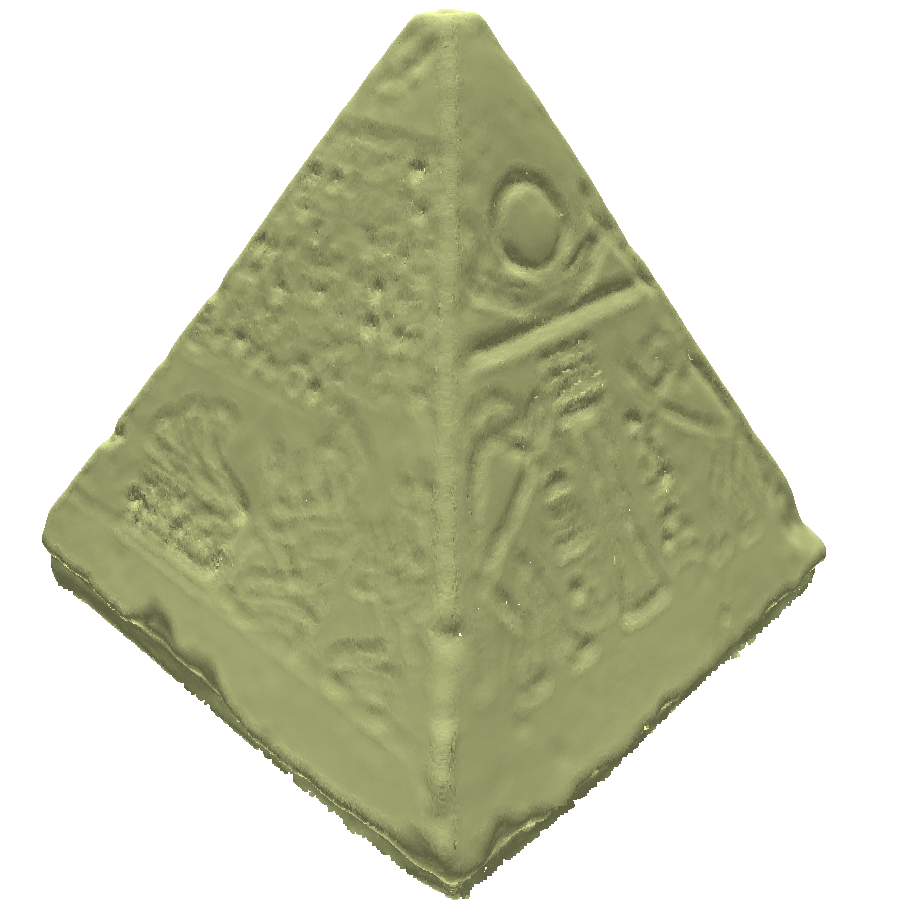}}
\subfigure[EAR]{\includegraphics[width=0.45\linewidth]{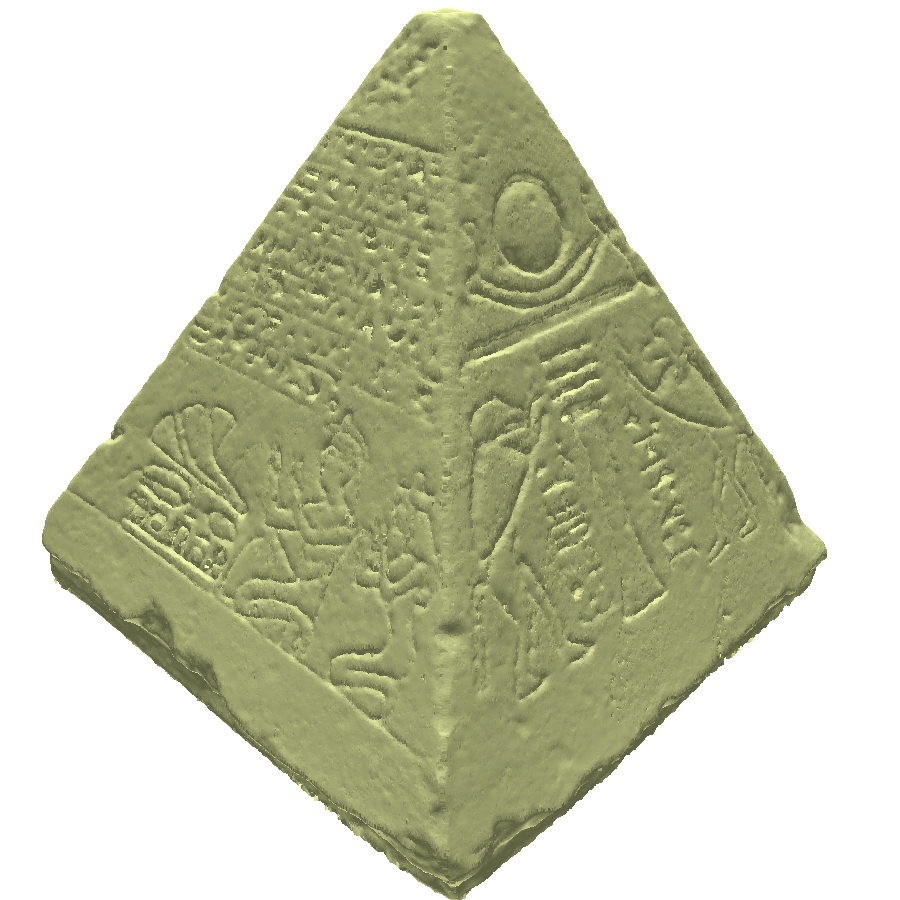}}
\subfigure[\textbf{LPF}]{\includegraphics[width=0.45\linewidth]{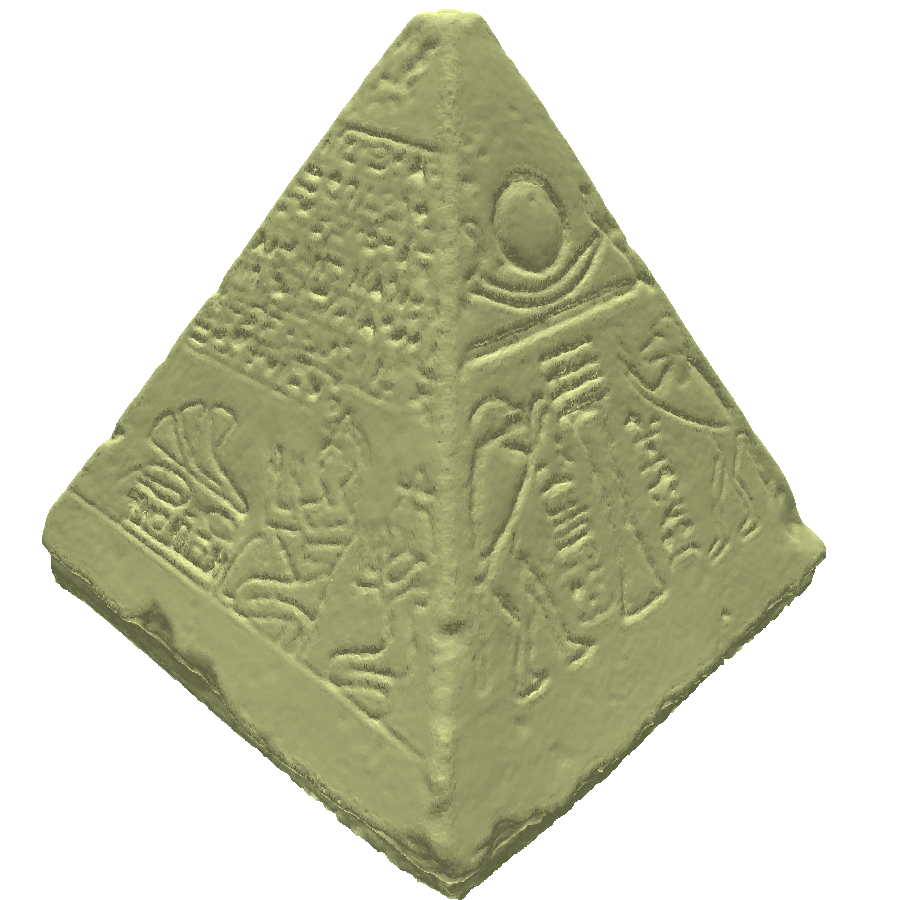}}
\caption{The pyramid (initial: top left, $280K$ points) resampled using standard
MLS (top right), EAR (bottom left) and our analysis framework (bottom right).
EAR and LPF both manage to recover the details. All
resampled point sets have the same number of points ($2.2M$ points), and
their normals are recomputed using Hoppe et al.'s method \protect 
\cite{surf_rec} with the same parameters ($r=0.8$, $M=793$, $d=64$, shape diagonal: $125.51$).}
\label{fig:pyramid}
\end{figure}

We compare it with EAR \cite{edge_aware_point_set_resampling} on a pyramid shape with
engravings (Fig.~\ref{fig:pyramid}) and sparse initial sampling.
EAR and LPF both outperform Moving Least Squares and permit to upsample the shape while recovering
the engravings.
On the contrary, on the cube with curve model or ship point set, EAR is bound
to fail at resampling faithfully the curve because of its manifold surface assumption (Figures
\ref{fig:ship}, \ref{fig:cubecurve}).

\begin{figure}
 \begin{center}

\begin{tikzpicture}
 \setlength{\fboxsep}{0pt}
 \setlength{\fboxrule}{1.5pt}

\node[anchor=south west,inner sep=0] at (0, 0)
{\fbox{\includegraphics[width=0.25\linewidth]{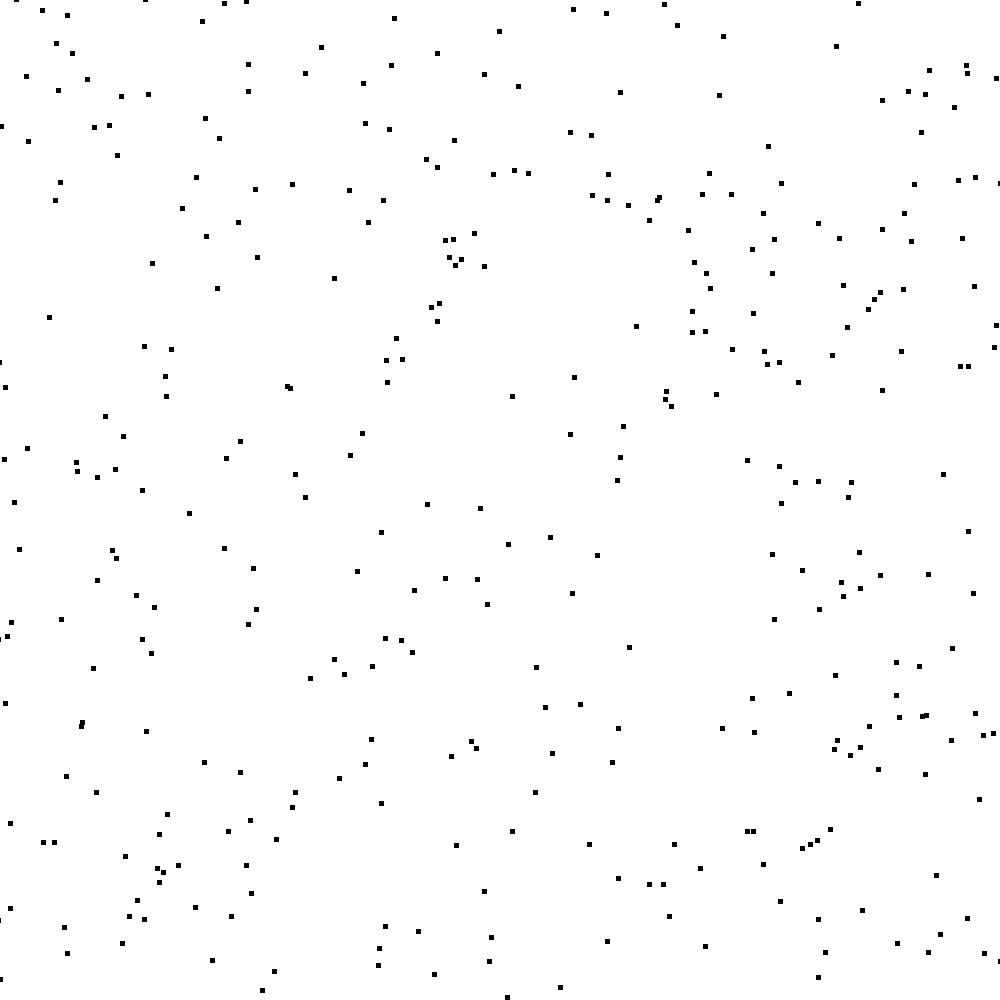}}};
\node[anchor=south west,inner sep=0] at (0.25\linewidth,0)
{\fbox{\includegraphics[width=0.25\linewidth]{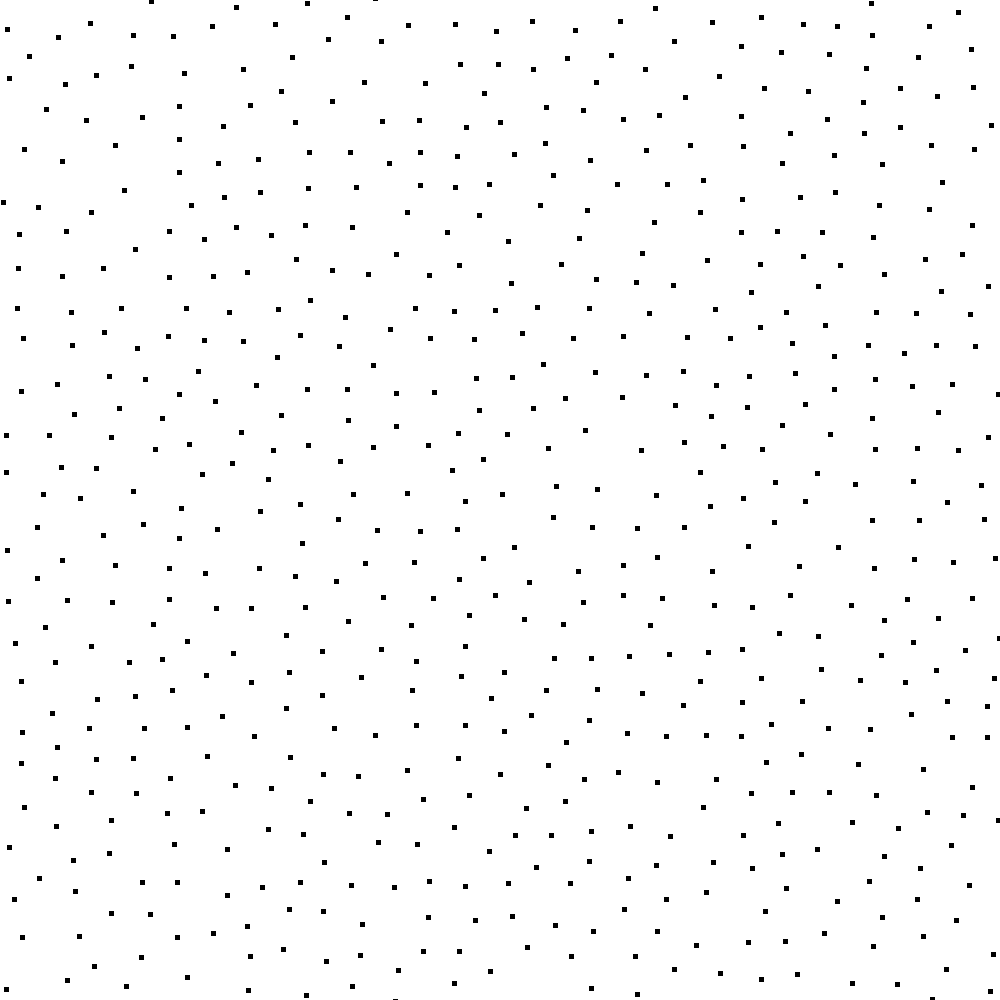}}};
\node[anchor=south west,inner sep=0] at (0.50\linewidth, 0)
{\fbox{\includegraphics[width=0.25\linewidth]{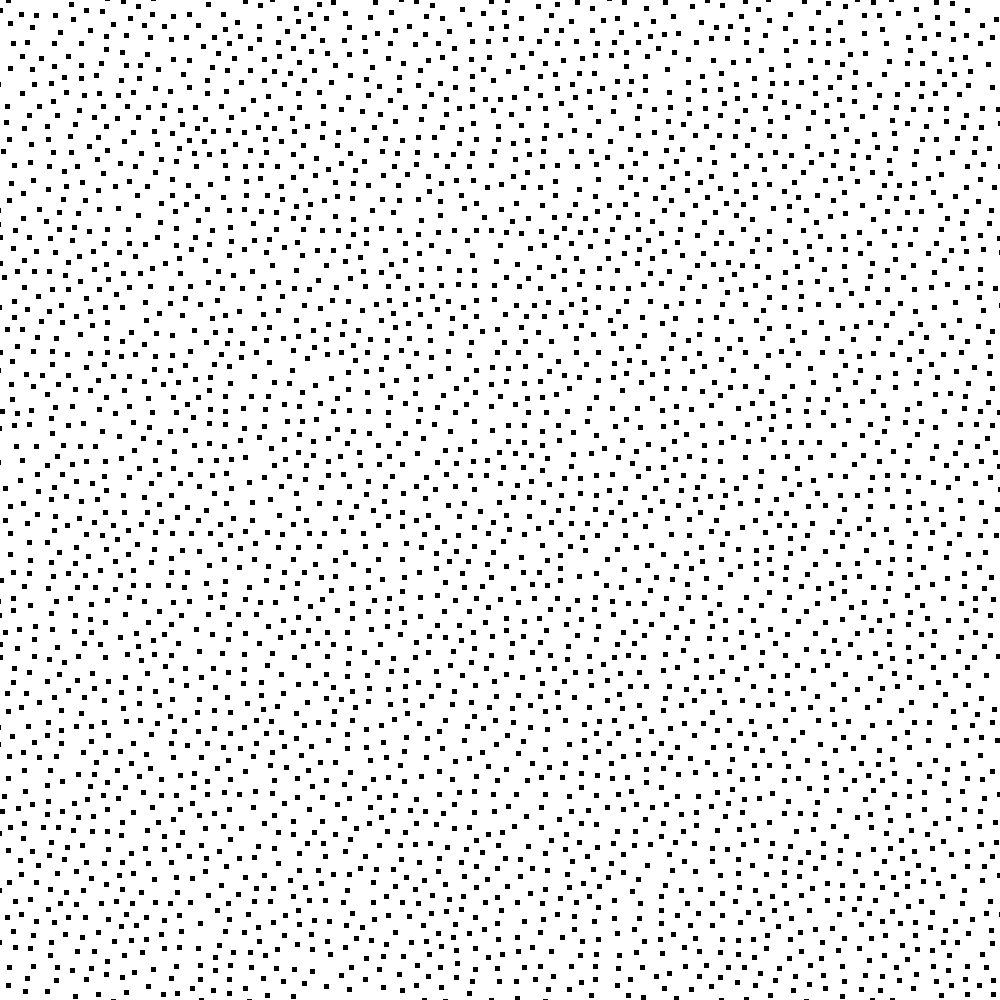}}};
\node[anchor=south west,inner sep=0] at (0.75\linewidth,0)
{\fbox{\includegraphics[width=0.25\linewidth]{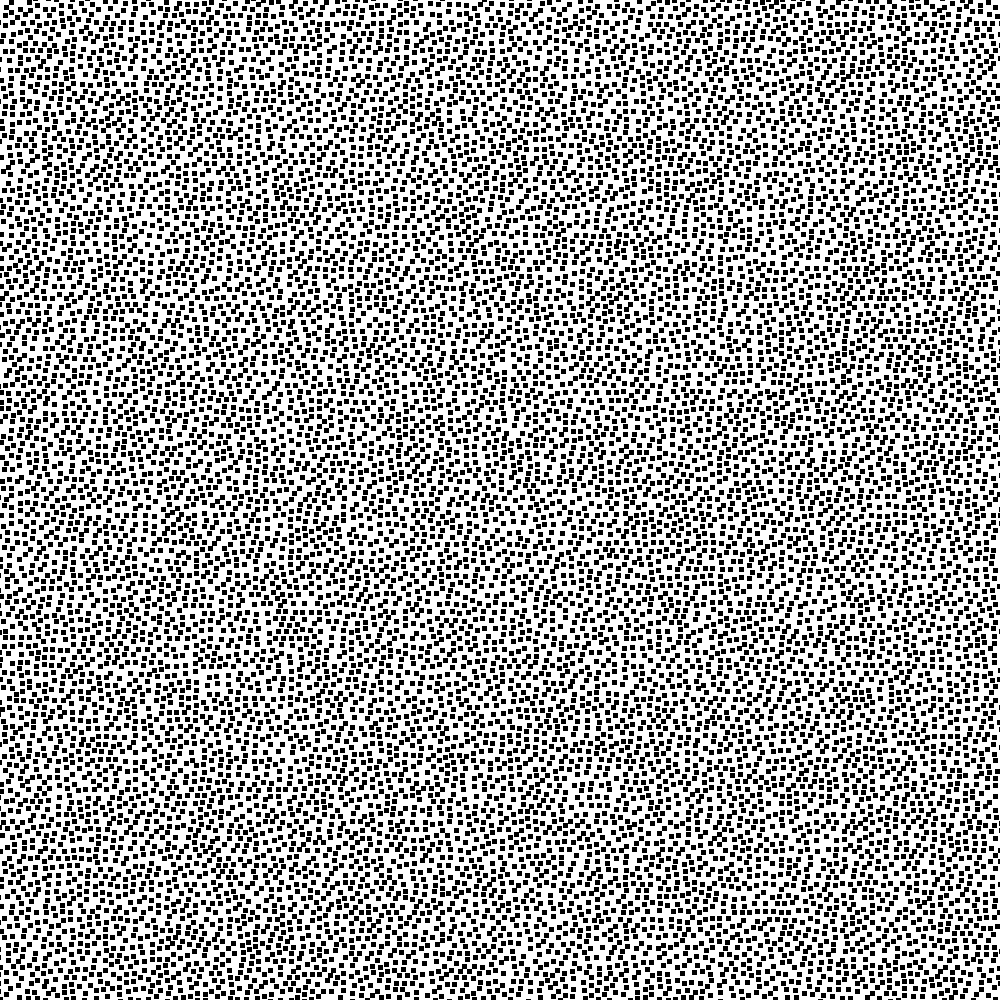}}};

\node[anchor=south west,inner sep=0] at (0.25\linewidth, 0)
{\includegraphics[width=0.25\linewidth]{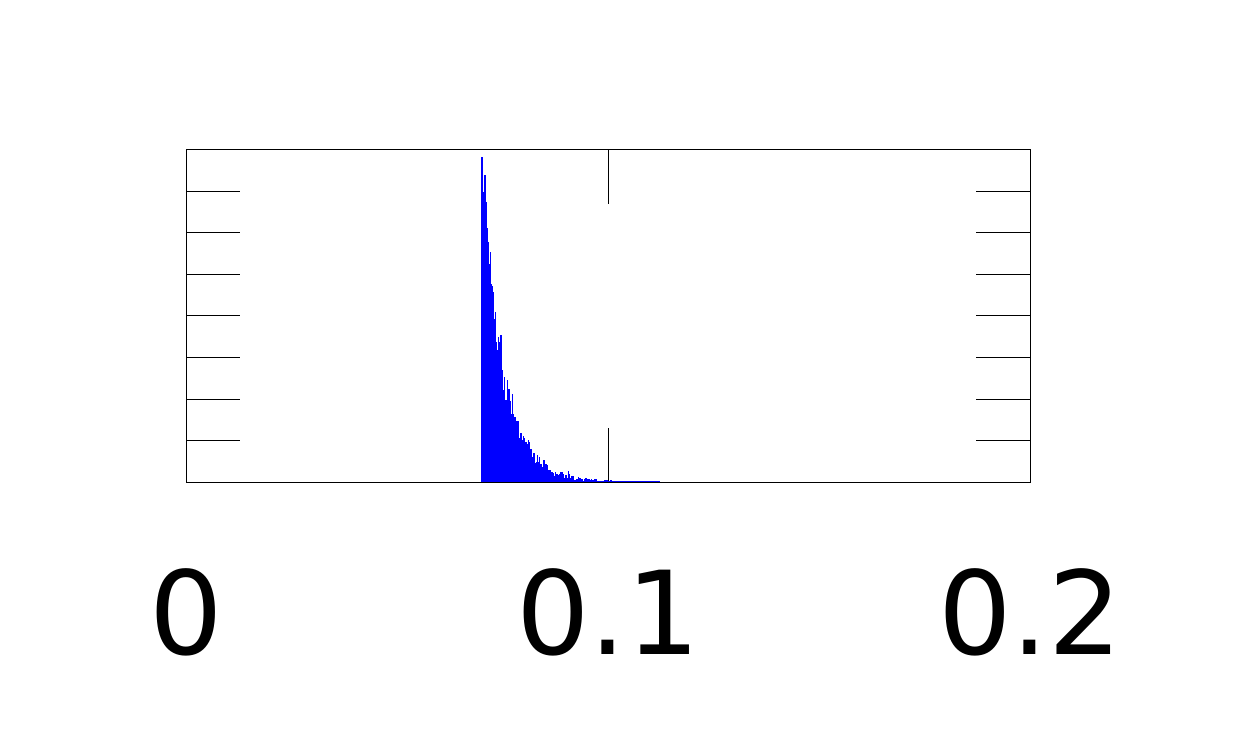}};
\node[anchor=south west,inner sep=0] at (0.50\linewidth, 0)
{\includegraphics[width=0.25\linewidth]{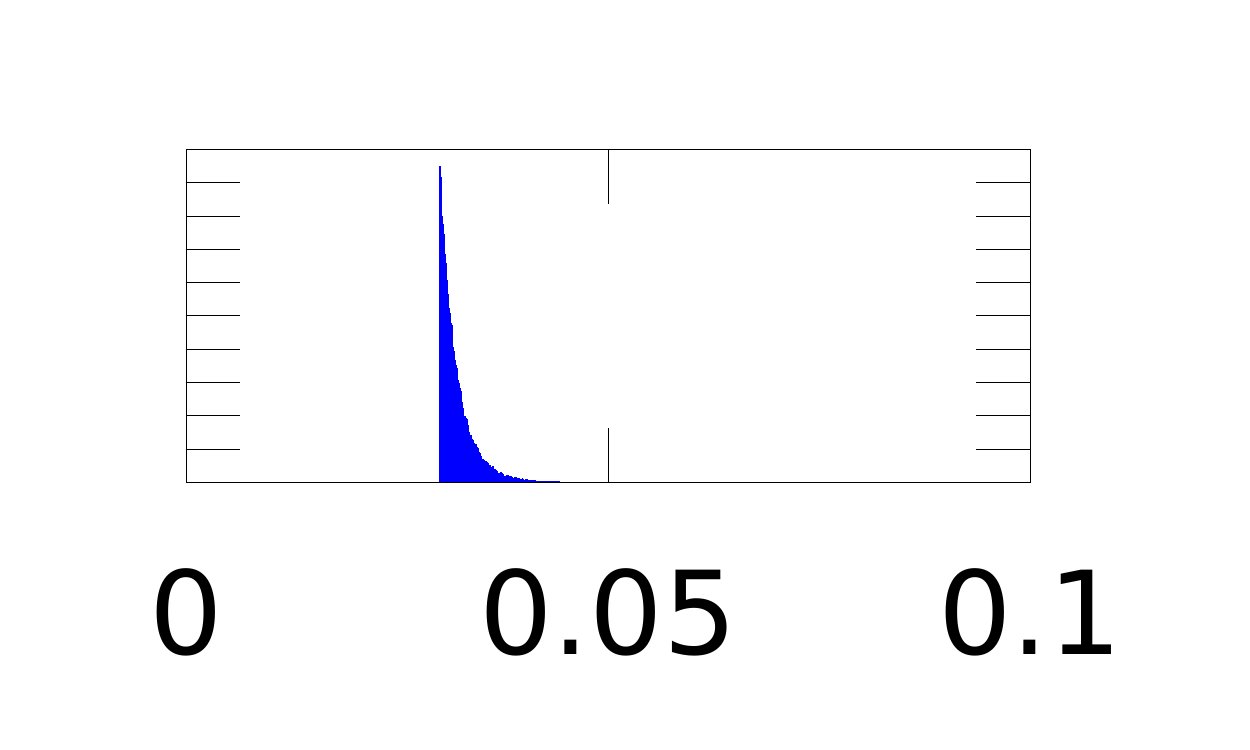}};
\node[anchor=south west,inner sep=0] at (0.75\linewidth, 0)
{\includegraphics[width=0.25\linewidth]{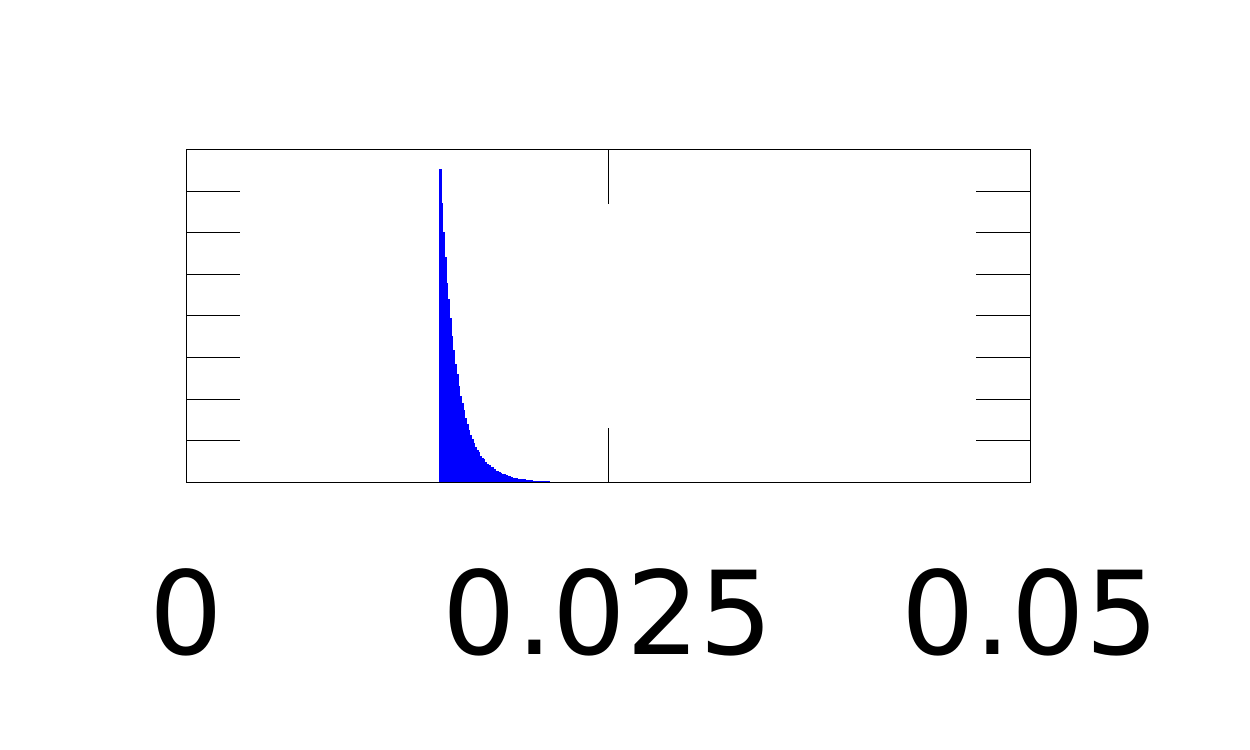}};
\end{tikzpicture}

\end{center}
\caption{Resampling a randomly sampled perfect plane. Effect of the sampling
rate of the pattern, the same radius $r$ is kept, but the
pattern grid size is changed. From left to right: original, $M=193, 793,
3205$. For each value, nearest neighbor distance histograms are also shown.}
\label{fig:plane}
\end{figure}

The obtained shape sampling rate is guided by the sampling rate of the pattern. This can experimentally be seen on Fig.~\ref{fig:plane}, where a planar
grid is used as pattern.
In this experiment of a perfect plane resampling, the mean distance between two
points is strongly linked to the distance between two points in the resampling
pattern. If the pattern is a $16\times 16$ grid (thus $M=193$
points once intersected with a sphere) and $r=1$, the grid step is $0.06$ and
the measured mean is $0.075$. Similarly for a $32\times 32$ grid and $M=793$
(\emph{resp.} a $64\times 64$ grid and $M=3205$) the grid step is $0.0312$
(\emph{resp.} $0.015$) and the measured mean $0.0320$ (\emph{resp.} $0.015$).
The histograms of the distance to the nearest neighbors are also shown on
Fig.~\ref{fig:plane}.
These distributions are very close to what Poisson disk sampling would do,
which is due to the consolidation process. One can see that the sampling
remains stable and the density is driven by the distance between points on the
pattern.

One important feature of our method is its ability to learn the local dimension from the shape,
which can be seen on the cube with curve point set (Fig.~\ref{fig:cubecurve}). EAR being
particularly well suited for edge recovery, it performs better on the edges of the cube but it
fails to recover the curve structure whenever it is too close to the shape. On the contrary,
$\lpf$
is able to resample it. This advantage of our method can also be seen on the
ship example (Fig.~\ref{fig:ship}). We have enhanced the MLS-based resampling
with an anisotropic behavior: we measure a local stretch ratio $r_s$, the ratio
between the two largest
covariance matrix eigenvalues. In the regions where $r_s$ is
below  a threshold ($3$ in our case), MLS sampling occurs as expected, while
regions where $r_s$ is above the threshold are sampled as curves.
This modification improves the sampling of the curves, but it still tends
to blur the riggings, and
creates spurious points. EAR fills the gaps between the ship's rigging, whereas $\lpf$ nicely
enhances them.

\subsection{Point Set Denoising} \label{sec:denoising}

The second application of our framework is point set denoising. Similarly to
\cite{elad2006}, we start from the initial point set $S_0$ and aim at
finding the denoised point set $S$ minimizing the following objective function:

\begin{equation}
\min_{S,D,\alpha} \| S - S_0 \|^2 + \gamma\sum _{j = 0}^{N-1} \| R_j(S) -
D\alpha_j \|_F^2 + \lambda \sum_{j = 0}^{N-1} \| \alpha_j \|_1
\label{eq:denoising_energy}
\end{equation}

where $R_j$ is the operator that extracts the $\lpf$ at position $s_j$
(\emph{i.e.} $R_j(S) = V_j$) and $\gamma,\lambda$ are two parameters
constraining
the data attachment and the sparsity.
$\|S - S_0\|^2$ corresponds to the squared distance between the initial noisy point set and the current point set, measured as the sum of squared distances from points in $S$ to their nearest neighbor in $S_0$.
This objective function is hard to minimize but is in fact closely linked to our
minimization defined in Equation (\ref{eq:energy}).
We solve it by splitting it into two steps that are iterated until the point set is stable

\begin{itemize}
 \item The first step learns the dictionary and coefficients as well as the
$\lpf$ positions and orientations.
 \item The second step uses the dictionary to find the best denoised point set.
\end{itemize}

The first step is exactly our shape analysis defined in Equation
(\ref{eq:energy}).
Then, once the dictionary and coefficients are fixed, minimizing
(\ref{eq:denoising_energy}) amounts to solving in the second step:

\begin{equation}
\min_{S} \| S - S_0 \|^2 + \gamma\sum _{j = 0}^{N-1} \| R_j(S) -
D\alpha_j \|_F^2.
\end{equation}

Minimizing this energy means finding for each point $q$ the best consensus
among all $\lpf$s that describe the vicinity of $q$. More precisely, each
initial point $q$ relates to a set of consolidated $\lpf$s providing \emph{better}
candidate positions for this point, i.e. positions that fit better the
dictionary decomposition.
A \emph{better} position estimation can thus be computed, similarly to the shape
resampling application: starting with an initial point position $q$, each
neighboring $\lpf$ proposes a new candidate position for $q$ by projecting $q$ onto
the pattern support plane and computing a position $q_i$ from the set of
$v_is$. All candidate positions are then averaged as:

\begin{equation}
\tilde q = \frac{\sum_i w(q,q_i) q_i}{\sum_i w(q,q_i)}
\text{ with } w(q,p) = \exp{-\frac{\|p-q\|^2}{2\tau_p^2}}.
\label{eq:denoising}
\end{equation}
The point $q$ is then moved toward the proposed position at a rate $\gamma$ : 
\begin{equation}
q_{denoised} = \frac{ q + \gamma \tilde q}{1+\gamma}
\end{equation}
and this process is repeated until the denoised positions are stable (or after
a chosen number of iterations).

\begin{figure}
\centering

\includegraphics[width=0.325\linewidth]{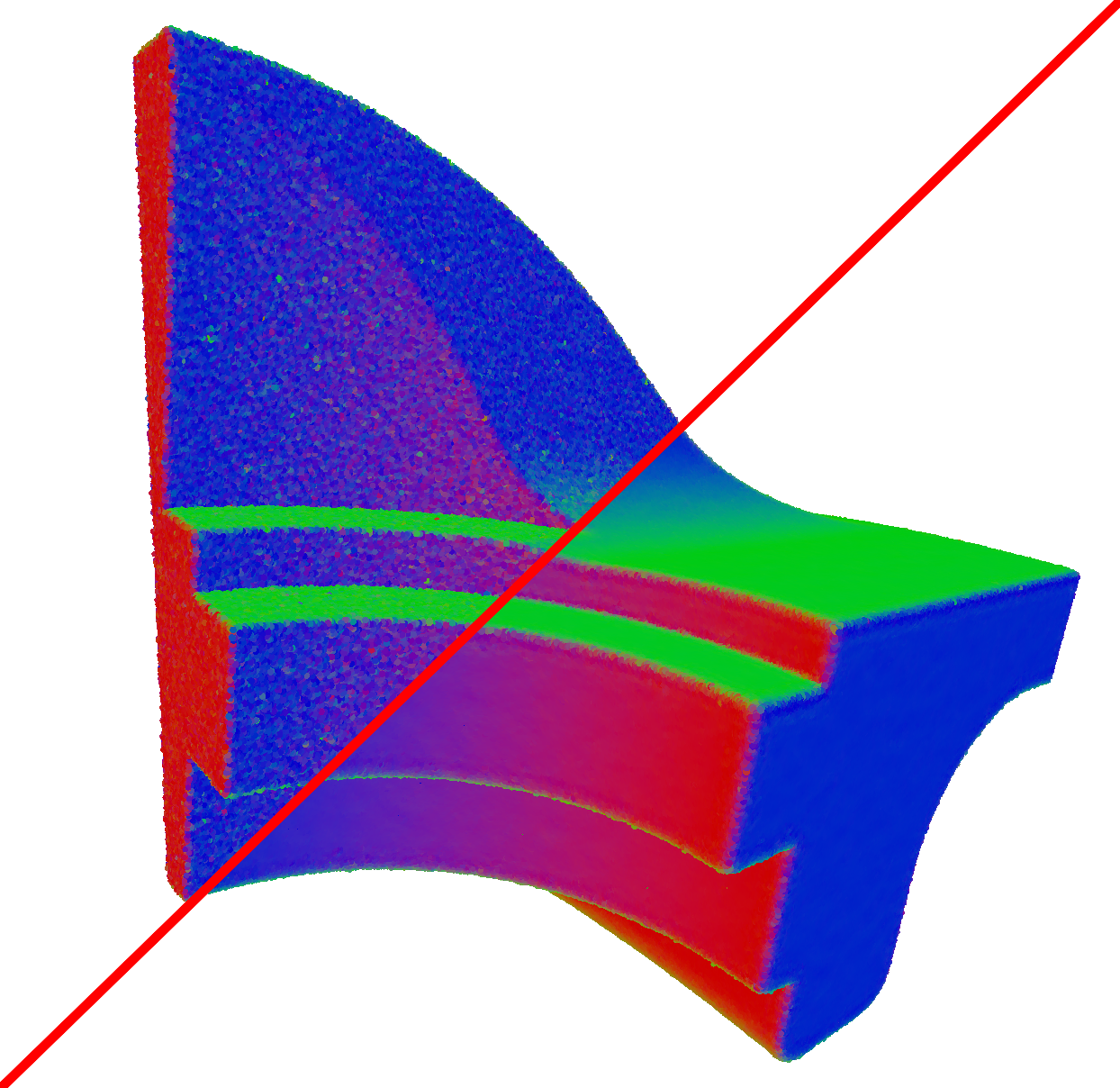}
\includegraphics[width=0.325\linewidth]{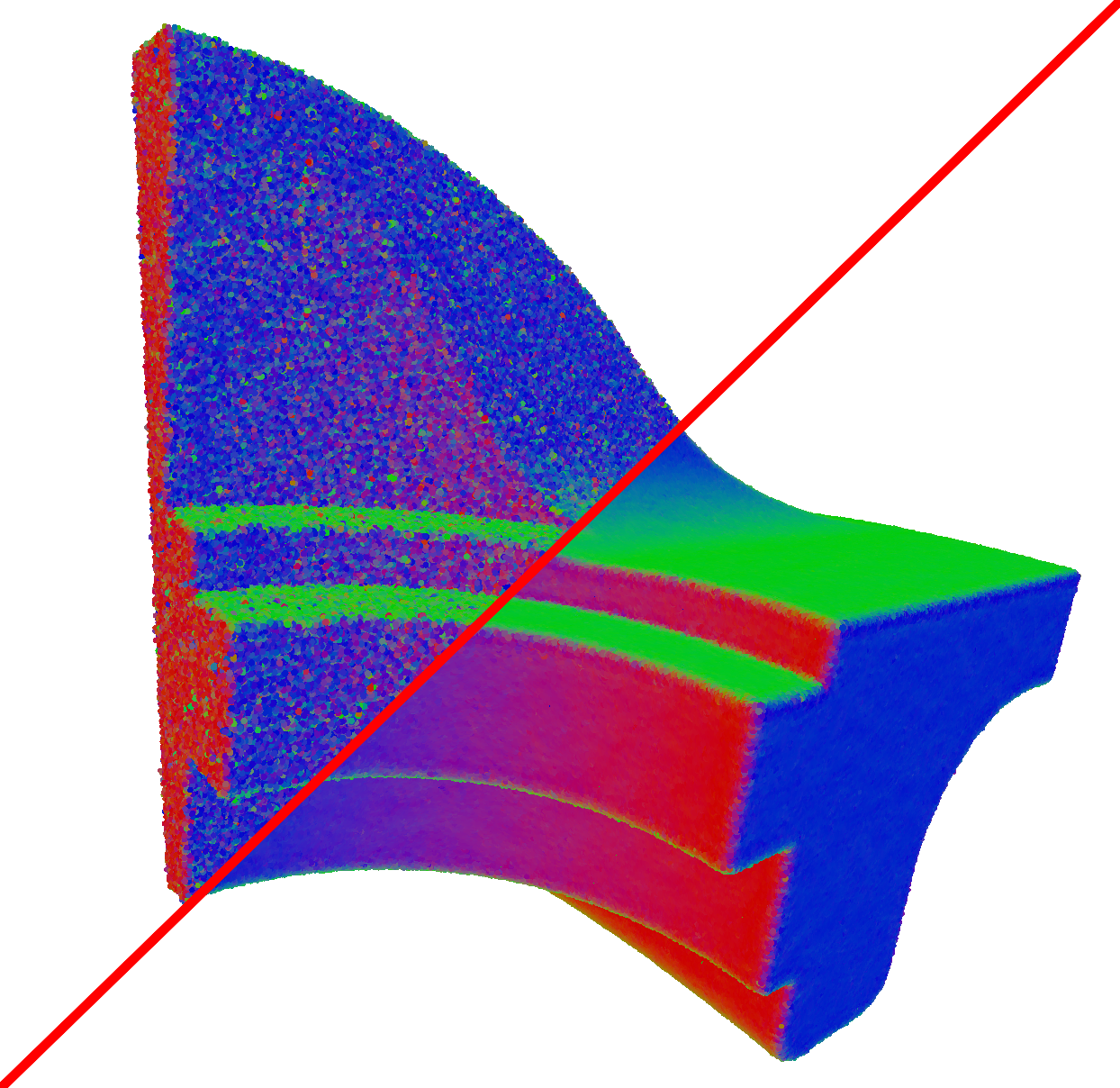}
\includegraphics[width=0.325\linewidth]{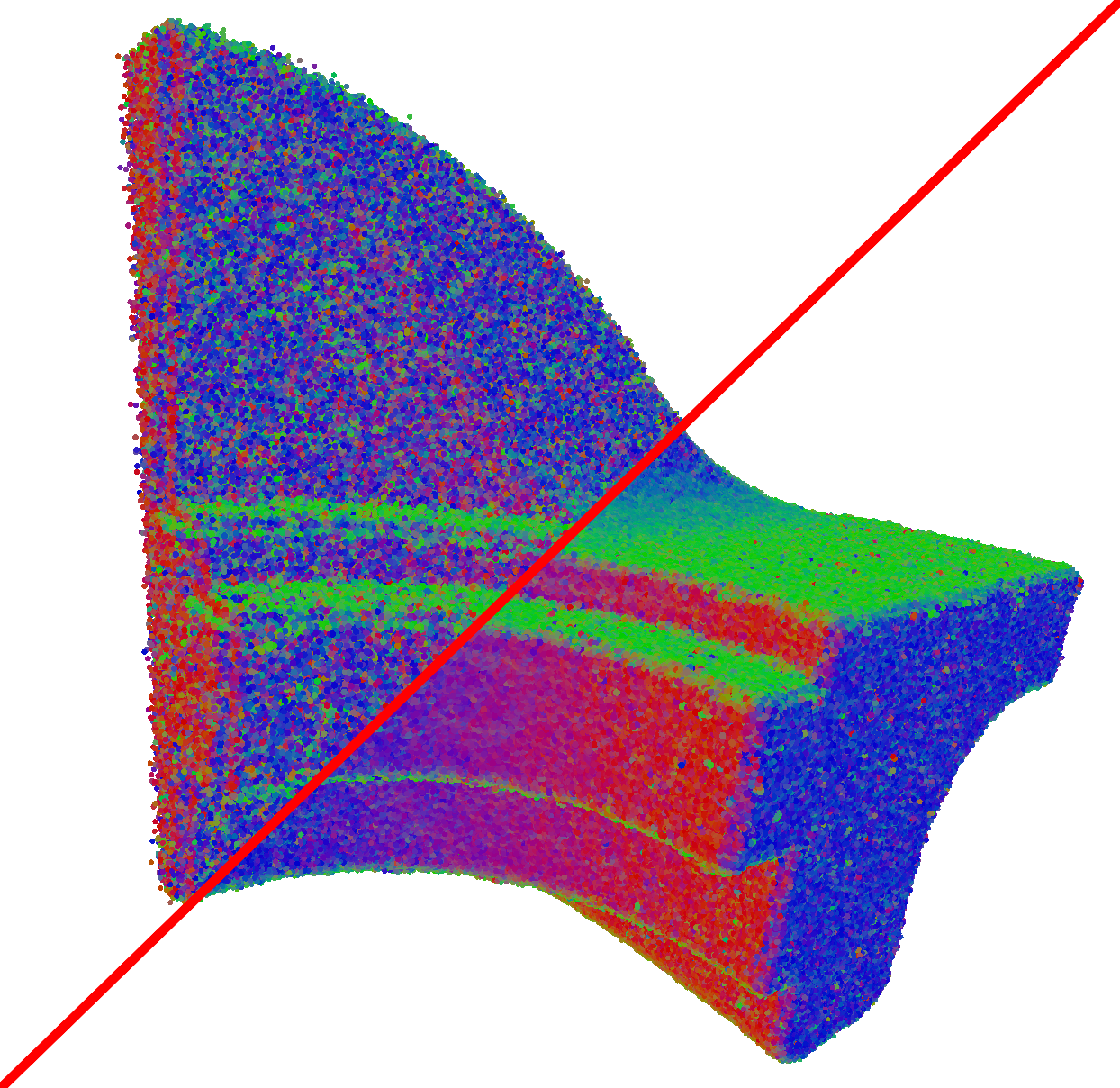}
\caption{Denoising the Fandisk point set ($600K$ points). From left to right:
noisy point sets (noise standard deviations: $1\%, 1.5\%, 2\%$ of the shape
diagonal)
and their corresponding denoised shapes. Normals are
computed by local quadric fitting ($r=0.2$, $M=193$, $d=8$, shape diagonal: $7.61$).}
\label{fig:fandisk_denoising}
\end{figure}

\begin{figure}
\centering
\subfigure[Noisy]{\includegraphics[width=0.325\linewidth]{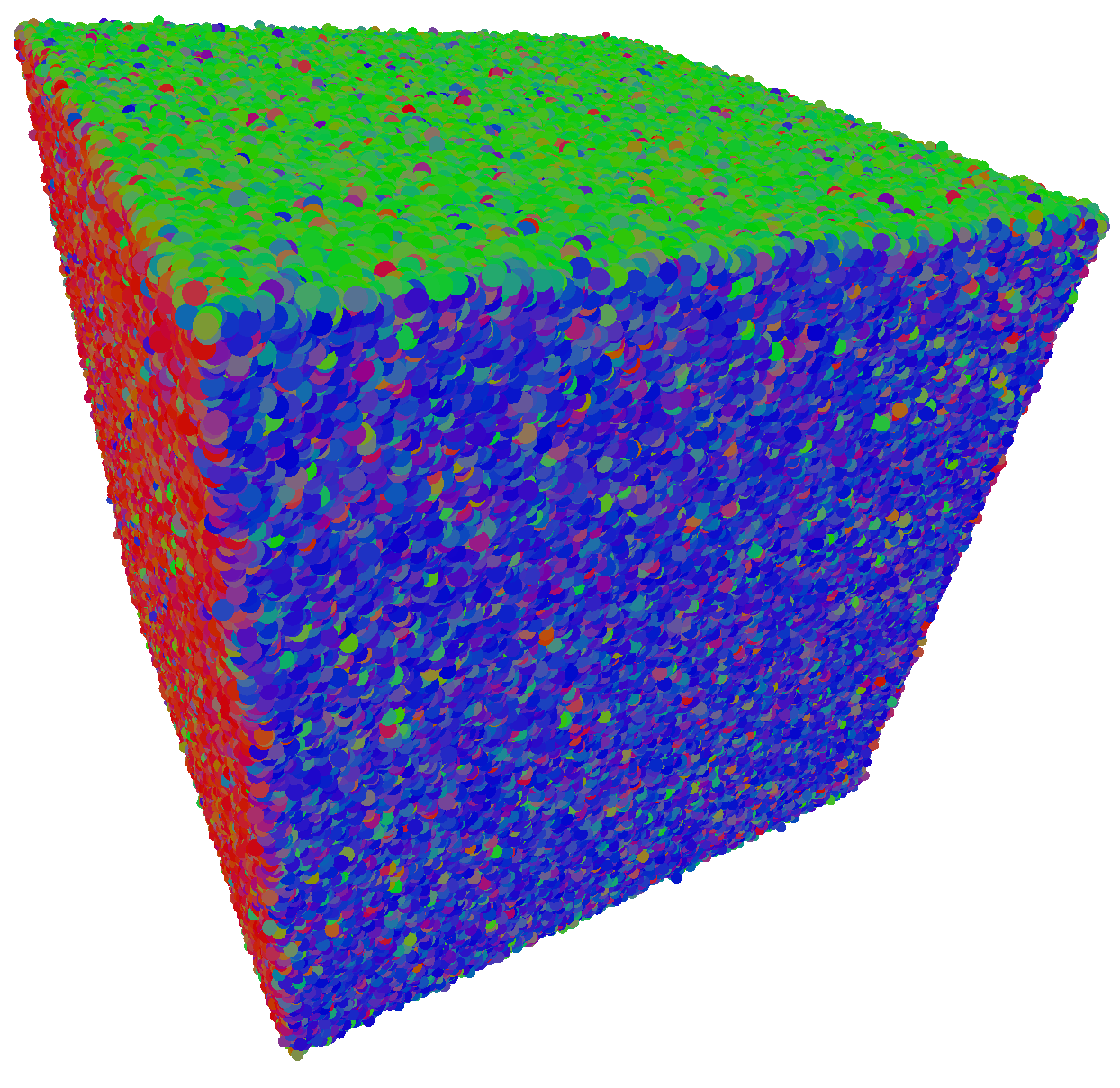}}
\subfigure[Bilateral]{\includegraphics[width=0.325\linewidth]{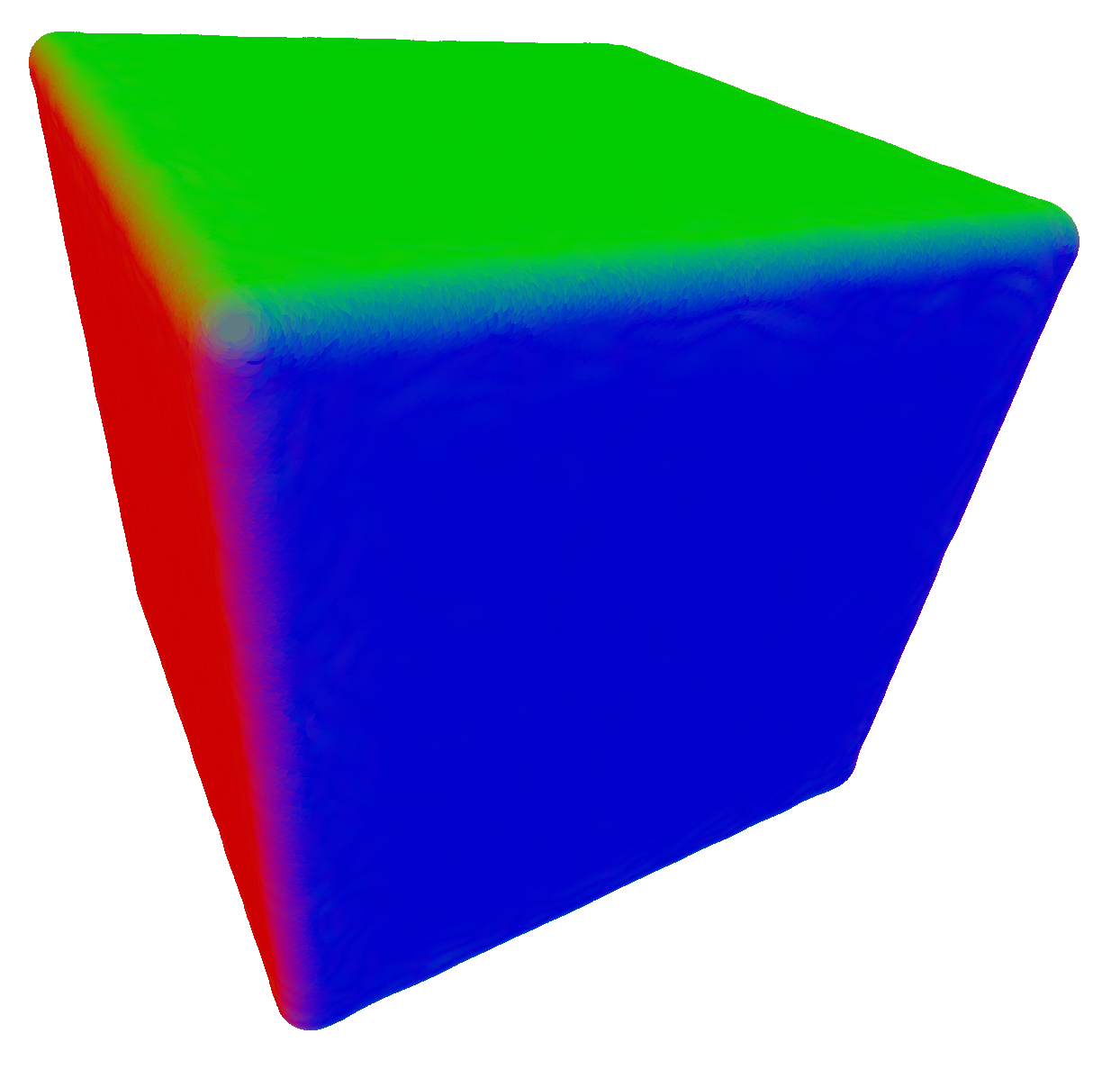}}
\subfigure[\textbf{LPF}]{\includegraphics[width=0.325\linewidth]{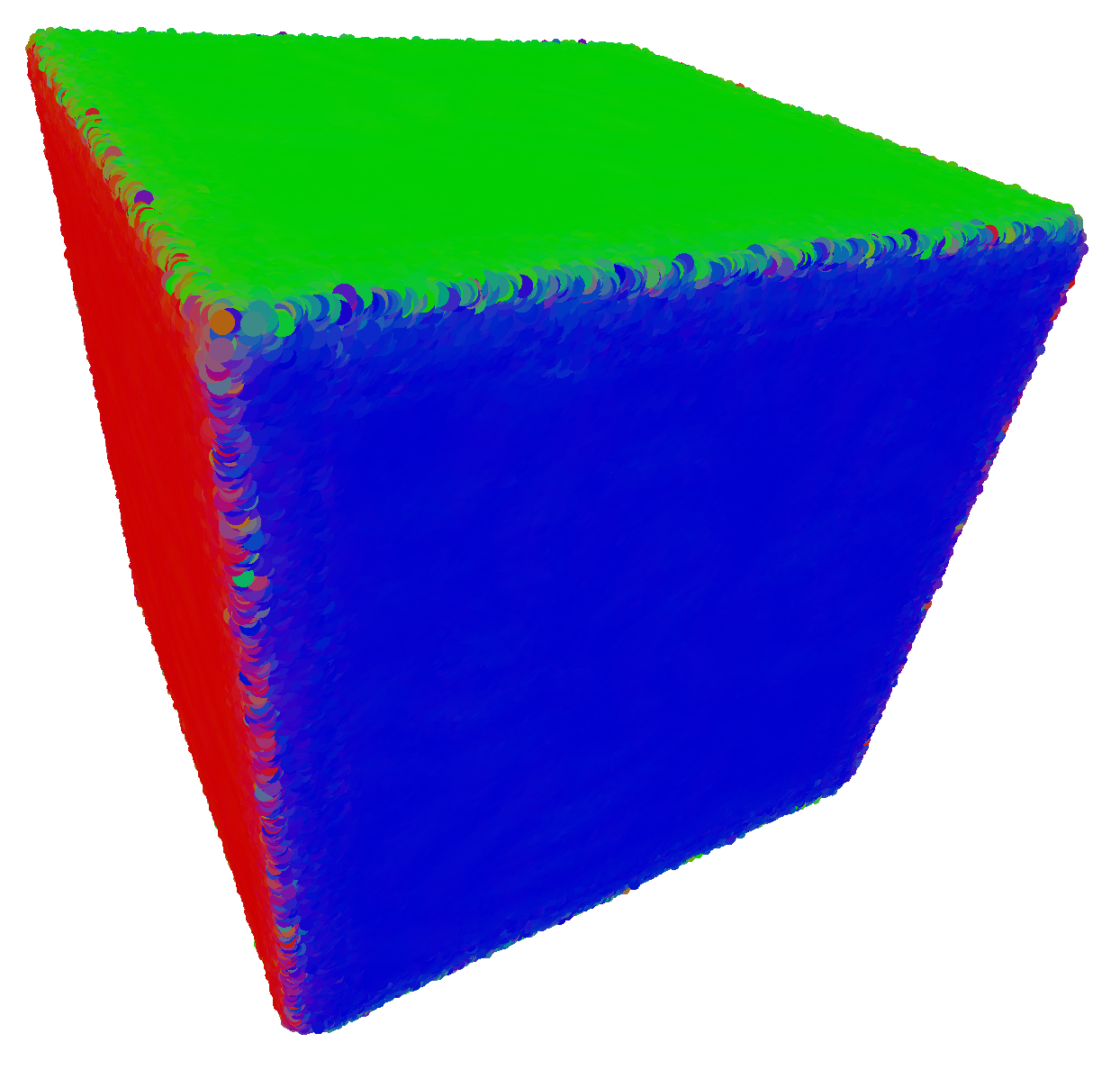}}\\
\vspace{0.25cm}
 \setlength{\fboxsep}{0pt}
 \setlength{\fboxrule}{1.5pt}
\fbox{\includegraphics[width=0.31\linewidth]{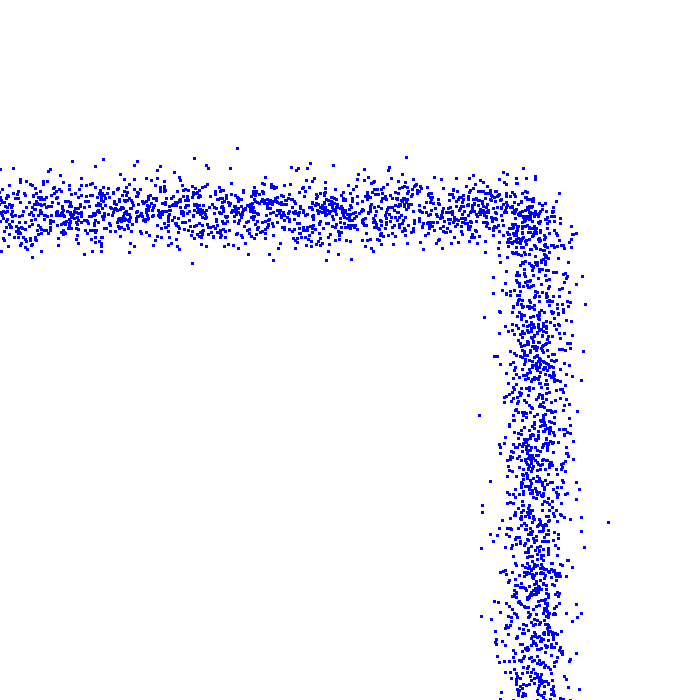}}
\fbox{\includegraphics[width=0.31\linewidth]{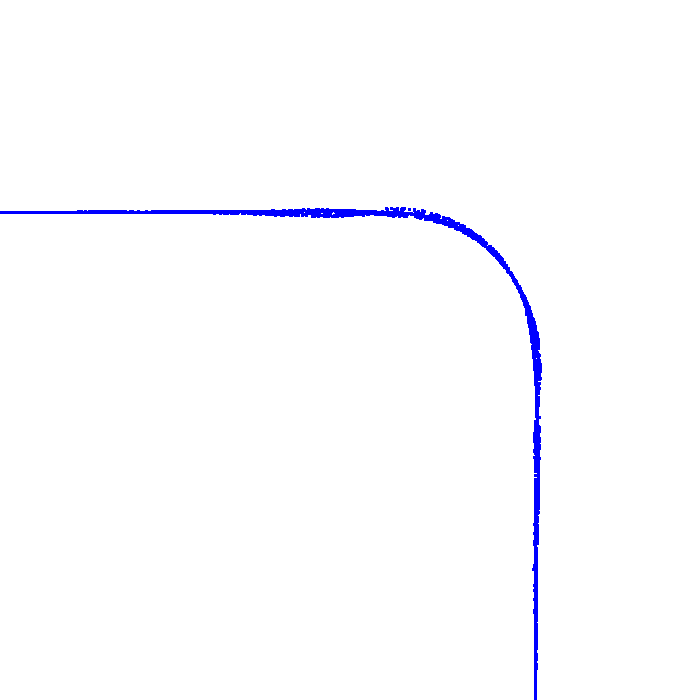}}
\fbox{\includegraphics[width=0.31\linewidth]{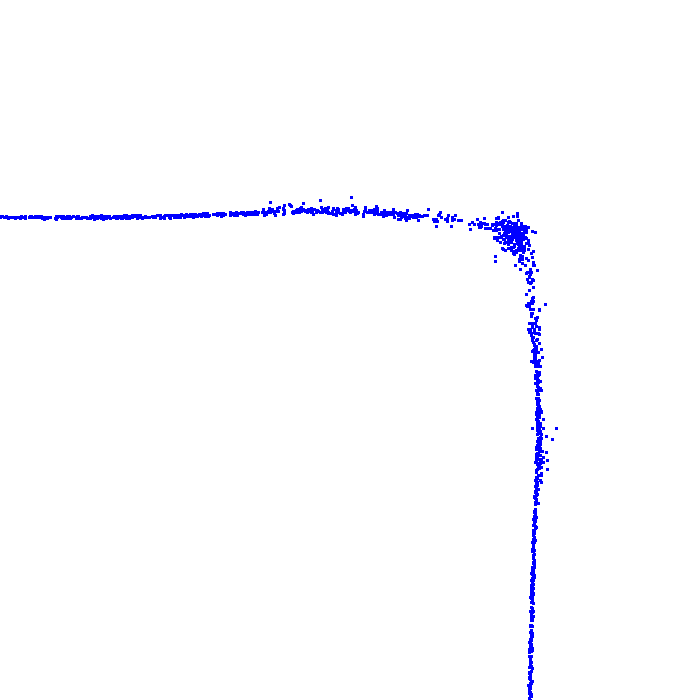}}
\caption{Denoising a noisy cube (left) and comparison between the Bilateral Filter
\protect\cite{bilateral} adapted to point sets (middle), and $\lpf$ framework (right).
Top row: point sets with normal computed by quadric fitting, bottom: close-up
on a slice of the point set.
The parameters for the bilateral filter were tuned to get the same error as the
$\lpf$ denoising
(error computed with respect to the noise-free model on the cube facets).
$r=0.2$, $d=3$, $M=193$, shape diagonal: $9.08$.}
\label{fig:cube_denoising}
\end{figure}

\begin{figure}
\centering
\subfigure[Noisy]{\includegraphics[width=0.3\linewidth]{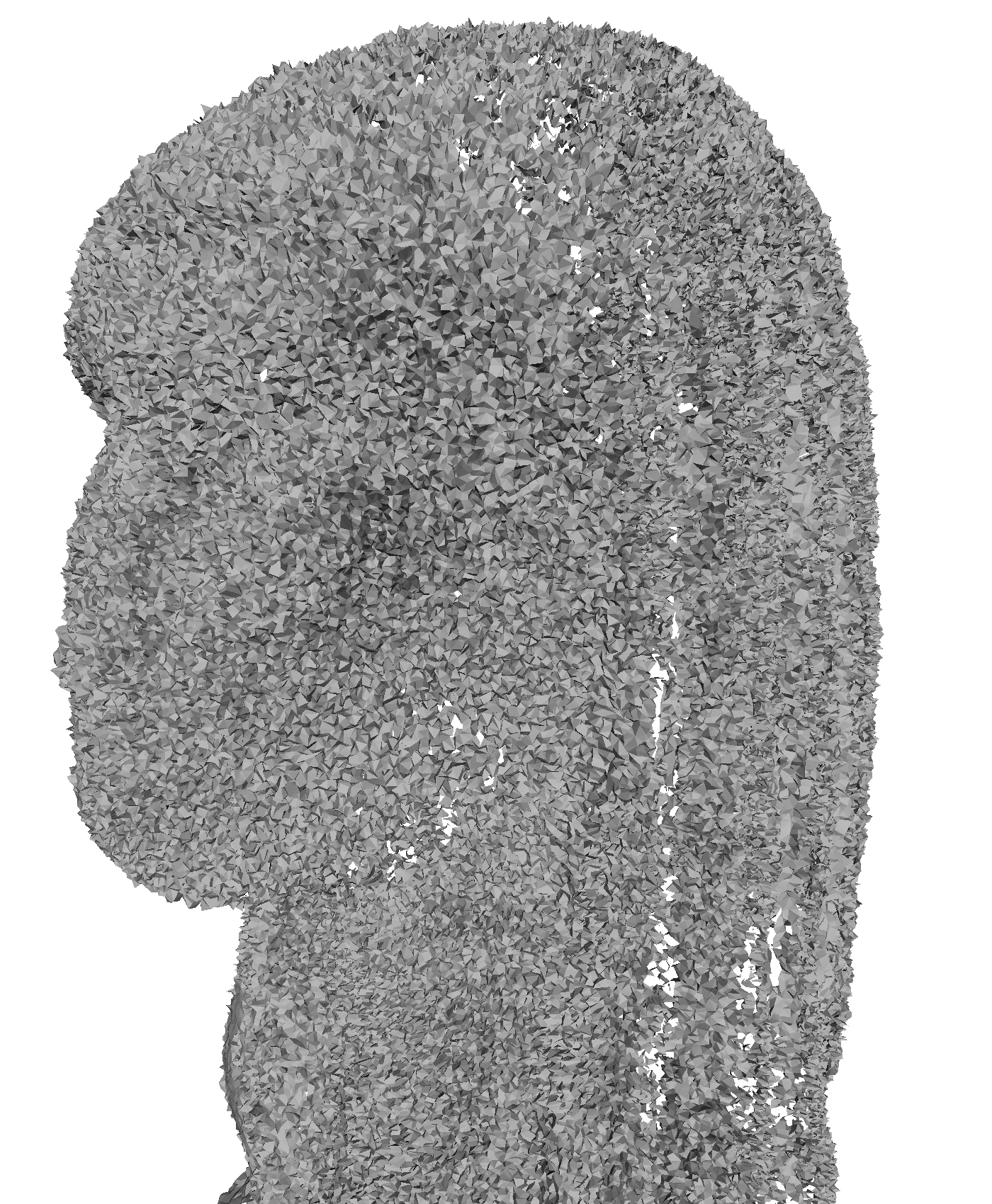}}
\subfigure[Bilateral]{\includegraphics[width=0.3\linewidth]{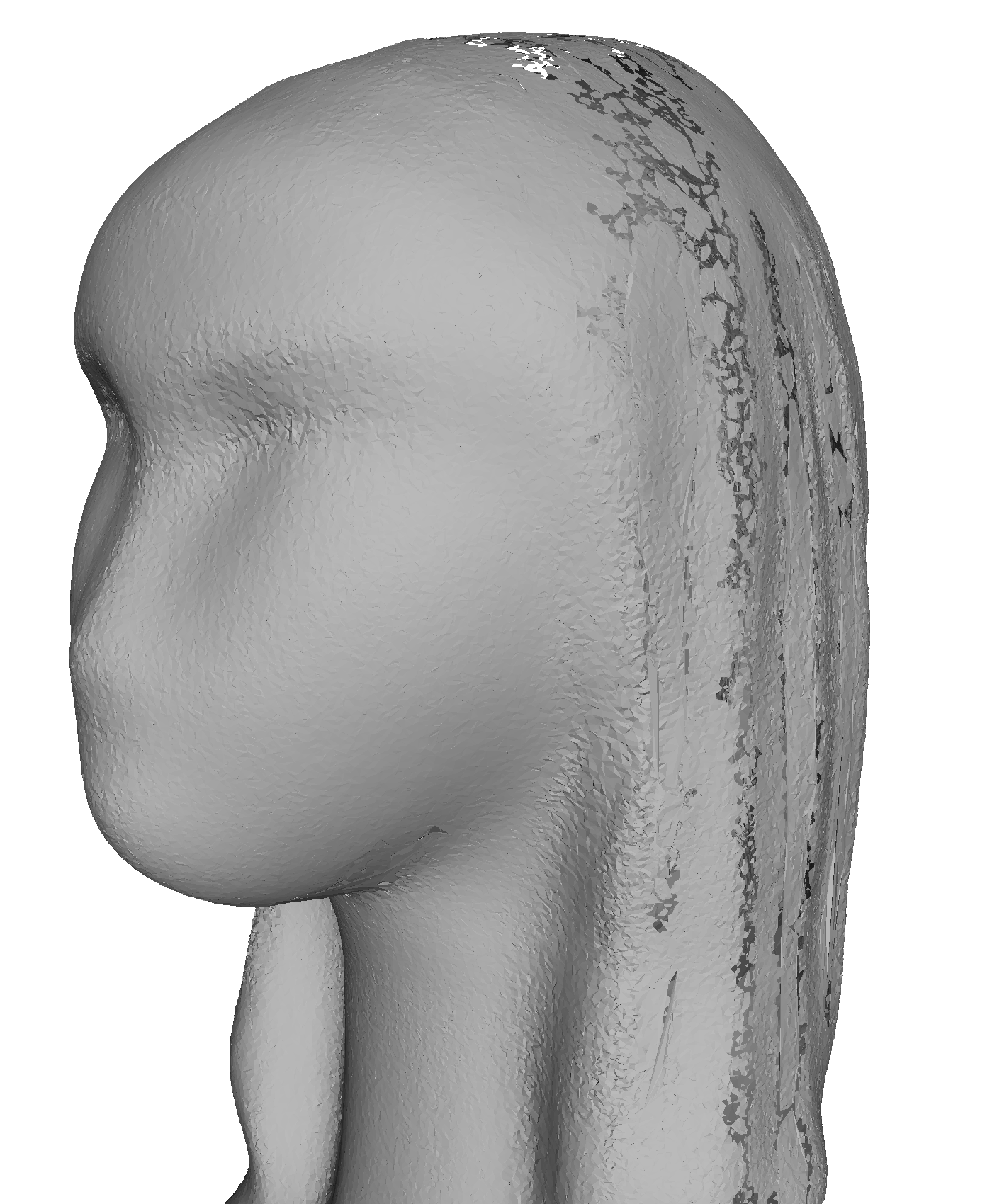}}
\subfigure[RIMLS]{\includegraphics[width=0.3\linewidth]{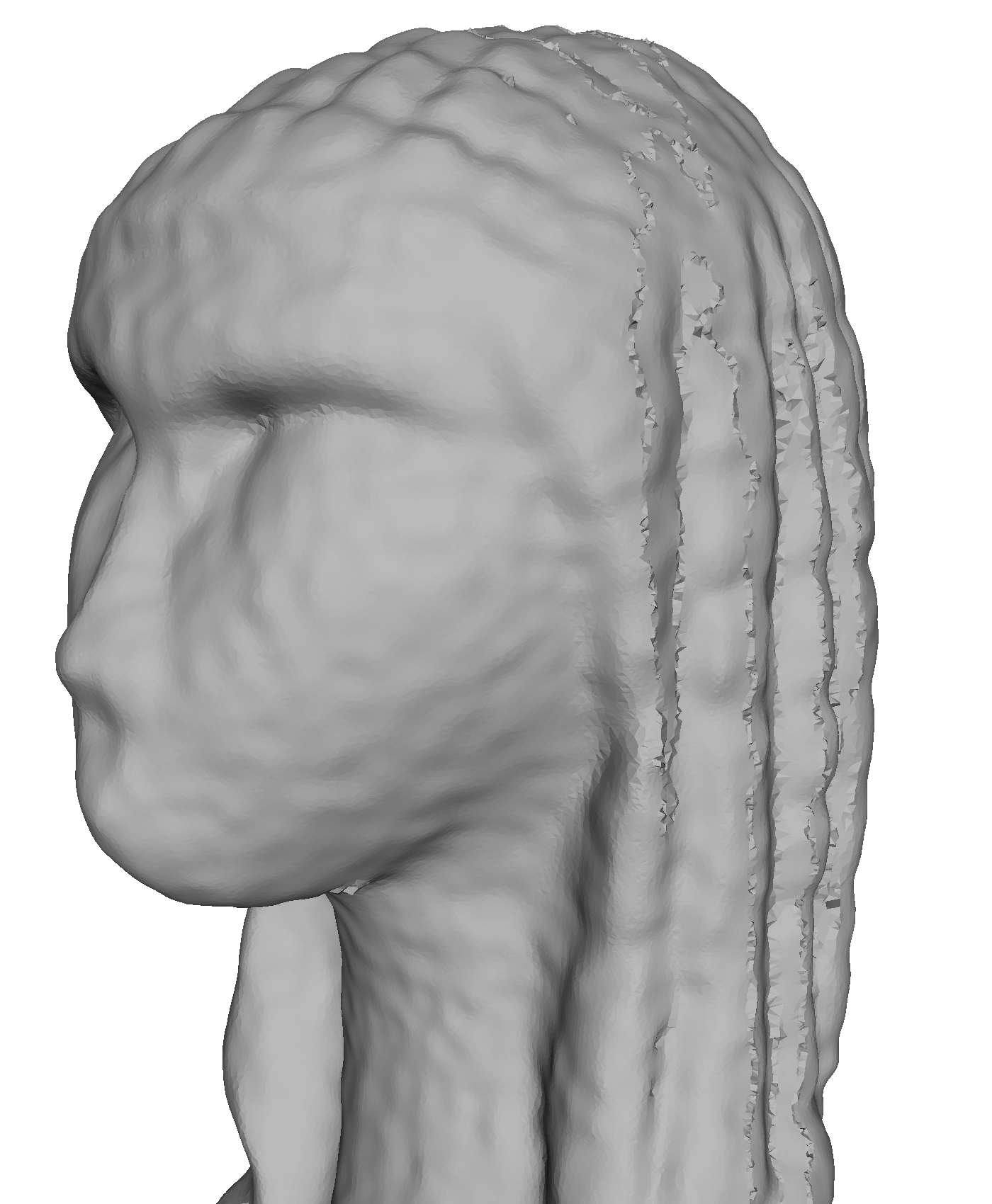}}
\subfigure[$\ell^0$ denoising]{\includegraphics[width=0.3\linewidth]{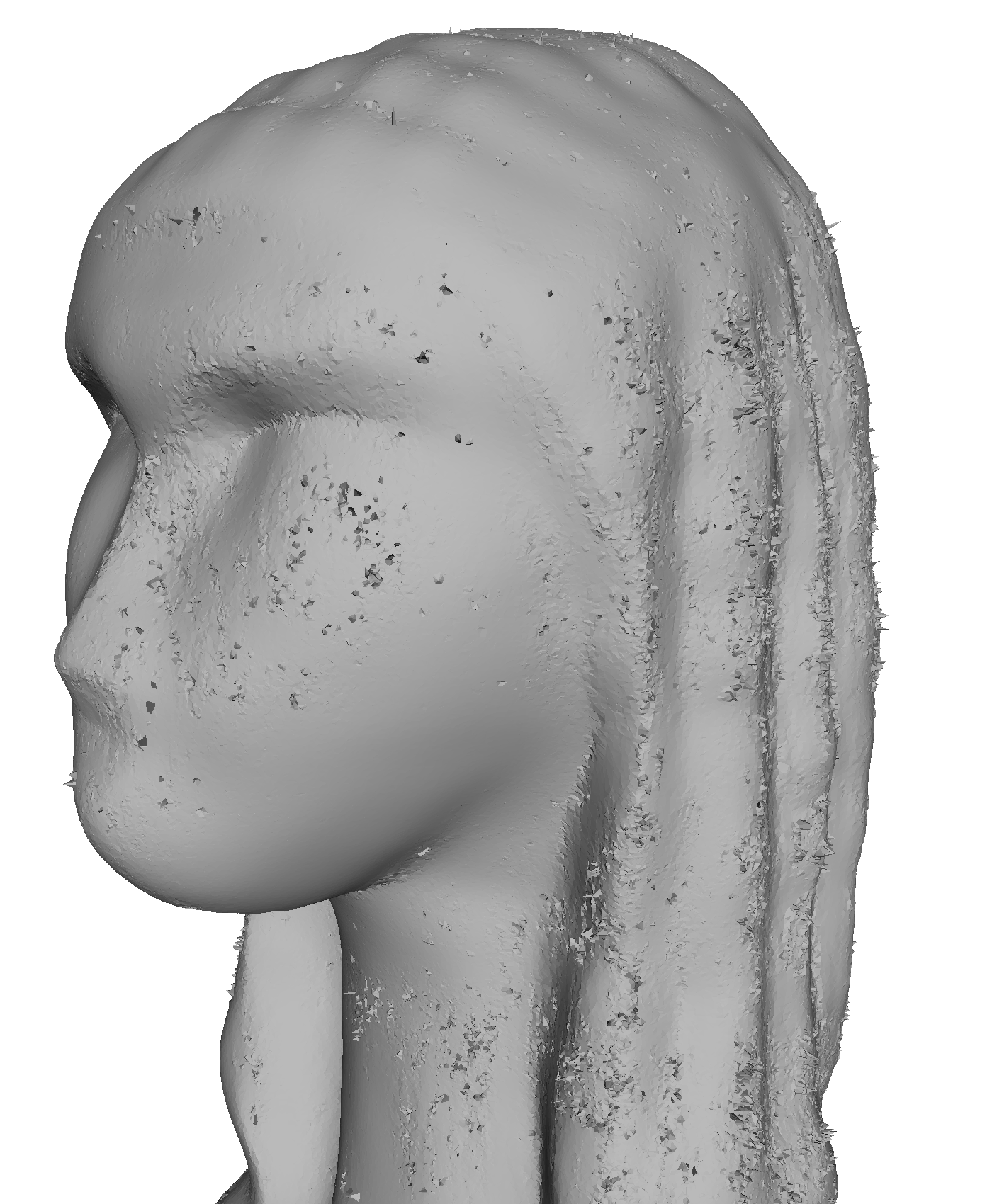}}
\subfigure[WLOP]{\includegraphics[width=0.3\linewidth]{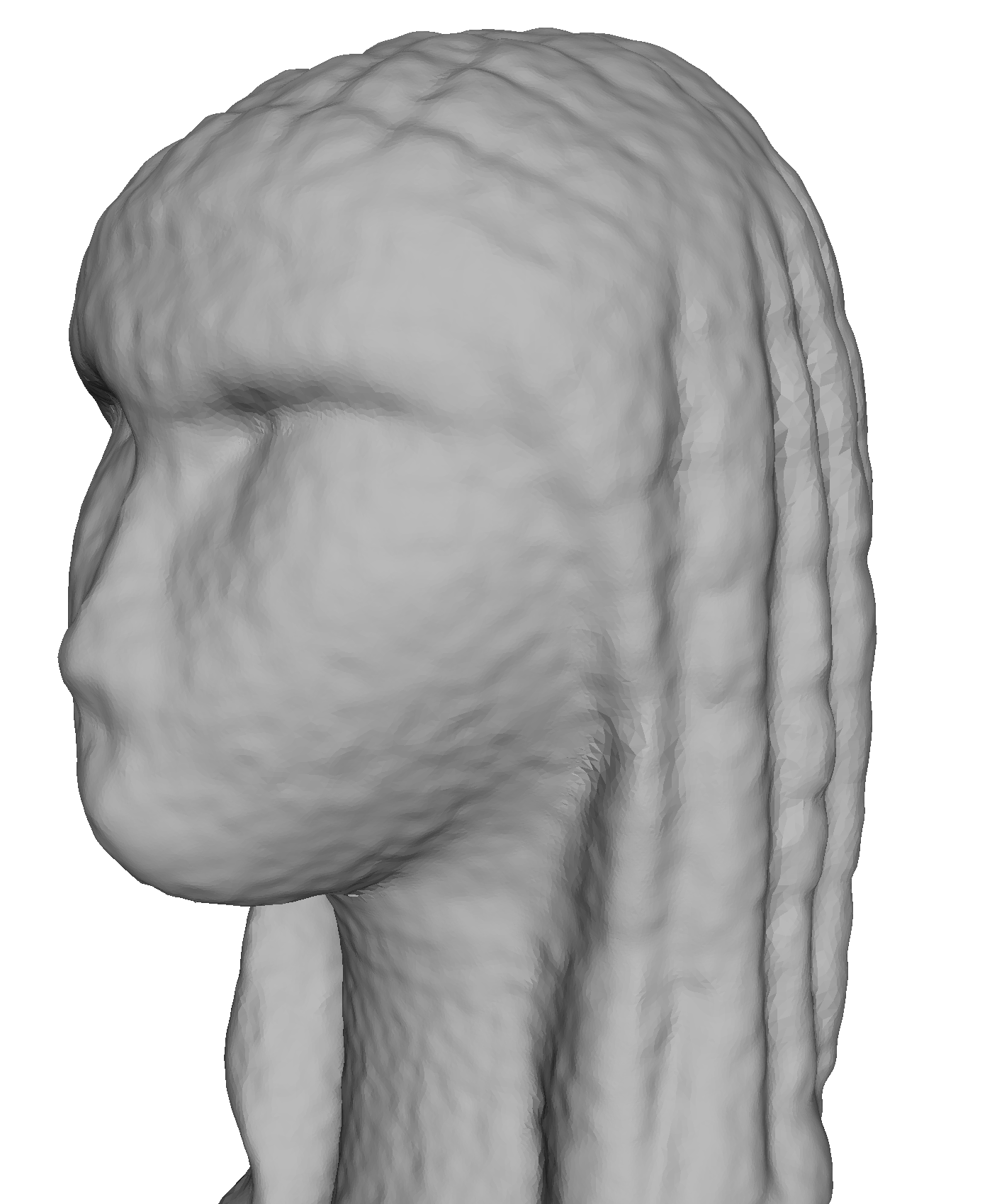}}
\subfigure[\textbf{LPF}]{\includegraphics[width=0.3\linewidth]{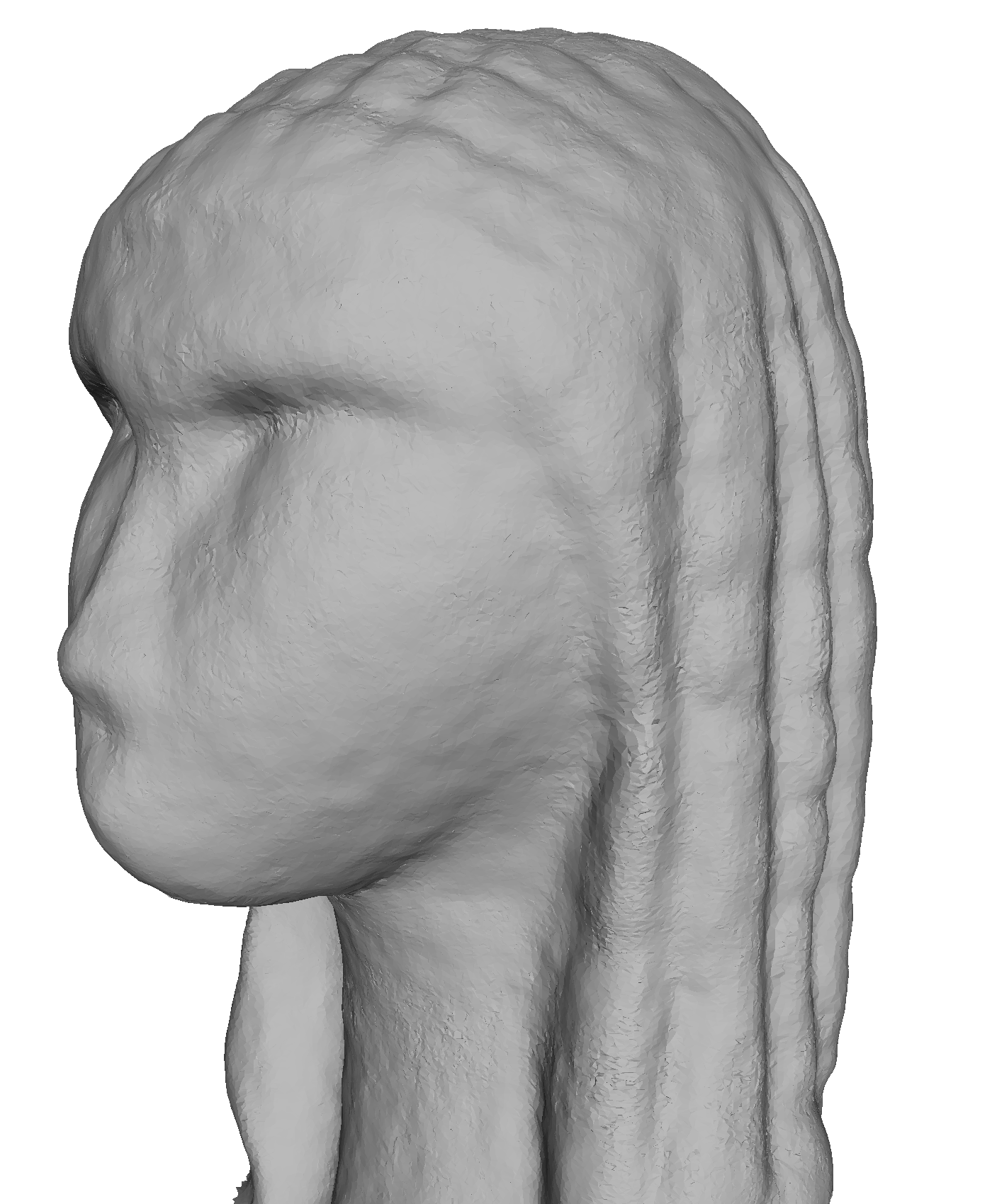}}
\caption{Denoising the Brassempouy point set with added noise of variance
$0.4\%$ of the diagonal. From left to right: initial shape, bilateral, RIMLS \protect\cite{featurepss}
(Meshlab implementation with depth $6$), $\ell^0$ denoising \protect\cite{l0denoising}, WLOP \protect\cite{wlop}
and LPF ($r=0.7$, $M=193$, $d=32$, shape diagonal $51.96$) . For visualization purpose, all shapes were reconstructed as interpolating meshes.}
\label{fig:brassempouy}
\end{figure}

In all denoising experiments, we initially set one $\lpf$ per sample point.
This way all points are represented equally. We also set $\gamma=0.5$ (moving
the point halfway toward its guessed position).
On Figure \ref{fig:fandisk_denoising}, we show the denoising iterations of the
shape learning process.
The method performs well at removing noise while preserving the features.
On figure \ref{fig:cube_denoising}, we compare our approach with the bilateral
filter \cite{bilateral} adapted to point clouds and show that the edge is less
smoothed out using our approach.
On Figure \ref{fig:brassempouy} we compare $\lpf$ denoising with several denoising methods.
One can see that the bilateral filter \cite{bilateral} tends to oversmooth the details, RIMLS \cite{featurepss} tends to create some low-frequency artifacts on the surface while still preserving the details well, $\ell^0$ denoising \cite{l0denoising} tends to create spurious artifacts. WLOP \cite{wlop} outputs a better compromise between feature preservation and noise removal, yet some low-frequency artifacts remain in areas that should be smooth. Our $\lpf$ based denoising permits to denoise featured and smooth areas without adding low-frequency artifacts.

Our method is particularly well suited for point sets that represent a mix
of surfaces and curves. In this setting normals are irrelevant and
all methods based on normal estimation fail. This is the case for the bilateral filter \cite{bilateral}, APSS \cite{algebraic_pss}, RIMLS
\cite{featurepss} as well as more complex sparsity-regularized denoising
methods \cite{l0denoising}.

\begin{figure}
 \begin{center}
  \subfigure[Original]{\includegraphics[width=0.49\linewidth]{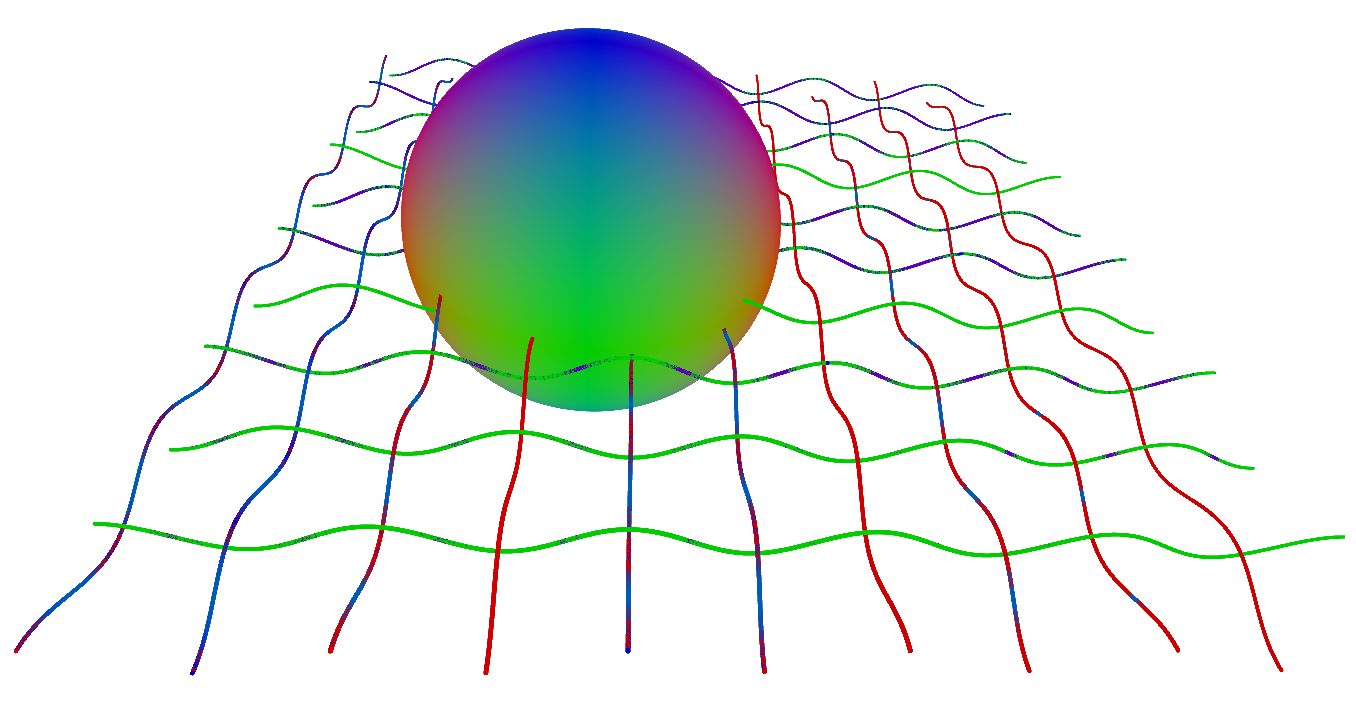}}
  \subfigure[Noisy ($RMSE=0.124$)]{\includegraphics[width=0.49\linewidth]{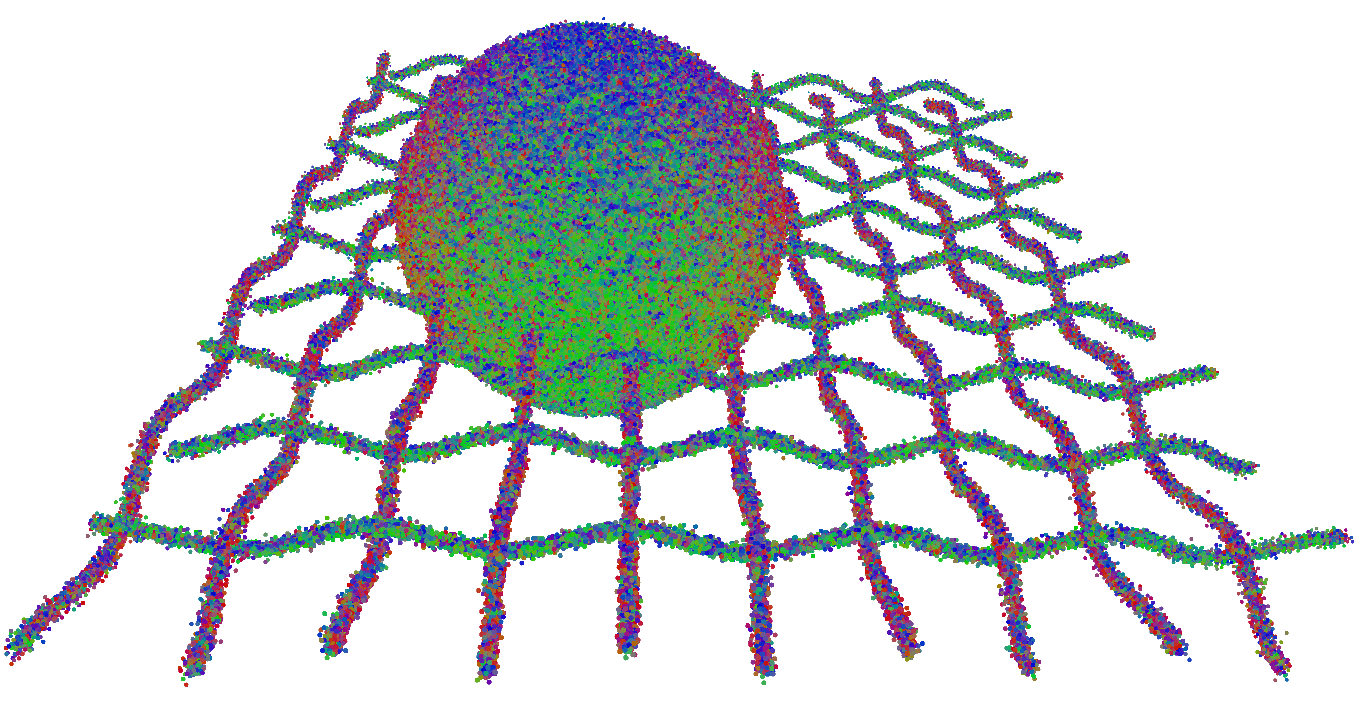}}
  \subfigure[APSS ($RMSE=0.085$)]{\includegraphics[width=0.49\linewidth]{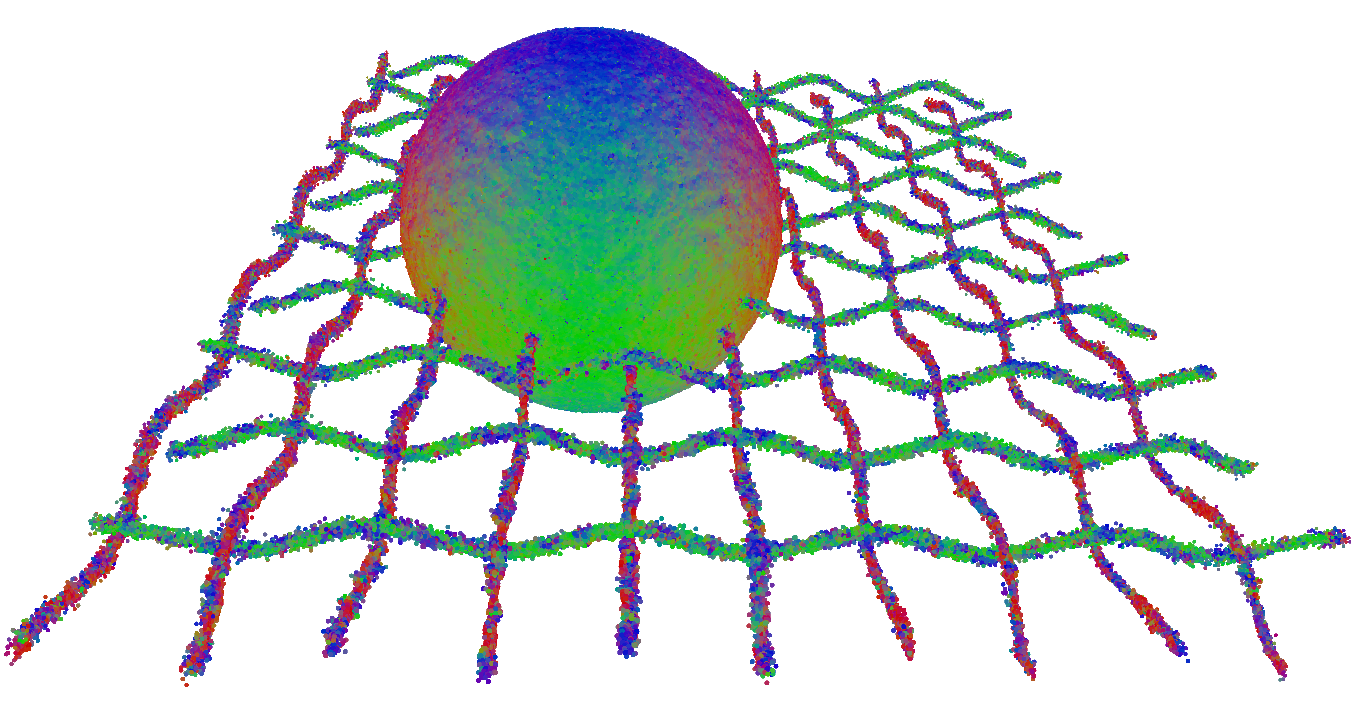}}
  \subfigure[RIMLS ($RMSE=0.081$)]{\includegraphics[width=0.49\linewidth]{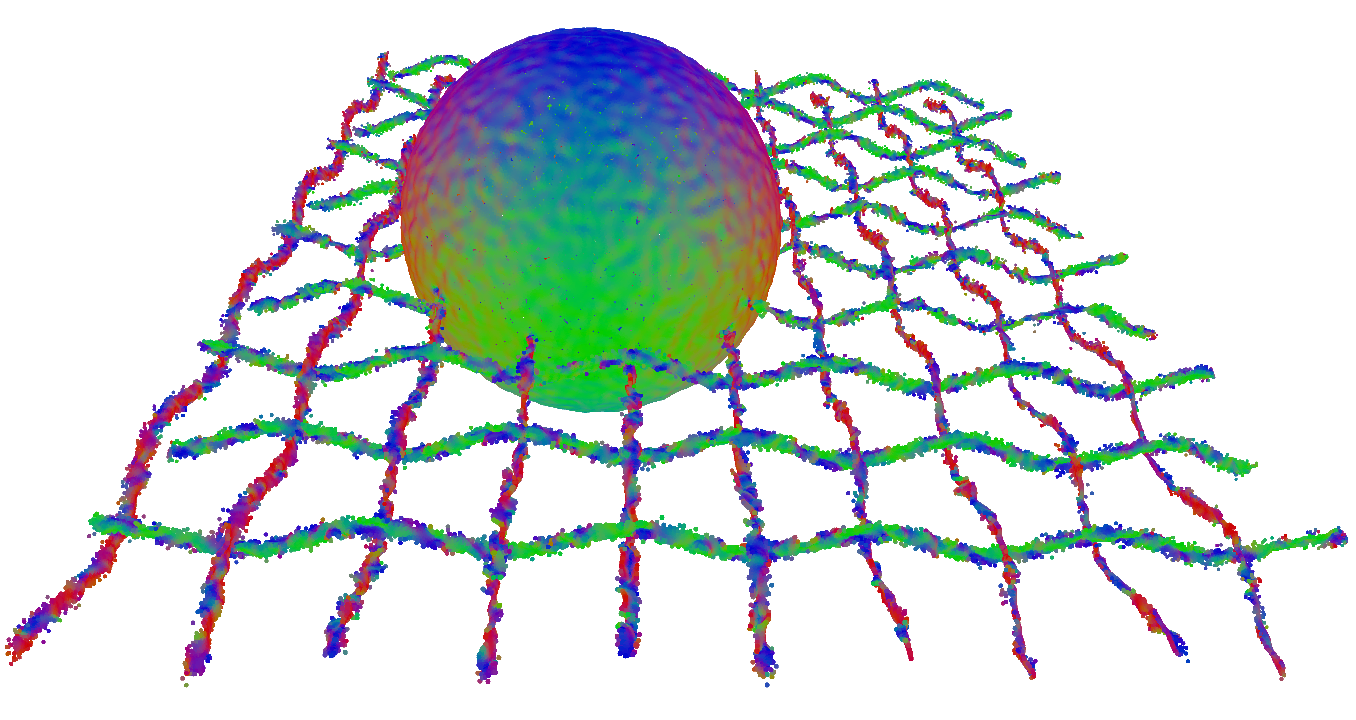}}
  \subfigure[Bilateral($RMSE=0.071$)]{\includegraphics[width=0.49\linewidth]{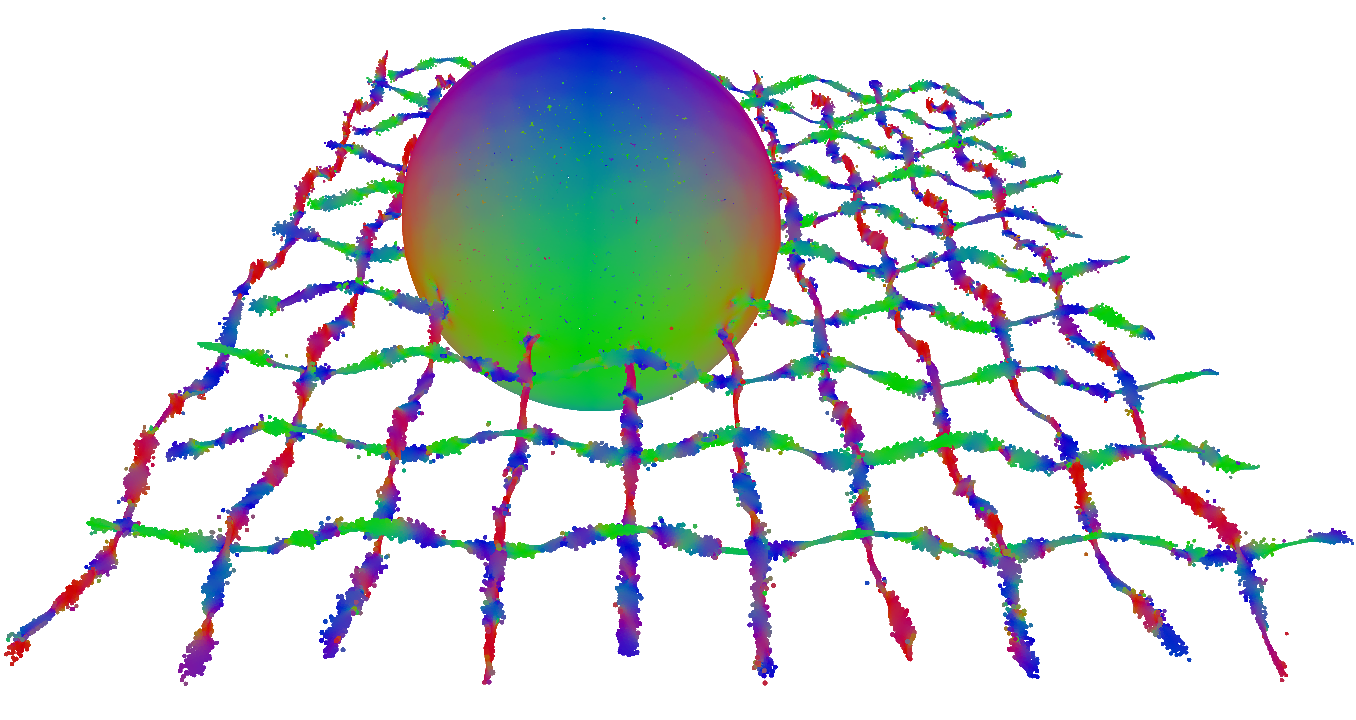}}
  \subfigure[WLOP ($RMSE=0.051$)]{\includegraphics[width=0.49\linewidth]{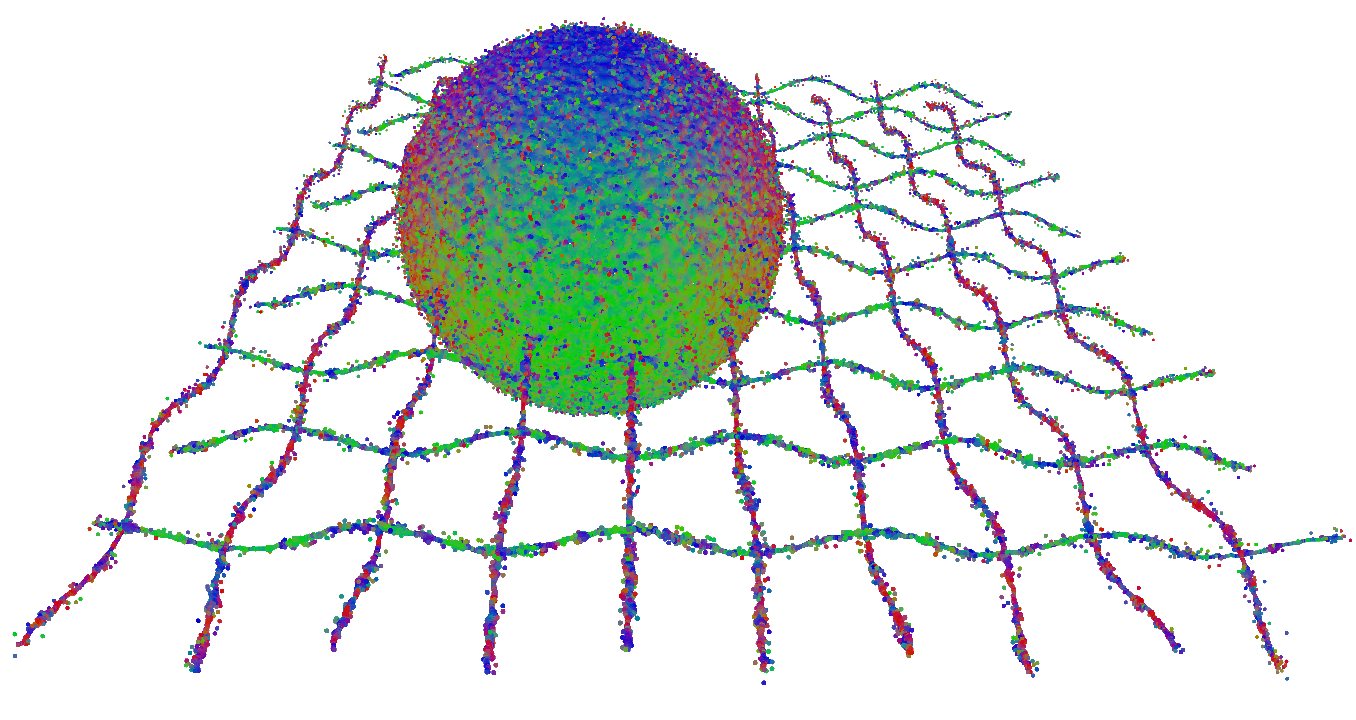}}
  \subfigure[\textbf{LPF} ($RMSE=0.017$)]{\includegraphics[width=0.49\linewidth]{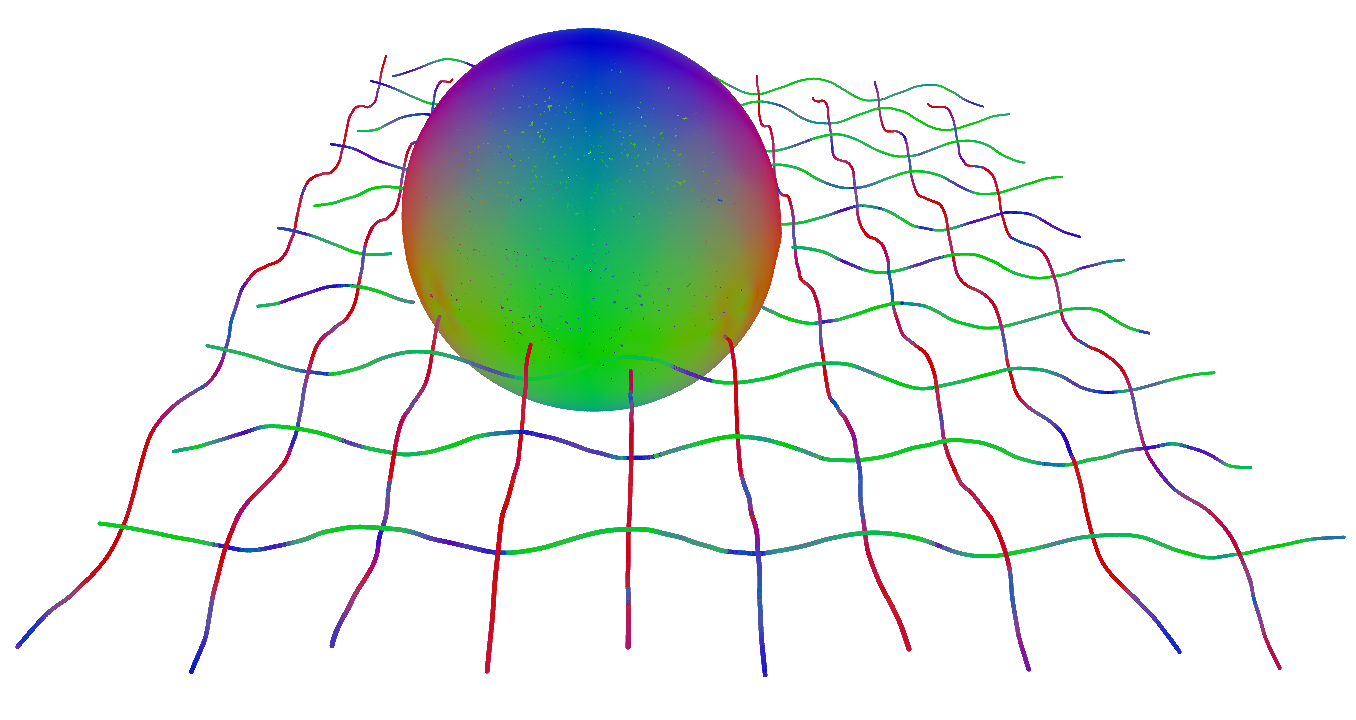}}
 \end{center}
\caption{Comparison with standard denoising methods on a synthetic
point set ($400K$ points). Only our algorithm is able to handle the curves, while algorithms relying on a manifold surface assumption fail to denoise the shape. Root Mean
Square Errors with respect to the noise-free synthetic data are given ($400K$ LPFs, $r=0.5$, $M=193$, $d=16$, shape diagonal $47.29$).}
\label{fig:synthetic_denoising}
\end{figure}

On Fig. \ref{fig:synthetic_denoising} and Table \ref{tab:denoising}, we compare visually and quantitatively the performance of our denoising algorithm on a synthetic point cloud.
Only our algorithm is able to denoise correctly both the surface and the curve net.
Notice that Weighted Locally Optimal Projection (WLOP) \cite{wlop} recovers the curves better than the other state-of-the art algorithms, although it is outperformed by our LPF approach.

\begin{table}
\begin{center}
 \begin{tabular}{|c||c|c|c|}
 \hline
 \backslashbox{Denoising\\ Method}{Initial Noise\\ Level} & $0.124$ & $0.25$ &
$0.38$ \\ \hhline{|=||=|=|=|}
  APSS & $0.085$ & $0.242$ & $0.301$ \\ \hline
  RIMLS & $0.081$ & $0.201$ & $0.272$ \\ \hline
  Bilateral & $0.071$ & $0.155$ & $0.231$ \\ \hline
  WLOP & $0.051$ & $0.214$ & $0.352$\\ \hline
  \textbf{LPF} & \textbf{$0.017$} & \textbf{$0.054$} & \textbf{$0.104$} \\ \hline
 \end{tabular}
\end{center}
 \caption{RMSE of the denoising results on the shape of Fig. \ref{fig:synthetic_denoising} using
APSS, RIMLS, the bilateral filter, WLOP and our LPF denoising. All errors are given in distance units.
The first column corresponds to the results shown on Figure \ref{fig:synthetic_denoising}.}
\label{tab:denoising}
\end{table}

\subsection{Implementation details}

Our algorithm was implemented in C++ using the Eigen library for matrix
representation. For a non-optimized code and $1.1$ million input point set,
covered by $100000$ $\lpf$s with a pattern containing $193$ points (a
pattern of $16\times16$ points intersected with a sphere) and a
dictionary of $32$ atoms, the initial $\lpf$ computation took $30s$ and the
shape learning process took $150s$ ($10$ iterations of dictionary computation,
pose estimation and recomputation) on a $4$-core Intel\textregistered
Core\texttrademark i7 laptop.

\section{Limitations} \label{sec:limitations}
The principal limitation of our method is that it learns shapes in a
statistical manner: an important feature that is not frequent enough will not
be accounted for in the shape dictionary.
As an example, no corner is present in the cube with curve dictionary (Figure \ref{fig:cubecurve_dict}).
To overcome this limitation one could devise a strategy that adds more $\lpf$s
where the representation error is largest. This would however require
further analysis to ensure that the added information does not
correspond to noise. Moreover, optimizing the shape of the template during analysis could also improve the results quality.
One should notice that the dictionary is learned on a specific shape, and analyzing another shape might not be done as efficiently using the same dictionary, except if both shapes share local similarities.

Outliers are also a limitation of our method. Indeed, if a LPF is centered at an outlier position then the LPF will represent a single point.
In the joint analysis this LPF will be replaced by a consolidated version which would have little sense.
This can be easily alleviated by adding a preprocessing step to remove outliers.

Finally, our method requires some parameters: the user has to choose both the scale and
the dictionary size. 
The choice of a scale can be argued for, since the analysis reveals different
similarities depending on the chosen scale.
The size of the dictionary is application and data-dependent; it should be
large enough to encode shape variations but small enough for similarities to
emerge.
It is simple to set for geometric shapes, such as cubes, but more difficult for
complex natural shapes.
Selecting a good dictionary size is still an open issue that we want to explore
in future works.

\section{Conclusion and Future Work}

We have introduced a pliable framework to analyze shapes by consolidating Local 
Probing Fields defined in the ambient space around the whole shape.
By jointly analyzing this set of descriptions we provided a new tool for sparse shape
description expressed as a dictionary learning problem. As demonstrated in our
experiments, this tool allows the shape to reveal its non-local similarities.
The efficiency of this framework is illustrated on shape resampling and point 
set denoising applications.
As a future work we mean to study the theoretical properties of the Local 
Probing Fields and in particular possible improvements of the statistical 
properties of the consolidated point distribution.
A whole set of applications of our framework remains to be explored, including shape compression, segmentation and 
registration, as well as the extension of our approach to shape collections.

\ifCLASSOPTIONcompsoc
  \section*{Acknowledgments}
\else
  \section*{Acknowledgment}
\fi

This work was supported by the ANR PAPS project, grant ANR-14-CE27-0003 of the
French Agence Nationale de la Recherche.

\ifCLASSOPTIONcaptionsoff
  \newpage
\fi

\bibliographystyle{IEEEtran}
\bibliography{biblio_ldm}

\begin{thebibliography}{10}
\providecommand{\url}[1]{#1}
\csname url@samestyle\endcsname
\providecommand{\newblock}{\relax}
\providecommand{\bibinfo}[2]{#2}
\providecommand{\BIBentrySTDinterwordspacing}{\spaceskip=0pt\relax}
\providecommand{\BIBentryALTinterwordstretchfactor}{4}
\providecommand{\BIBentryALTinterwordspacing}{\spaceskip=\fontdimen2\font plus
\BIBentryALTinterwordstretchfactor\fontdimen3\font minus
  \fontdimen4\font\relax}
\providecommand{\BIBforeignlanguage}[2]{{%
\expandafter\ifx\csname l@#1\endcsname\relax
\typeout{** WARNING: IEEEtran.bst: No hyphenation pattern has been}%
\typeout{** loaded for the language `#1'. Using the pattern for}%
\typeout{** the default language instead.}%
\else
\language=\csname l@#1\endcsname
\fi
#2}}
\providecommand{\BIBdecl}{\relax}
\BIBdecl

\bibitem{symmetry_in_3d_geometry}
N.~J. Mitra, M.~Pauly, M.~Wand, and D.~Ceylan, ``Symmetry in 3d geometry:
  Extraction and applications,'' \emph{Computer Graphics Forum}, vol.~32,
  no.~6, pp. 1--23, 2013.

\bibitem{discovering_structural_regularity}
M.~Pauly, N.~J. Mitra, J.~Wallner, H.~Pottmann, and L.~J. Guibas, ``Discovering
  structural regularity in 3d geometry,'' \emph{ACM Transactions on Graphics,
  Proc SIGGRAPH'08}, no.~3, pp. 43:1--43:11, 2008.

\bibitem{non_local_algorithm_for_image_denoising}
A.~Buades, B.~Coll, and J.-M. Morel, ``A non-local algorithm for image
  denoising,'' in \emph{IEEE Conference on Computer Vision and Pattern
  Recognition}, 2005, pp. 60--65.

\bibitem{Shechtman07}
E.~Shechtman and M.~Irani, ``Matching local self-similarities across images and
  videos,'' \emph{2007 IEEE Conference on Computer Vision and Pattern
  Recognition}, pp. 1--8, 2007.

\bibitem{meshNLmeans}
S.~Yoshizawa, A.~Belyaev, and H.-P. Seidel, ``Smoothing by example: Mesh
  denoising by averaging with similarity-based weights,'' in \emph{SMI
  '06}.\hskip 1em plus 0.5em minus 0.4em\relax IEEE, 2006, p.~9.

\bibitem{similarity_based_filtering}
J.~Digne, ``Similarity based filtering of point clouds.'' in \emph{CVPR
  Workshops}.\hskip 1em plus 0.5em minus 0.4em\relax IEEE, 2012, pp. 73--79.

\bibitem{Hamdi2017}
A.~Hamdi-Cherif, J.~Digne, and R.~Chaine, ``Super-resolution of point set
  surfaces using local similarities,'' \emph{to appear in Computer Graphics
  Forum}, 2017.

\bibitem{non_local_point_set_surfaces}
T.~Guillemot, A.~Almansa, and T.~Boubekeur, ``Non local point set surfaces,''
  in \emph{Proc. 3DIMPVT 2012}, 2012, pp. 324--331.

\bibitem{context_based_surface_completion}
A.~Sharf, M.~Alexa, and D.~Cohen-Or, ``Context-based surface completion,''
  \emph{ACM Transactions on Graphics}, vol.~23, no.~3, pp. 878--887, 2004.

\bibitem{non_local_scan_consolidation}
Q.~Zheng, A.~Sharf, G.~Wan, Y.~Li, N.~J. Mitra, D.~Cohen-Or, and B.~Chen,
  ``Non-local scan consolidation for 3d urban scenes,'' \emph{ACM Transactions
  on Graphics, Proc. SIGGRAPH 10}, 2010.

\bibitem{Aharon2006}
M.~Aharon, M.~Elad, and A.~Bruckstein, ``\rm k -svd: An algorithm for designing
  overcomplete dictionaries for sparse representation,'' \emph{IEEE
  Transactions on Signal Processing}, vol.~54, 2006.

\bibitem{online_dictionary_learning_for_sparse_coding}
J.~Mairal, F.~Bach, J.~Ponce, and G.~Sapiro, ``Online dictionary learning for
  sparse coding,'' in \emph{International Conference on Machine
  Learning}.\hskip 1em plus 0.5em minus 0.4em\relax ACM, 2009, pp. 689--696.

\bibitem{surf_dico}
S.~Xiong, J.~Zhang, J.~Zheng, J.~Cai, and L.~Liu, ``Robust surface
  reconstruction via dictionary learning,'' \emph{ACM Transactions On Graphics,
  Proc. SIGGRAPH Asia 2014}, vol.~33, 2014.

\bibitem{Litman2014}
R.~Litman, A.~Bronstein, M.~Bronstein, and U.~Castellani, ``Supervised learning
  of bag-of-features shape descriptors using sparse coding,'' \emph{Computer
  Graphics Forum}, vol.~33, no.~5, pp. 127--136, 2014.

\bibitem{digne_eg2014}
J.~Digne, R.~Chaine, and S.~Valette, ``{Self-similarity for accurate
  compression of point sampled surfaces},'' \emph{{Computer Graphics Forum}},
  vol.~33, no.~2, pp. 155--164, 2014, proceedings of Eurographics 2014.

\bibitem{Xu2015}
L.~Xu, R.~Wang, J.~Zhang, Z.~Yang, J.~Deng, F.~Chen, and L.~Liu, ``Survey on
  sparsity in geometric modeling and processing,'' \emph{Graphical Models},
  vol.~82, pp. 160 -- 180, 2015.

\bibitem{diff_coord}
Y.~Lipman, O.~Sorkine, D.~Cohen-Or, D.~Levin, C.~Rossi, and H.-P. Seidel,
  ``Differential coordinates for interactive mesh editing,'' in \emph{Proc. SMI
  2004}.\hskip 1em plus 0.5em minus 0.4em\relax IEEE, 2004, pp. 181--190.

\bibitem{laplacian_surf_editing}
O.~Sorkine, D.~Cohen-Or, Y.~Lipman, M.~Alexa, C.~R\"{o}ssl, and H.-P. Seidel,
  ``Laplacian surface editing,'' \emph{Computer Graphics Forum (Proc. SGP
  2004)}, pp. 179--188, 2004.

\bibitem{Wang2016}
R.~Wang, L.~Liu, Z.~Yang, K.~Wang, W.~Shan, J.~Deng, and F.~Chen,
  ``Construction of manifolds via compatible sparse representations,''
  \emph{ACM Transactions on Graphics}, vol.~35, no.~2, pp. 14:1--14:10, 2016.

\bibitem{lowe2004sift}
D.~G. Lowe, ``Distinctive image features from scale-invariant keypoints,''
  \emph{International journal of computer vision}, vol.~60, no.~2, pp. 91--110,
  2004.

\bibitem{spin}
A.~E. Johnson and M.~Hebert, ``Using spin images for efficient object
  recognition in cluttered 3d scenes,'' \emph{IEEE Transactions on Pattern
  Analysis and Machine Intelligence}, vol.~21, pp. 433--449, 1999.

\bibitem{Tombari2010}
F.~Tombari, S.~Salti, and L.~Di~Stefano, ``Unique signatures of histograms for
  local surface description,'' in \emph{IEEE European Conference on Computer
  Vision}.\hskip 1em plus 0.5em minus 0.4em\relax Springer-Verlag, 2010, pp.
  356--369.

\bibitem{shape_contexts}
S.~Belongie, J.~Malik, and J.~Puzicha, ``Shape matching and object recognition
  using shape contexts,'' \emph{IEEE Transactions on Pattern Analysis and
  Machine Intelligence}, vol.~24, pp. 509--522, 2002.

\bibitem{Kokkinos2012}
I.~Kokkinos, M.~M. Bronstein, R.~Litman, and A.~M. Bronstein, ``Intrinsic shape
  context descriptors for deformable shapes,'' in \emph{IEEE Conference on
  Computer Vision and Pattern Recognition}, June 2012, pp. 159--166.

\bibitem{mesh-hog}
A.~Zaharescu, E.~Boyer, K.~Varanasi, and R.~Horaud, ``Surface feature detection
  and description with applications to mesh matching,'' in \emph{IEEE
  Conference on Computer Vision and Pattern Recognition}, June 2009, pp.
  373--380.

\bibitem{concise_ms_signature}
J.~Sun, M.~Ovsjanikov, and L.~Guibas, ``A concise and provably informative
  multi-scale signature based on heat diffusion,'' \emph{Computer Graphics
  Forum (Proc. SGP 2009)}, pp. 1383--1392, 2009.

\bibitem{Li2016}
N.~Li, S.~Wang, M.~Zhong, Z.~Su, and H.~Qin, ``Generalized local-to-global
  shape feature detection based on graph wavelets,'' \emph{IEEE Transactions on
  Visualization and Computer Graphics}, vol.~22, no.~9, pp. 2094--2106, Sept
  2016.

\bibitem{shapegoogle}
A.~M. Bronstein, M.~M. Bronstein, L.~J. Guibas, and M.~Ovsjanikov, ``Shape
  google: Geometric words and expressions for invariant shape retrieval,''
  \emph{ACM Transactions on Graphics}, vol.~30, pp. 1--20, 2011.

\bibitem{mattausch14eg}
O.~Mattausch, D.~Panozzo, C.~Mura, O.~Sorkine-Hornung, and R.~Pajarola,
  ``Object detection and classification from large-scale cluttered indoor
  scans,'' \emph{Computer Graphics Forum}, vol.~33, no.~2, pp. 11--21, 2014.

\bibitem{Wei15}
L.~Wei, Q.~Huang, D.~Ceylan, E.~Vouga, and H.~Li, ``Dense human body
  correspondences using convolutional networks,'' \emph{CoRR}, vol.
  abs/1511.05904, 2015.

\bibitem{Masci15}
J.~Masci, D.~Boscaini, M.~M. Bronstein, and P.~Vandergheynst, ``Geodesic
  convolutional neural networks on riemannian manifolds,'' in \emph{ICCV
  Workshops 2015}.\hskip 1em plus 0.5em minus 0.4em\relax IEEE, 2015.

\bibitem{Boscaini16}
D.~Boscaini, J.~Masci, E.~Rodol{\`{a}}, and M.~M. Bronstein, ``Learning shape
  correspondence with anisotropic convolutional neural networks,'' \emph{CoRR},
  vol. abs/1605.06437, 2016.

\bibitem{Huang2009}
H.~Huang, D.~Li, H.~Zhang, U.~Ascher, and D.~Cohen-Or, ``Consolidation of
  unorganized point clouds for surface reconstruction,'' \emph{ACM Transactions
  on Graphics}, vol.~28, pp. 176:1--176:78, 2009.

\bibitem{edge_aware_point_set_resampling}
H.~Huang, S.~Wu, M.~Gong, D.~Cohen-Or, U.~Ascher, and H.~R. Zhang, ``Edge-aware
  point set resampling,'' \emph{ACM Transactions on Graphics}, vol.~32, no.~1,
  pp. 9:1--9:12, Feb. 2013.

\bibitem{lop}
Y.~Lipman, D.~Cohen-Or, D.~Levin, and H.~Tal-Ezer, ``Parameterization-free
  projection for geometry reconstruction,'' \emph{ACM Transactions on
  Graphics}, vol.~26, 2007.

\bibitem{wlop}
H.~Huang, D.~Li, H.~Zhang, U.~Ascher, and D.~Cohen-Or, ``Consolidation of
  unorganized point clouds for surface reconstruction,'' in \emph{SIGGRAPH Asia
  2009}.\hskip 1em plus 0.5em minus 0.4em\relax ACM, 2009, pp. 176:1--176:7.

\bibitem{pss}
M.~Alexa, J.~Behr, D.~Cohen-Or, S.~Fleishman, D.~Levin, and C.~T. Silva,
  ``Point set surfaces,'' in \emph{Proc. Vis '01}.\hskip 1em plus 0.5em minus
  0.4em\relax IEEE, 2001, pp. 21--28.

\bibitem{defining_pss}
N.~Amenta and Y.~J. Kil, ``Defining point-set surfaces,'' in \emph{SIGGRAPH
  '04}.\hskip 1em plus 0.5em minus 0.4em\relax USA: ACM Press, 2004, pp.
  264--270.

\bibitem{rmls}
S.~Fleishman, D.~Cohen-Or, and C.~T. Silva, ``Robust moving least-squares
  fitting with sharp features,'' \emph{ACM Transactions on Graphics}, vol.~24,
  no.~3, pp. 544--552, 2005.

\bibitem{algebraic_pss}
G.~Guennebaud and M.~Gross, ``Algebraic point set surfaces,'' \emph{ACM
  Transactions on Graphics}, vol.~26, 2007, proc. SIGGRAPH 2007.

\bibitem{featurepss}
A.~C. Oztireli, G.~Guennebaud, and M.~Gross, ``Feature preserving point set
  surfaces based on non-linear kernel regression,'' \emph{Computer Graphics
  Forum}, vol.~28, pp. 493--501(9), 2009.

\bibitem{deep_points}
S.~Wu, H.~Huang, M.~Gong, M.~Zwicker, and D.~Cohen-Or, ``Deep points
  consolidation,'' \emph{ACM Transactions on Graphics}, vol.~34, no.~6, pp.
  176:1--176:13, 2015.

\bibitem{Wang2014}
R.~Wang, Z.~Yang, L.~Liu, J.~Deng, and F.~Chen, ``Decoupling noise and features
  via weighted $\ell_1$-analysis compressed sensing,'' \emph{ACM Transactions
  on Graphics}, vol.~33, no.~2, pp. 18:1--18:12, Apr. 2014.

\bibitem{l0denoising}
Y.~Sun, S.~Schaefer, and W.~Wang, ``Denoising point sets via ${L}_0$
  minimization,'' \emph{Computer Aided Geometric Design}, vol. 35–36, pp. 2
  -- 15, 2015, geometric Modeling and Processing 2015.

\bibitem{Zhang15}
H.~Zhang, C.~Wu, J.~Zhang, and J.~Deng, ``Variational mesh denoising using
  total variation and piecewise constant function space,'' \emph{IEEE
  Transactions on Visualization and Computer Graphics}, vol.~21, no.~7, pp.
  873--886, July 2015.

\bibitem{Lu16}
X.~Lu, Z.~Deng, and W.~Chen, ``A robust scheme for feature-preserving mesh
  denoising,'' \emph{IEEE Transactions on Visualization and Computer Graphics},
  vol.~22, no.~3, pp. 1181--1194, March 2016.

\bibitem{Wei15b}
M.~Wei, J.~Yu, W.~M. Pang, J.~Wang, J.~Qin, L.~Liu, and P.~A. Heng, ``Bi-normal
  filtering for mesh denoising,'' \emph{IEEE Transactions on Visualization and
  Computer Graphics}, vol.~21, no.~1, pp. 43--55, Jan 2015.

\bibitem{icp}
P.~J. Besl and N.~D. McKay, ``A method for registration of 3-d shapes,''
  \emph{IEEE Transactions Pattern Analysis and Machine Intelligence}, vol.~14,
  no.~2, pp. 239--256, 1992.

\bibitem{online_learning_for_matrix_factor}
J.~Mairal, F.~Bach, J.~Ponce, and G.~Sapiro, ``Online learning for matrix
  factorization and sparse coding,'' \emph{J. Mach. Learn. Res.}, vol.~11, pp.
  19--60, 2010.

\bibitem{lars}
B.~Efron, T.~Hastie, I.~Johnstone, and R.~Tibshirani, ``Least angle
  regression,'' \emph{Annals of Statistics}, vol.~32, pp. 407--499, 2004.

\bibitem{screened_poisson}
M.~Kazhdan and H.~Hoppe, ``Screened poisson-surface reconstruction,'' \emph{ACM
  Transactions on Graphics}, 2013.

\bibitem{surf_rec}
H.~Hoppe, T.~DeRose, T.~Duchamp, J.~McDonald, and W.~Stuetzle, ``Surface
  reconstruction from unorganized points,'' in \emph{SIGGRAPH '92}, 1992, pp.
  71--78.

\bibitem{elad2006}
M.~Elad and M.~Aharon, ``Image denoising via sparse and redundant
  representations over learned dictionaries,'' \emph{IEEE Transactions on Image
  Processing}, vol.~15, no.~12, pp. 3736--3745, Dec 2006.

\bibitem{bilateral}
S.~Fleishman, I.~Drori, and D.~Cohen-Or, ``Bilateral mesh denoising,''
  \emph{ACM Transactions on Graphics}, vol.~22, no.~3, pp. 950--953, 2003.

\end{thebibliography}

\begin{IEEEbiography}[{\includegraphics[width=1in,height=1.25in,clip,keepaspectratio]{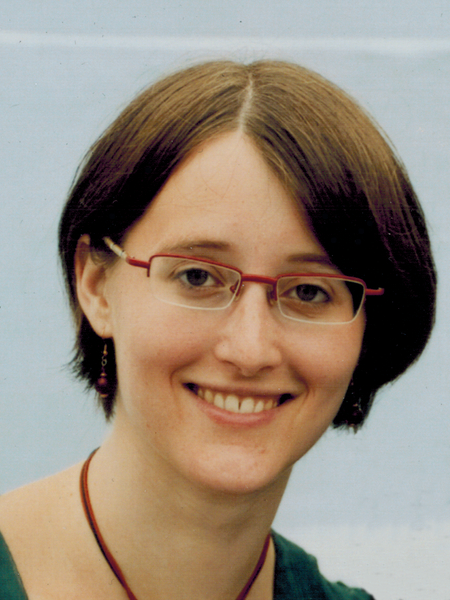}}]{Julie Digne}
Julie Digne is a junior CNRS researcher at LIRIS laboratory working in Geometry Processing. She obtained her PhD in 2010 in applied mathematics at Ecole Normale Supérieure de Cachan in 2010, which was awarded the Jacques Hadamard Foundation for Mathematics PhD award. She was a post-doc at INRIA. She joined the GeoMod team at LIRIS, University of Lyon in 2012, her current work focuses on point set surface processing including denoising and super-resolution.
\end{IEEEbiography}

\begin{IEEEbiography}[{\includegraphics[width=1in,height=1.25in,clip,keepaspectratio]{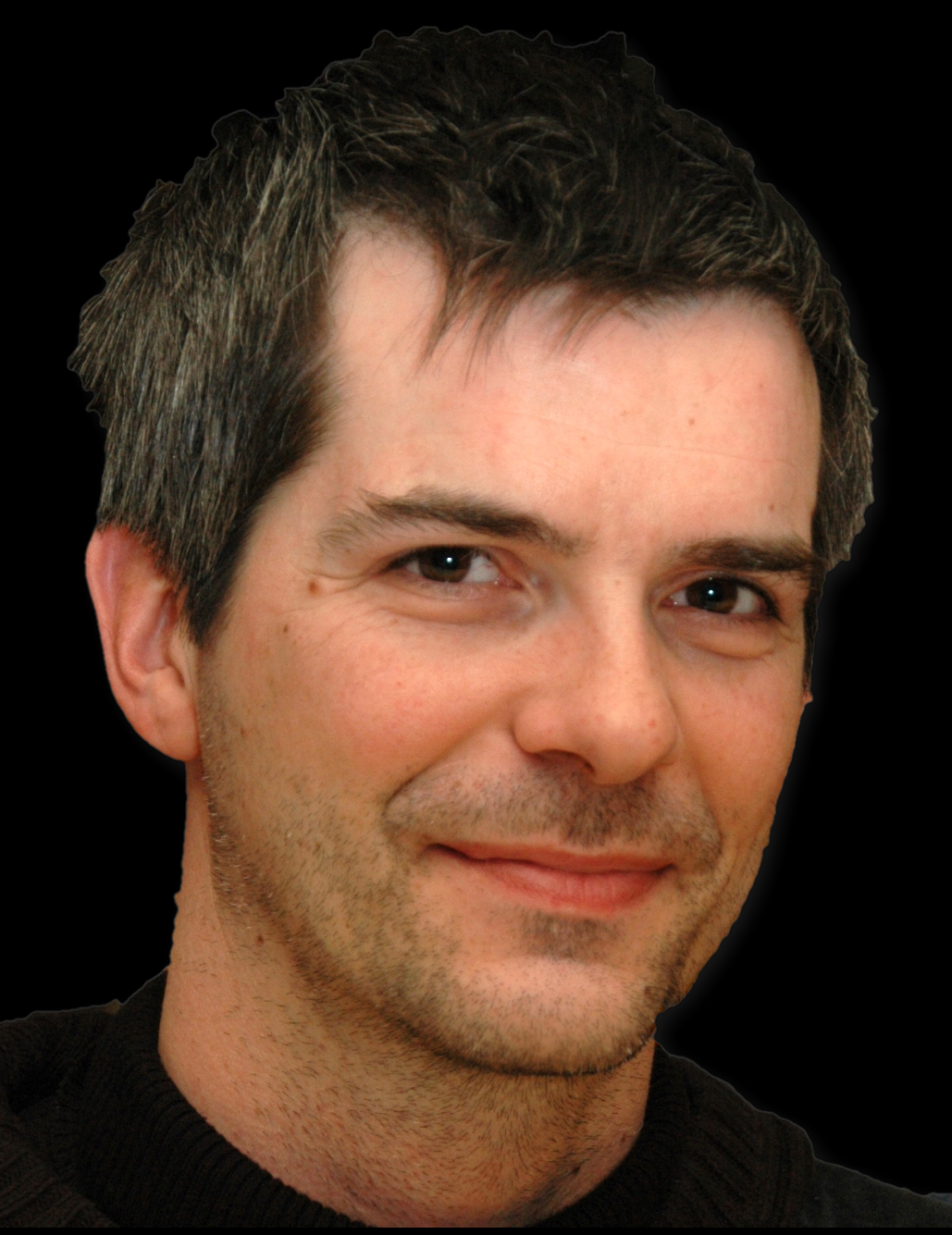}}]{S\'ebastien Valette}
S\'{e}bastien Valette is a CNRS Researcher at CREATIS, Lyon, France. He recieved the Electrical Engineering degree and PhD Degree at INSA-Lyon in 2002. His
research interests include geometry processing, data compression and image analysis.
\end{IEEEbiography}

\begin{IEEEbiography}[{\includegraphics[width=1in,height=1.25in,clip,keepaspectratio]{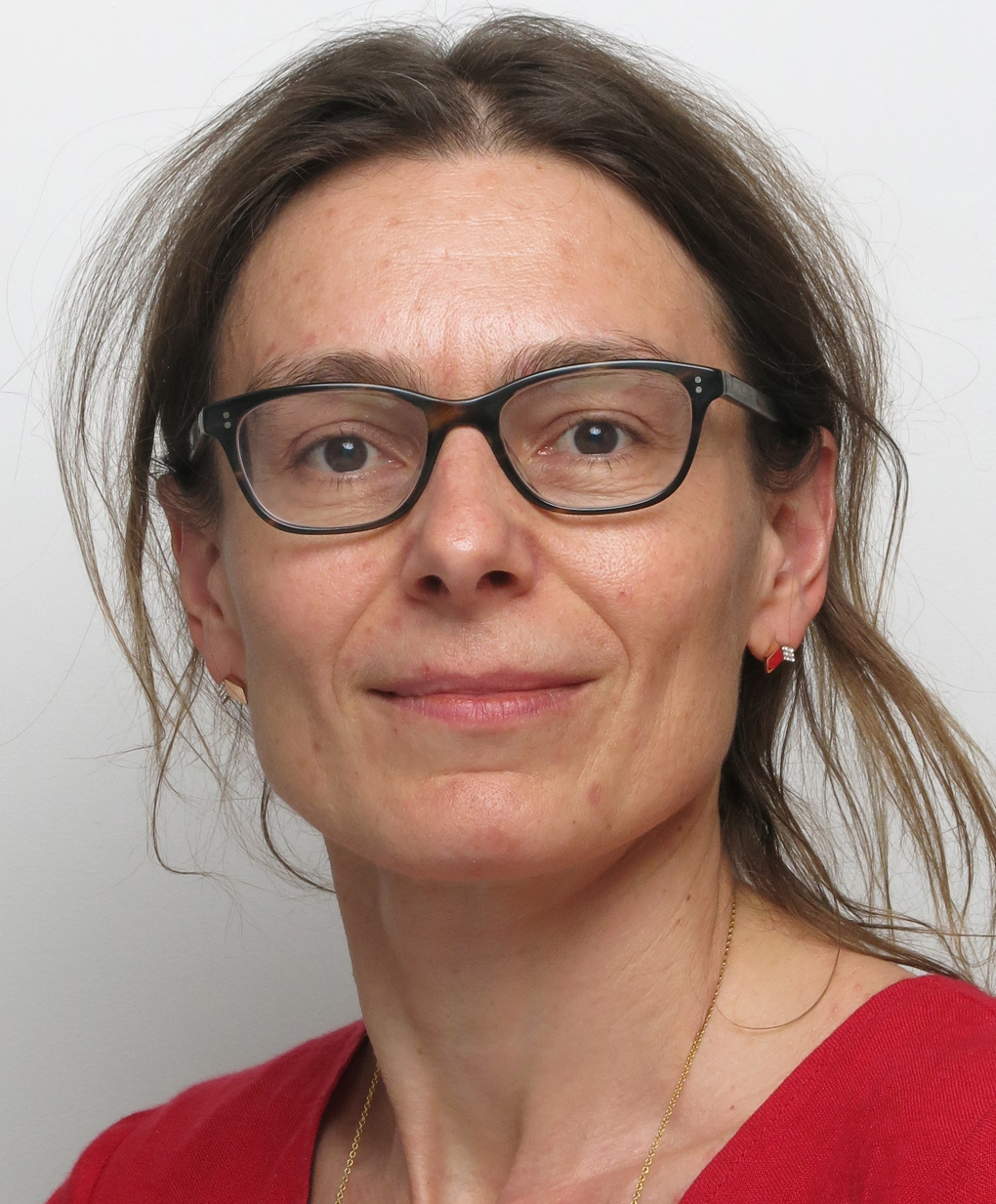}}]{Rapha\"elle Chaine}'s research interests include geometry processing, geometric modeling and virtual sculpture. She graduated with the M.S. Degree at Ecole Centrale de Lyon in 1995. After completing her PhD in computer science from the Universit\'e Claude Bernard in France, in the domain of point set surfaces analysis, she joined the PRISME Geometrica team at INRIA Sophia Antipolis in 2000. There, she focused her work on 3D reconstruction and computational geometry. In 2003 Rapha\"elle joined the LIRIS (CNRS) laboratory where she was promoted to a professor position in 2013 at Lyon 1 University.
\end{IEEEbiography}

\end{document}